\documentclass[10pt,aps,prd,superscriptaddress,nofootinbib,floatfix]{revtex4-2}

\usepackage{amsmath}
\usepackage{amssymb}
\usepackage{graphicx}
\usepackage{epstopdf}
\usepackage[utf8]{inputenc}
\usepackage{dcolumn}
\usepackage{color}

\definecolor{BlueViolet}{rgb}{0.2, 0.00, 0.7}
\definecolor{Blue}{rgb}{0.15, 0.00, 0.9}
\usepackage[colorlinks=true,linkcolor=Blue,citecolor=Blue,urlcolor=BlueViolet]{hyperref}
\usepackage[hyphenbreaks]{breakurl}

\newcommand{\be}{\begin{equation}}
\newcommand{\ee}{\end{equation}}
\newcommand{\bea}{\begin{eqnarray}}
\newcommand{\eea}{\end{eqnarray}}

\begin{document}

\title{Dispersive analysis of the experimental data on the electromagnetic form factor\\[2mm] of charged pions at spacelike momenta}

\author{Silvano Simula}
\email[]{silvano.simula@roma3.infn.it}
\affiliation{Istituto Nazionale di Fisica Nucleare, Sezione di Roma Tre, Via della Vasca Navale 84, I-00146 Rome, Italy}

\author{Ludovico Vittorio}
\email[]{ludovico.vittorio@lapth.cnrs.fr}
\affiliation{LAPTh, Université Savoie Mont-Blanc and CNRS, F-74941 Annecy, France}

\begin{abstract}
\noindent
The experimental data on the electromagnetic (em) form factor of charged pions available at spacelike momenta are analyzed using the Dispersive Matrix (DM) approach\,\cite{DiCarlo:2021dzg}, which describes the momentum dependence of hadronic form factors without introducing any explicit parameterization and includes properly the constraints coming from unitarity and analyticity. The unitary bound is evaluated nonperturbatively making use of the results of lattice QCD simulations of suitable two-point correlation functions contributing to the Hadronic Vacuum Polarization term of the muon. Thanks to the DM method we determine the pion charge radius from existing spacelike data in a completely model-independent way and consistently with the unitary bound, obtaining $\langle r_\pi \rangle_{DM} = 0.703 \pm 0.027$ fm. This finding differs by $\simeq 1.6$ standard deviations from the latest PDG\,\cite{ParticleDataGroup:2022pth} value $\langle r_\pi \rangle_{PDG} = 0.659 \pm 0.004$ fm, which is dominated by the very precise results of dispersive analyses of timelike data coming from measurements of the cross section  of the $e^+ e^- \to \pi^+ \pi^-$ process. We have analyzed the spacelike data using also traditional $z$-expansions, like the Boyd-Grinstein-Lebed  (BGL) or Bourrely-Caprini-Lellouch (BCL) fitting functions and adopting a simple procedure that incorporates {\it ab initio} the non-perturbative unitary bound in the fitting process. We get $\langle r_\pi \rangle_{BGL} = 0.711 \pm 0.039$ fm and $\langle r_\pi \rangle_{BCL} = 0.709 \pm 0.028$ fm in nice agreement with the DM result. A detailed comparison in a wide range of spacelike momenta between the results of the BGL/BCL fitting procedures and those of the DM method indicates that unitarity must be imposed not only on the fitting function but also on the input data.
We have addressed also the issue of the onset of perturbative QCD (pQCD) by performing a sensitivity study of the pion form factor at large spacelike momenta, based only on experimental spacelike data and unitarity. Hence, although the leading pQCD behaviour is found to set in only at very large momenta, our DM bands may provide information about the pre-asymptotic effects related to the scale dependence of the pion distribution amplitude.
\end{abstract}

\maketitle

\section{Introduction}

Since the pion is the lightest bound state in QCD, its physical properties carry important information about the way quark and gluon degrees of freedom govern the low-energy dynamics. Therefore, its precise determination represents an important test for our fundamental theory of the strong interactions and requires nonperturbative theoretical approaches, like QCD simulations on the lattice.
The electromagnetic (em) form factor of a (charged) pion, $F_\pi^V(Q^2)$, is defined in pure QCD by the matrix element
\be
    \label{eq:FVpion}
    \langle \pi^+(p^\prime) | J_\mu^{em} | \pi^+(p) \rangle = \left( p + p^\prime \right)_\mu F_\pi^V(Q^2) ~ , ~
\ee
where $q  = p - p^\prime$ is the 4-momentum transfer, $Q^2 \equiv- q^2$ and $J_\mu^{em}$ is the em current operator, namely
\be
    \label{eq:Jmu_em}
    J_\mu^{em}(x) = \sum_{f = u, d, s , ...} q_f \, \bar{\psi}_f(x) \gamma_\mu \psi_f(x)
\ee 
with $q_f$ being the electric charge of the quark with flavor $f$ in units of the electron charge.

For spacelike values of the squared 4-momentum transfer ($Q^2 \geq 0$ or, equivalently, $q^2 \leq 0$) the em pion form factor contains information on the distribution of its charged constituents, namely valence and sea light quarks, while for timelike values it has a branch cut starting at the annihilation threshold $4 M_\pi^2$. For $Q^2 \leq -4 M_\pi^2$ ($q^2 \geq 4 M_\pi^2$) it becomes complex and its modulus is a crucial quantity governing the $2 \pi$ contribution to the hadronic vacuum polarization (HVP) of the muon anomalous magnetic moment (see, e.g., Ref.\,\cite{Aoyama:2020ynm}). As well known, the muon HVP has long played an important role for testing the Standard Model of particle physics.

The experimental information on the em pion form factor is quite rich.
At spacelike values of $Q^2$ the form factor has been determined using electron-pion scattering experiments\,\cite{Dally:1982zk, NA7:1986vav, SELEX:2001fbx} and pion production off nucleons\,\cite{Bebek:1977pe, Ackermann:1977rp, Brauel:1979zk, JeffersonLabFpi:2000nlc, JeffersonLabFpi-2:2006ysh, JeffersonLabFpi:2007vir, JeffersonLab:2008jve, JeffersonLab:2008gyl}. 
In the timelike region the modulus of the pion form factor has been extensively measured using the cross section of the process $e^+ e^- \to \pi^+ \pi^-$ (see Ref.\,\cite{Aoyama:2020ynm} for a recent compilation) as well as data on the hadronic $\tau$ decays in the limit of isospin symmetry.
Concerning the extraction of the em pion form factor from experimental data and its analysis in terms of dispersion methods a consistent treatment of radiative corrections (due to both vacuum polarization and final-state radiation effects) must be guaranteed, as described in Refs.\,\cite{Ananthanarayan:2016mns, Colangelo:2018mtw}. In the case of the spacelike data the radiative corrections considered  in the experiments include already the subtraction of vacuum polarization effects\,\cite{Kahane:1964zz, Adylov:1977kj}. Thus, the dispersive treatment can be applied to the spacelike data for the em pion form factor without any adjustment.

An important quantity characterizing the em pion form factor is its slope at $Q^2 = 0$, more precisely the pion charge radius, $\langle r_\pi \rangle$, defined as
$\langle r_\pi \rangle \equiv \sqrt{\langle r_\pi^2 \rangle}$ with
\be
    \label{eq:rpi2}
     \langle r_\pi^2 \rangle \equiv -6 \frac{dF_\pi^V(Q^2)}{dQ^2} \Big|_{Q^2 = 0} = \frac{6}{\pi} \int_{4M_\pi^2}^\infty dt ~ \frac{\mbox{Im}F_\pi^V(t)}{t^2} ~ , ~
\ee
where the rightmost formula can be obtained via dispersion relations.
In the latest PDG review\,\cite{ParticleDataGroup:2022pth} the result for $\langle r_\pi \rangle$ reads
\be
    \label{eq:rpi_PDG}
    \langle r_\pi \rangle_{PDG} = 0.659 \pm 0.004 ~ \mbox{fm} ~ , ~
\ee
coming from an average of four different results: $\langle r_\pi \rangle = 0.656 \pm 0.005$ fm representing a suitable average of the analyses of timelike ($e^+ e^-$) and spacelike\,\cite{NA7:1986vav} data made in Refs.\,\cite{Colangelo:2018mtw, Ananthanarayan:2017efc}, $\langle r_\pi \rangle = 0.663 \pm 0.023$ fm using the spacelike data from the F2 experiment at FNAL\,\cite{Dally:1982zk}, $\langle r_\pi \rangle = 0.663 \pm 0.006$ fm using the spacelike data from the NA7 experiment at CERN\,\cite{NA7:1986vav} and $\langle r_\pi \rangle = 0.65 \pm 0.08$ fm using the spacelike data from the SELEX experiment at FNAL\,\cite{SELEX:2001fbx}.
The first result from Refs.\,\cite{Colangelo:2018mtw, Ananthanarayan:2017efc} is based on a dispersive representation of the em pion form factor, which properly satisfies unitarity and analyticity. On the contrary, the other three results, based only on spacelike data, are obtained by fitting the data with a simple monopole Ansatz, which may introduce a disturbing model dependence and may be inconsistent with unitarity (see similar remarks made in Ref.\,\cite{Masjuan:2008fv}, where Pad\'{e} approximants are employed).

The aim of the present work is to describe the $Q^2$-dependence of the experimental data on the em pion form factor at spacelike momenta without introducing any explicit parameterization and fulfilling at the same time the constraints coming from unitarity and analyticity. This will allow us to determine the pion charge radius from existing spacelike data in a completely model-independent way, while fulfilling unitarity.
This goal can be achieved by adopting the Dispersive Matrix (DM) approach developed in Ref.\,\cite{DiCarlo:2021dzg}, and already applied successfully to the description of the hadronic form factors relevant in semileptonic B-meson weak decays in Refs.\,\cite{Martinelli:2021frl, Martinelli:2021myh, Martinelli:2021onb, Martinelli:2022tte, Martinelli:2022xir}. In this work the unitary bound on $F_\pi^V(Q^2)$ will be imposed using for the first time a nonperturbative determination of the relevant transverse vector susceptibility obtained using the results of lattice QCD (LQCD) simulations of suitable two-point correlation function contributing also to the muon HVP. Our result is
\be
    \label{eq:rpi_DM_intro}
    \langle r_\pi \rangle_{DM} = 0.703 \pm 0.027 ~ \mbox{fm} ~ , ~
 \ee
 which differs by $\simeq 1.6$ standard deviations from the PDG value\,(\ref{eq:rpi_PDG}) with an uncertainty much larger (by a factor $\simeq 4.5$) than the one quoted in the experimental work of Ref.\,\cite{NA7:1986vav}.
 We have analyzed the spacelike data using also traditional $z$-expansions, like the renowned Boyd-Grinstein-Lebed  (BGL)\,\cite{Boyd:1997kz}  or Bourrely-Caprini-Lellouch (BCL)\,\cite{Bourrely:2008za} fitting functions and adopting a simple procedure that easily incorporates {\it ab initio} the non-perturbative unitary bound in the fitting process. We get $\langle r_\pi \rangle_{BGL} = 0.711 \pm 0.039$ fm and $\langle r_\pi \rangle_{BCL} = 0.709 \pm 0.028$ fm in nice agreement with the DM result\,(\ref{eq:rpi_DM_intro}).

The analysis of the em pion form factor using the basic features of the DM approach is not new at all. In Refs.\,\cite{Ananthanarayan:2012tn, Ananthanarayan:2013dpa, Ananthanarayan:2013zua, Ananthanarayan:2017efc, Ananthanarayan:2018nyx} the authors adopted a matrix approach similar to the DM one, characterized by the use of one timelike and one spacelike constraint at the same time, and by a subsequent suitable averaging procedure of the results corresponding to different pairs of input data. In this work we introduce a new procedure, the {\it unitary sampling procedure}, valid for any number of data points. In this way we demonstrate that the DM approach is an easy and very effective tool for analyzing even large sets of data points (more than 50 in this work) fulfilling exactly the unitarity and analyticity constraints.

The structure of this work is as follows.

In Section\,\ref{sec:DM} and Appendix\,\ref{sec:appA} we summarize the main features of the DM approach applied to the em pion form factor. In particular, we elucidate the meaning of the DM unitary filter, which allows to select in a model-independent way only the subset of input data that can be reproduced exactly by a unitary $z$-expansion. This feature is not guaranteed by approaches based on explicit $z$-expansions and it becomes more important as the impact of the unitary filter is more severe. In these approaches the attention is focused only on the fitting function and not also on the fitted data (either experimental or theoretical ones). Even if the fitting function is constructed to satisfy unitarity, the fitting procedure is applied to all input data regardless whether the latter ones satisfy unitarity or not (i.e., regardless whether the input data can be exactly reproduced by a unitary $z$-expansion). We point out that fitting non-unitary input data might introduce distortions in unitary $z$-expansions (see Section\,\ref{sec:BGLfit} for a numerical evidence). Up to our knowledge this potential problem is avoided only in the DM method.

In Section\,\ref{sec:susceptibility} we discuss the non-perturbative determination of the unitary bound used in this work, namely the transverse vector susceptibility $\chi_T$, obtained using the results of lattice QCD (LQCD) simulations of suitable two-point correlation functions contributing to the HVP term of the muon.

In Sections\,\ref{sec:JLABpi} and\,\ref{sec:sampling} we apply the DM method to the electroproduction JLAB-$\pi$ data\,\cite{JeffersonLab:2008jve}. Since the unitary bound turns out to be extremely selective as the number of data points increases, we develop an efficient procedure to generate a distribution of values for the pion form factor satisfying unitarity, i.e.~to get a set of {\it unitary} input data, valid for any number of data points. The unitary sampling procedure is described in detail in the case of the electroproduction JLAB-$\pi$ data and it can be easily generalized to any set of hadronic form factors, which must satisfy unitary bounds.

In Section\,\ref{sec:CERN+JLABpi} the unitary sampling method is applied to both the CERN~\cite{NA7:1986vav} and electroproduction JLAB-$\pi$~\cite{JeffersonLab:2008jve} data for a total of more than 50 data points. The DM band for the em pion form factor is positively compared with the results obtained in Ref.\,\cite{Colangelo:2018mtw} by means of a unitary analysis of both timelike $e^+ e^-$ and spacelike CERN data. A difference is observed at small values of $Q^2$, which translates into the value\,(\ref{eq:rpi_DM_intro}) of the pion charge radius w.r.t.~the result $\langle r_\pi \rangle = 0.655 \pm 0.003$ fm from Ref.\,\cite{Colangelo:2018mtw}. 

In Sections\,\ref{sec:BGLfit} and\,\ref{sec:BCLfit} we analyze the spacelike data using the BGL\,\cite{Boyd:1997kz} and BCL\,\cite{Bourrely:2008za} $z$-expansions, respectively.
We adopt a simple procedure that incorporates {\it ab initio} the non-perturbative unitary bound, described in Appendices\,\ref{sec:appB} and\,\ref{sec:appC}. A detailed comparison among the unitary BGL fitting procedure and the DM method is performed, showing explicitly that distortions are produced at large spacelike values of $Q^2$ by fitting non-unitary input data.

In Section\,\ref{sec:Q02} we investigate the role of the auxiliary quantity $\overline{Q}_0^2$, at which the transverse susceptibility $\chi_T(\overline{Q}_0^2)$ is evaluated, on the unitary DM filter and on the corresponding DM band for the em pion form factor.

In Section\,\ref{sec:pQCD} we address the issue of the onset of perturbative QCD (pQCD) at large spacelike values of $Q^2$. As well known, for $Q^2 \to \infty$ the leading behaviour of $F_\pi^V(Q^2)$ predicted by pQCD\,\cite{Lepage:1979zb, Efremov:1979qk, Chernyak:1977as, Farrar:1979aw} is given by $8 \pi f_\pi^2 \alpha_s(Q^2) / Q^2$, where $f_\pi \simeq 130$ MeV is the pion decay constant and $\alpha_s(Q^2)$ is the running strong coupling. We perform a sensitivity study and present the DM predictions for $Q^2 \gtrsim 5$ GeV$^2$ based only on unitarity and experimental data available at spacelike momenta. Although the leading pQCD behaviour is found to set in only at very large momenta, our DM bands may provide information about the pre-asymptotic effects related to the scale dependence of the pion distribution amplitude.

Our conclusions are summarized in Section\,\ref{sec:conclusions}.

We point out that the DM approach is equally well suited to be applied also to available results of LQCD calculations of $F_\pi^V(Q^2)$ and, more generally, to experimental plus LQCD data on $F_\pi^V(Q^2)$. In this work, since tensions are present among $e^+ e^-$ experiments (further exacerbated by the recent results from the CMD-3 Collaboration\,\cite{CMD-3:2023alj}), we are interested in the analysis of the experimental data available at spacelike momenta without any mixing with timelike data, allowing in this way an interesting comparison with the results of Ref.\,\cite{Colangelo:2018mtw}, which are based almost totally on timelike data. We leave the DM analysis of LQCD data as well as of timelike plus spacelike data to future separate works.

\section{The DM approach for the em pion form factor}
\label{sec:DM}

The DM approach is a non-perturbative method for computing hadronic form factors in a model-independent way in their full kinematical range\,\cite{DiCarlo:2021dzg, Lellouch:1995yv}.

The starting point is a dispersive bound that, for a generic form factor $f$, can be written as~\cite{Boyd:1994tt, Boyd:1997kz, Caprini:1997mu} 
\be
    \label{eq:JQ2z}
    \frac{1}{2\pi i } \oint_{\vert z\vert =1} \frac{dz}{z}   \vert\phi(z) f(z)\vert^2 \leq \chi\, ,
\ee
where $\phi(z)$ is a kinematical function dependent on the specific spin-parity channel and $\chi$ is the so-called susceptibility, related to the derivative of the Fourier transform of a suitable Green function of bilinear quark operators~\cite{Boyd:1997kz}. The conformal variable $z(t)$ is  defined as
\be
    \label{eq:conformal}
    z(t) = \frac{\sqrt{t_+ - t} - \sqrt{t_+ - t_0}}{\sqrt{t_+ - t} + \sqrt{t_+ - t_0}} ~ , ~
\ee
where $t = q^2 = - Q^2$ is the squared 4-momentum transfer and, for the case of interest in this work, $t_+ = 4 M_\pi^2$ and $t_0 = 0$\footnote{We anticipate here that the DM band for the form factor $f(z)$, given by Eqs.\,(\ref{eq:bounds})-(\ref{eq:di_final}), at a generic value of $z$ does not depend upon the value of the variable $t_0$ (see the proof in Appendix\,\ref{sec:appA}). This is at variance with BGL or BCL $z$-expansions and it is consistent with the fact that the DM method does not use explicitly any $z$-expansion.}.

In the case of sub-threshold bound-state poles located at $t_i= M_{R_i}^2 < t_+$, the requirement of analyticity can be fulfilled by modifying the kinematical function $\phi(z)$ through the so-called Blaschke factors, namely\,\cite{Lellouch:1995yv}
\be
    \phi(z) \to \phi(z) \cdot \prod_i \frac{z - z(M_{R_i}^2)}{1 - \bar{z}(M_{R_i}^2) z} ~ , ~ \nonumber
\ee
where $\bar{z}(t)$ is the complex conjugate of the conformal variable $z(t)$. In the case of the em pion form factor no sub-threshold pole is present.

By introducing the inner product~\cite{Bourrely:1980gp, Lellouch:1995yv}
\be
     \label{eq:inpro}
    \langle g\vert h\rangle =\frac{1}{2\pi i } \oint_{\vert z\vert=1 } \frac{dz}{z}   \bar {g}(z) h(z)\, , \nonumber
\ee
where $\bar{g}(z)$ is the  complex conjugate of the function $g(z)$, Eq.~(\ref{eq:JQ2z}) can be also written as
\be
    \label{eq:JQinpro}
    0 \leq \langle \phi f \vert \phi  f\rangle \leq \chi\, .
\ee
Following Refs.~\cite{Bourrely:1980gp,Lellouch:1995yv} we introduce the set of functions
\be
g_t(z) \equiv \frac{1}{1-\bar{z}(t) z}\, , \nonumber
\ee
so that the use of  Cauchy's theorem yields
\bea
\langle g_t | \phi f \rangle  & = & \phi(z(t))\, f\left(z(t)\right)\, , \nonumber \\[2mm]
\langle g_{t_m} | g_{t_l} \rangle  & = & \frac{1}{1- \bar{z}(t_l) z(t_m)} \, . \nonumber
\eea
The central ingredient of the DM method is the matrix~\cite{Bourrely:1980gp,Lellouch:1995yv}
\be
\label{eq:matrix}
\mathbf{M} \equiv \left(
\begin{array}{ccccc}
\langle\phi f | \phi f \rangle  & \langle\phi f | g_t \rangle  & \langle\phi f | g_{t_1} \rangle  &\cdots & \langle\phi f | g_{t_N}\rangle  \\[2mm]
\langle g_t | \phi f \rangle  & \langle g_t |  g_t \rangle  & \langle  g_t | g_{t_1} \rangle  &\cdots & \langle g_t | g_{t_N}\rangle  \\[2mm]
\langle g_{t_1} | \phi f \rangle  & \langle g_{t_1} | g_t \rangle  & \langle g_{t_1} | g_{t_1} \rangle  &\cdots & \langle g_{t_1} | g_{t_N}\rangle  \\[2mm]
\vdots & \vdots & \vdots & \vdots & \vdots \\[2mm] 
\langle g_{t_N} | \phi f \rangle  & \langle g_{t_N} | g_t \rangle  & \langle g_{t_N} | g_{t_1} \rangle  &\cdots & \langle g_{t_N} | g_{t_N} \rangle 
\end{array} \right)  ~ , ~
\ee
where $t_1, \ldots, t_N$ are the values of the squared 4-momentum transfer at which the form factor $f(z)$ is known. 
Note that the DM method can be applied not only to a series of theoretical values $f(z(t_i))$ (with $i = 1, 2, ... N$), but also directly to experimental data (as done in this work and in Ref.\,\cite{Simula:2021yvm}). 

The important feature of the matrix $\mathbf{M}$ is that, thanks to the positivity of the inner products, its determinant is positive semidefinite, i.e.\,$\det \mathbf{M} \geq 0$. This property is not modified when the matrix element $\langle\phi f | \phi f \rangle$ is replaced by the upper bound given by the susceptibility $\chi$ through Eq.~(\ref{eq:JQinpro}). Thus, the original matrix~(\ref{eq:matrix}) can  be replaced  by
\be
\label{eq:matrix_final}
\mathbf{M}_{\chi} = \left( 
\begin{tabular}{ccccc}
   $\chi$ & $\phi f$                            & $\phi_1 f_1$                             & $...$ & $\phi_N f_N$ \\[2mm]
   $\phi f$     & $\frac{1}{1 - z^2}$     & $\frac{1}{1 - z z_1}$      & $...$ & $\frac{1}{1 - z z_N}$ \\[2mm]
   $\phi_1 f_1$ & $\frac{1}{1 - z_1 z}$  & $\frac{1}{1 - z_1^2}$     & $...$ & $\frac{1}{1 - z_1 z_N}$ \\[2mm]
   $... $  & $...$                           & $...$                              & $...$ & $...$ \\[2mm]
   $\phi_N f_N$ & $\frac{1}{1 - z_N z}$ & $\frac{1}{1 - z_N z_1}$ & $...$ & $\frac{1}{1 - z_N^2}$
\end{tabular}
\right) ~ , ~
\ee
where $\phi_i f_i \equiv \phi(z_i) f(z_i)$ (with $i = 1, 2, ... N$) represent the known values of $\phi(z) f(z)$ corresponding to the given set of values $z_i$.

By imposing the positivity of the determinant of the matrix~(\ref{eq:matrix_final}) it is possible to explicitly compute the lower and upper bounds that unitarity imposes on the form factor $f(z)$ for a generic value of $z$ on the real axis, namely~\cite{DiCarlo:2021dzg}
\be
    \label{eq:bounds}
    \beta(z) - \sqrt{\gamma(z)} \leq f(z) \leq \beta(z) + \sqrt{\gamma(z)} ~ , ~
\ee 
where
\bea
      \label{eq:beta_final}
      \beta(z) & \equiv & \frac{1}{\phi(z) d(z)} \sum_{i = 1}^N \phi_i f_i d_i \frac{1 - z_i^2}{z - z_i} ~ , ~ \\
      \label{eq:gamma_final}
      \gamma(z) & \equiv &  \frac{1}{1 - z^2} \frac{1}{\phi^2(z) d^2(z)} \left( \chi - \chi_\text{DM} \right) ~ , ~ \\
      \label{eq:chi0_final}
      \chi_\text{DM} & \equiv & \sum_{i, j = 1}^N \phi_i f_i \phi_j  f_j d_i d_j \frac{(1 - z_i^2) (1 - z_j^2)}{1 - z_i z_j} ~ , ~ \\
      \label{eq:dz_final}
     d(z) & \equiv & \prod_{m = 1}^N \frac{1 - z z_m}{z - z_m}  ~ , ~ \\
     \label{eq:di_final}
     d_i & \equiv & \prod_{m \neq i = 1}^N \frac{1 - z_i z_m}{z_i- z_m}  ~ . ~ 
\eea

When $z \to z_i$ one has $d(z) \propto 1 / (z - z_i)$ and, therefore, $\beta(z) \to f_i$ and $\gamma (z) \to 0$. In other words, Eq.\,(\ref{eq:bounds}) exactly reproduces the set of input data $\{ f_i \}$. In a frequentist language this corresponds to a vanishing value of the $\chi^2$-variable.

Unitarity is satisfied only when $\gamma(z) \geq 0$, which implies the condition $\chi \geq \chi_\text{DM}$. Such a condition depends on the set of input data $\{ f_i \}$ and it is  independent on any parameterization or fitting Ansatz of the input data. 

The meaning of the DM {\em filter} $\chi \geq \chi_\text{DM}$ is clearer in terms of explicit $z$-expansions, like the BGL ones~\cite{Boyd:1997kz}.
When $\chi \geq \chi_\text{DM}$, it is guaranteed the existence of (at least) one BGL fit (either truncated or untruncated) that satisfies unitarity and, at the same time, reproduces exactly the input data.
On the contrary, when $\chi < \chi_\text{DM}$, a unitary $z$-expansion passing through the data does not exist, since the {\em input data} do not satisfy unitarity.
The important feature of the DM approach is that only the {\it unitary} input data are eligible for consideration, while those data that do not satisfy the unitary filter $\chi \geq \chi_\text{DM}$ are discarded. 

We want to elucidate better the relevance of the above feature of the DM approach. 
Let us consider a sample of input data corresponding to known values of the form factor at a series of points $z_i$ generated according to a given covariance matrix.
For each event we can apply the DM filter $\chi \geq \chi_\text{DM}$ and, consequently, we can divide the original sample into two disjoint subsets: the one corresponding to input data satisfying the DM filter and the one made of non-unitary events. 
In what follows we will refer to the first subset as the {\it unitary} input data and to the second one as the {\it non-unitary} input data.
%We stress again that within the DM approach only the unitary input data are considered, while the subset of input data that do not satisfy the unitary filter $\chi \geq \chi_\text{DM}$ is discarded. 

The DM approach provides a band of values for the form factor $f(z)$ which are consistent with unitarity without making use of any explicit fitting Ansatz.
More precisely, the DM band is given by the convolution of the uniform distribution corresponding to Eqs.~(\ref{eq:bounds})-(\ref{eq:gamma_final}) with the distribution of the unitary input  data $\{ f_j \}$, i.e.\,only those fulfilling the condition $\chi \geq \chi_\text{DM}$ (see later Section\,\ref{sec:sampling}). 
In other words, the DM approach automatically provides the {\em envelope} of the results of all possible (either truncated or untruncated) $z$-expansions, which satisfy unitarity and at the same time exactly reproduce the unitary input data. 

We stress again that separating the input data in the two disjoint subsets corresponding to either $\chi \geq \chi_\text{DM}$ or $\chi < \chi_\text{DM}$ is an important feature of the DM approach, which is not guaranteed by approaches based on explicit $z$-expansions (including the one of Ref.\,\cite{Bigi:2017njr} and also the recent Bayesian approach of Ref.\,\cite{Flynn:2023qmi}). Indeed, in these approaches the attention is focused only on the fitting function and not also on the fitted data (either experimental or theoretical ones). Even if the fitting function is constructed to satisfy unitarity, the fitting procedure is applied to all the input data regardless whether they satisfy unitarity or not (i.e., regardless whether the input data can be exactly reproduced by a unitary $z$-expansion).
In the case of the unitary subset of input data it is always possible to find a suitable BGL fit, that satisfies unitarity and at the same time exactly reproduces the input data. This corresponds to the possibility to reach a null value of the $\chi^2$-variable by increasing the order of the truncation of the BGL fit (up to the number of data points).
On the contrary, when the input data do not satisfy the unitary filter, it is not possible to find a fitting $z$-expansion that satisfies unitarity and at the same time exactly reproduces the input data. This corresponds to a non-vanishing value of the $\chi^2$-variable, which depends on the impact of the non-unitary  input data.
The above considerations applies equally well also to the case of explicit $z$-expansions like the BCL ones\,\cite{Bourrely:2008za}.

It is clear that the application of a fitting function (even if unitary) to a subset of input data that do not satisfy unitarity may lead to a distortion of the fitting results related directly to the impact of the non-unitary effects present in the input data. In particular, such a distortion may be relevant when the fitting function extrapolates the form factor in a kinematical region not covered by the input data. 
Thus, the application of the DM method is simpler and more general w.r.t.~other approaches, like the BGL or BCL $z$-expansions, particularly when the number of input data increases and the unitary constraint becomes more selective. In these cases a BGL or BCL truncated expansion would require to take into account a large number of fitting parameters with no guarantee of avoiding the non-unitary effects possibly present in the input data.
This issue will become evident in Sections\,\ref{sec:BGLfit} and \ref{sec:BCLfit}, where we apply unitary BGL or BCL approaches to analyze the spacelike data for the pion form factor. In Section\,\ref{sec:BGLfit} we address explicitly the issue of the impact of non-unitary input data on a unitary BGL fit.

Let us now consider explicitly the case of the em pion form factor assuming that it is known for a series of ($N + 1$) values $Q_i^2$, namely
\be
    \label{eq:data}
    F_i \equiv F_\pi^V(Q_i^2) \qquad \qquad \qquad \mbox{for i = 0, 1, ..., N} ~ , ~
\ee
where we have added the value $i = 0$ to include the absolute normalization condition $F_\pi^V(Q_0^2 = 0) = 1$.
The susceptibility relevant for the em pion form factor is the one of the transverse vector channel $\chi_T$ (see next Section for its explicit definition) and we introduce an auxiliary variable $\overline{Q}_0^2$ at which the susceptibility $\chi_T$ is evaluated.

Denoting by $z_i$ the value of the conformal $z$-variable corresponding to $Q_i^2$, i.e.~
\be
    \label{eq:zetai}
    z_i \equiv \frac{\sqrt{1 + Q_i^2 / 4M_\pi^2 } - 1}{\sqrt{1 + Q_i^2 / 4M_\pi^2 } + 1} \simeq \frac{Q_i^2}{16 M_\pi^2} + 
        {\cal{O}}\left( \frac{Q_i^4}{M_\pi^4} \right) ~ , ~ \nonumber 
\ee
the constraint due to unitarity and analyticity on the values $F_i$ can be written in the form
\be
    \label{eq:filter}
    4 M_\pi^2 \chi_T(\overline{Q}_0^2) \geq \chi_{DM}(\overline{Q}_0^2) ~ , ~
\ee
where 
\bea
     \label{eq:chiDM}
     \chi_{DM}(\overline{Q}_0^2) & = & \sum_{i, j = 0}^N F_i F_j \frac{\phi_i(\overline{Q}_0^2) d_i (1 - z_i^2) ~ 
                                                            \phi_j(\overline{Q}_0^2) d_j (1 - z_j^2)}{1 - z_i z_j} ~ , ~ \\[2mm]
     \label{eq:di}
     d_i & = &  \prod_{m \neq i = 0}^N \frac{1 - z_i z_m}{z_i - z_m} ~ 
\eea
and the kinematical factor $\phi_i(\overline{Q}_0^2)$ is explicitly given by~\cite{Boyd:1997kz, Buck:1998kp}
\bea
    \label{eq:phi}
    \phi_i(\overline{Q}_0^2) & = & \frac{1}{\sqrt{48 \pi}} \frac{1 + \frac{Q_i^2}{4M_\pi^2}}{\left[ 1 + \sqrt{1+\frac{Q_i^2}{4M_\pi^2}} \right]^{5/2}} ~ 
                                                    \left[ \frac{1 + \sqrt{1+\frac{Q_i^2}{4M_\pi^2}} }{\sqrt{1+\frac{\overline{Q}_0^2}{4M_\pi^2}} + 
                                                    \sqrt{1+\frac{Q_i^2}{4M_\pi^2}} } \right]^3 ~  \nonumber \\[2mm]
                                           & = & \frac{1}{\sqrt{1536 \pi}} (1 + z_i)^2 \sqrt{1 - z_i} \left( \frac{1 - \overline{z}_0}{1 - \overline{z}_0 z_i} \right)^3 ~
\eea
with
\be
    \label{eq:z0bar}
    \overline{z}_0 = \frac{\sqrt{1 + \overline{Q}_0^2 / 4M_\pi^2 } - 1}{\sqrt{1 + \overline{Q}_0^2 / 4M_\pi^2 } + 1} ~ . ~
\ee
For a generic value of $z$ on the real axis, when the unitary filter\,(\ref{eq:filter}) is satisfied, the pion form factor $F_\pi^V(z)$ is limited by the bounds
\bea
      \label{eq:bounds_pion}
      && \beta(z) - \sqrt{\gamma(z)} \leq F_\pi^V(z) \leq \beta(z) + \sqrt{\gamma(z)} ~ , ~ \\[2mm]
      \label{eq:beta_pion}
      && \beta(z) = \frac{1}{\phi(z, \overline{Q}_0^2) d(z)} \sum_{i = 0}^N \phi_i F_i d_i \frac{1 - z_i^2}{z - z_i} ~ , ~ \\[2mm]
      \label{eq:gamma_pion}
      && \gamma(z) = \frac{1}{(1 - z^2) \phi^2(z, \overline{Q}_0^2) d^2(z)} \left[ 4 M_\pi^2 \chi_T(\overline{Q}_0^2) - 
                                \chi_\text{DM}(\overline{Q}_0^2) \right] ~ , ~
\eea
where
\bea
     \label{eq:dz}
     d(z) & = & \prod_{m = 0}^N \frac{1 - z z_m}{z - z_m}  ~ , ~ \\
      \label{eq:phiz}    
      \phi(z, \overline{Q}_0^2) & = & \frac{1}{\sqrt{1536 \pi}} (1 + z)^2 \sqrt{1 - z} \left( \frac{1 - \overline{z}_0}{1 - \overline{z}_0 z} \right)^3 ~ , ~
\eea
while $\chi_\text{DM}(\overline{Q}_0^2)$ and $\overline{z}_0$ are given by Eqs.\,(\ref{eq:chiDM}) and (\ref{eq:z0bar}), respectively.

We stress once more that the important feature of the DM method is the possibility to predict the value of the form factor $F_\pi^V(Q^2)$ at a generic value of $Q^2$ using only the knowledge of the pion form factor at the series of values $Q_i^2$ without any reference to a specific parameterization, provided the unitary filter~(\ref{eq:filter}) is fulfilled.
We will describe the DM procedure in detail later on in Section~\ref{sec:sampling}, while in the next Section we address the non-perturbative determination of the  transverse vector susceptibility  $\chi_T(\overline{Q}_0^2)$.

\section{Non-perturbative determination of the transverse vector susceptibility $\chi_T(\overline{Q}_0^2)$}
\label{sec:susceptibility}

In QCD the transverse vector susceptibility $\chi_T(\overline{Q}_0^2)$ is given by~\cite{DiCarlo:2021dzg} 
\be
    \label{eq:chiT}
    \chi_T(\overline{Q}_0^2) \equiv \frac{1}{4} \int_0^\infty d\tau \tau^4 \frac{j_1(\overline{Q}_0 \tau)}{\overline{Q}_0 \tau} V_{2\pi}(\tau) ~ ,
\ee
where $V_{2\pi}(\tau)$ is the $2\pi$ contribution to the Euclidean vector-vector current correlator $V(\tau)$, $\tau$ is the Euclidean time distance and $j_1(x)$ is an ordinary Bessel spherical function.
Note that at $\overline{Q}_0^2 = 0$ the susceptibility $\chi_T(\overline{Q}_0^2 = 0)$ is proportional to the fourth moment of the correlator $V_{2\pi}(\tau)$, which contributes to the Hadronic Vacuum Polarization term of the muon ($g-2$) (see Ref.~\cite{Aoyama:2020ynm}).

As well known, the Euclidean correlator $V(\tau)$ can be obtained by taking the Fourier transform of the spatial components of the HVP tensor, which in turn is related via dispersion  relations to the (one photon) $e^+ e^-$ annihilation cross section into hadrons, namely (see, e.g., Ref.\,\cite{Giusti:2017jof})
\be
    \label{eq:Vtau}
    V(\tau) = \frac{1}{12 \pi^2} \int_{2M_\pi}^\infty d\omega \omega^2 R_{had}(\omega) e^{- \omega \tau} ~ , ~
\ee
where
\be
   \label{eq:Rhad}
   R_{had}(\omega) = \frac{3 \omega^2}{4 \pi \alpha_{em}^2} \sigma_{had}(\omega)
\ee
with $\omega$ being the center-of-mass energy.

In QCD, neglecting the electron mass, the $2\pi$ contribution $ R_{2\pi}(\omega)$ to $R_{had}(\omega)$ is given by (see, e.g., Ref.\,\cite{Colangelo:2018mtw}) 
\be
    R_{2\pi}(\omega) = \frac{1}{4} \left( 1 - \frac{4 M_\pi^2}{\omega^2}\right)^{3/2} | F_\pi^V(\omega) |^2 ~ , ~
\ee
where $F_\pi^V(\omega)$ is the em pion form factor in the time-like region $q^2 = \omega^2 \geq 4M_\pi^2$.
The $2\pi$ contribution to the Euclidean correlator reads as
\be
    \label{eq:V2pi}
    V_{2\pi}(\tau) = \frac{1}{48\pi^2} \int_{2M_\pi}^\infty d \omega \, \omega^2 \left( 1 - \frac{4M_\pi^2}{\omega^2} \right)^{3/2}
                             |F_\pi^V(\omega)|^2 \, e^{-\omega \tau} ~ . ~
\ee
Since
\be
      \frac{1}{4} \int_0^\infty d\tau \tau^4 \frac{j_1(\overline{Q}_0 \tau)}{\overline{Q}_0 \tau} = 
      \frac{2\omega}{\left( \overline{Q}_0^2 + \omega^2 \right)^3} ~ , ~
\ee
the transverse vector susceptibility $\chi_T(\overline{Q}_0^2)$ is given by
\be
    \label{eq:chiT_FV}
    \chi_T(\overline{Q}_0^2) = \frac{1}{24\pi^2}  \int_{2M_\pi}^\infty d \omega \, \omega^{-3}  \left( 1 - \frac{4M_\pi^2}{\omega^2} \right)^{3/2} 
                                               \frac{1}{\left( 1+\overline{Q}_0^2 / \omega^2 \right)^3} \, |F_\pi^V(\omega)|^2 ~ . ~
\ee
Note that:
\begin{itemize}

\item the integrand in the r.h.s.~of Eq.\,(\ref{eq:chiT_FV}) is positive definite at all energies;

\item for large values of $\overline{Q}_0^2$ the transverse susceptibility drops down as fast as $1 / \overline{Q}_0^6$
\be
     \label{eq:chiT_FV_asympt}
      \overline{Q}_0^6 ~ \chi_T(\overline{Q}_0^2) ~_{\overrightarrow{\overline{Q}_0^2 \to \infty} } ~ \frac{1}{24\pi^2}  
          \int_{2M_\pi}^\infty d \omega \, \omega^3  \left( 1 - \frac{4M_\pi^2}{\omega^2} \right)^{3/2} \, |F_\pi^V(\omega)|^2 ~ ; ~
\ee

 \item Eq.\,(\ref{eq:chiT_FV}) can be cast in the form of the unitary bound\,(\ref{eq:JQ2z}) as a strict equality. 
Indeed, the relation between the conformal variable $z$ on the unit circle $|z| = 1$ and the center-of-mass energy $\omega$ is 
\be
    z = e^{i \alpha} = \frac{i \sqrt{\omega^2 / 4 M_\pi^2 - 1} - 1}{i \sqrt{\omega^2 / 4 M_\pi^2 - 1} + 1} ~ , ~ 
\ee
so that one has $\omega = 2 M_\pi \sqrt{2 / (1 - \mbox{cos}\alpha)}$. It follows that
\bea
     4M_\pi^2 \chi_T(\overline{Q}_0^2) & = & \frac{1}{48 \pi^2} \int_{-\pi}^{+\pi} d\alpha \, \frac{\left| \mbox{sin}\alpha \right|}{4} 
                                                                      \left[ \frac{1 + \mbox{cos}\alpha}{2} \right]^{3/2} 
                                                                      \frac{|F_\pi^V|^2}{\left[ 1 + \frac{\overline{Q}_0^2}{8M_\pi^2}  (1 - \mbox{cos}\alpha) \right]^3} 
                                                                      \quad \nonumber \\[2mm]
                                                            & = & \frac{1}{2\pi i} \int_{|z|=1} \frac{dz}{z} \frac{1}{1536 \pi} \left| (1 + z)^4 (1 - z) 
                                                                      \left( \frac{1 - \overline{z}_0}{1 - \overline{z}_0 z} \right)^6 \right| |F_\pi^V|^2 \nonumber \\[2mm]
                                                            & = & \frac{1}{2\pi i} \int_{|z|=1} \frac{dz}{z} \left| \phi(z, \overline{Q}_0^2) F_\pi^V \right|^2 ~ , ~
\eea
where $\phi(z, \overline{Q}_0^2)$ is the kinematical function given in Eq.\,(\ref{eq:phiz}).

\end{itemize}

\begin{figure}[htb!]
\begin{center}
\includegraphics[scale=0.60]{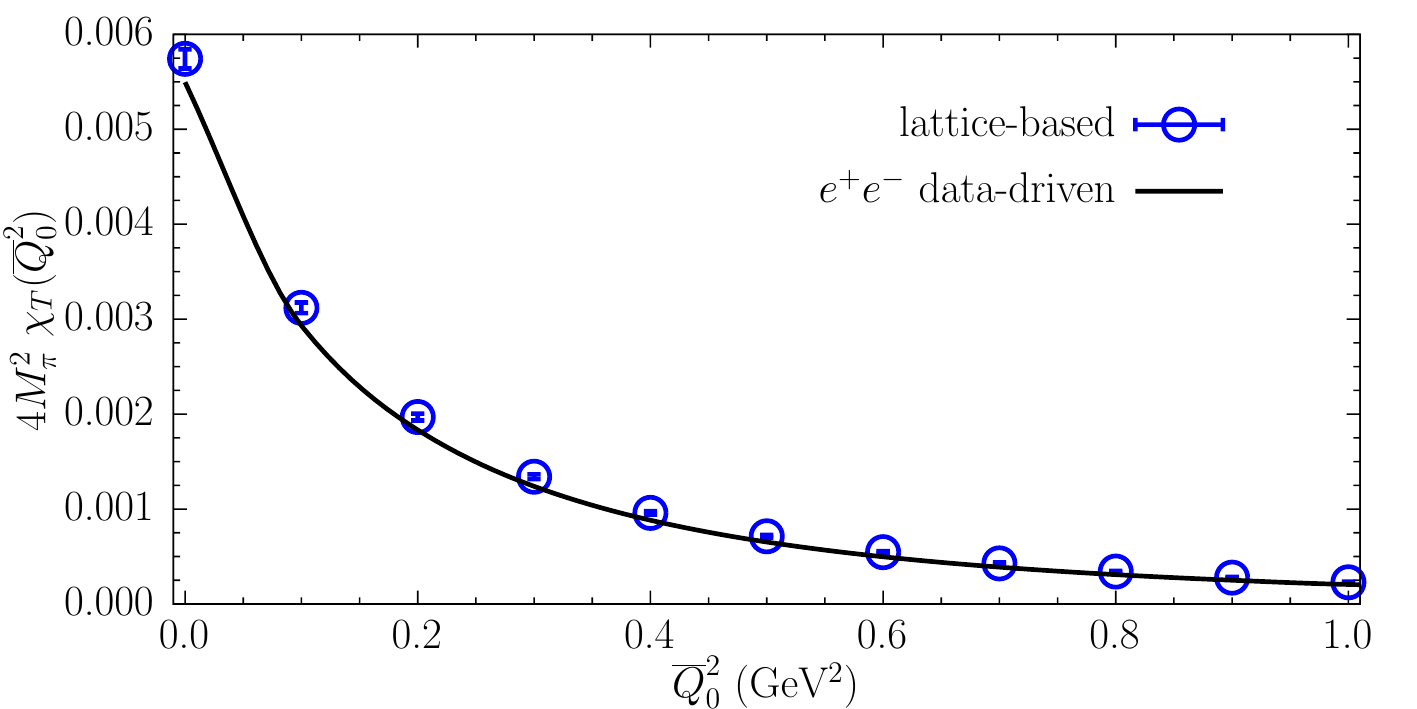}
\end{center}
\vspace{-0.5cm}
\caption{\it \small Blue dots: transverse vector susceptibility $4M_\pi^2 \chi_T(\overline{Q}_0^2)$ versus $\overline{Q}_0^2$ evaluated using Eq.~(\ref{eq:chiT}) and adopting the $2\pi$ correlator $V_{2\pi}(\tau)$ obtained in Ref.~\cite{Giusti:2018mdh} using LQCD simulations (see text). Black line: the susceptibility $4M_\pi^2 \chi_T(\overline{Q}_0^2)$ evaluated using Eq.\,(\ref{eq:chiT_FV}) adopting for $|F_\pi^V(\omega)|$ the results of the dispersive analysis of the $e^+ e^-$ data available from Ref.\,\cite{Colangelo:2018mtw} up to $\omega = 1$ GeV and putting $|F_\pi^V(\omega)| = 0$ for $\omega > 1$ GeV. The uncertainties of the data-driven results are hardly visible on the vertical scale being of the order of $\simeq 0.6 \%$.}.
\label{fig:chiT}
\end{figure}

The Euclidean correlator of two em currents has been evaluated on the lattice by several collaborations (see, e.g., Ref.~\cite{Aoyama:2020ynm}). In particular, the $2\pi$ contribution $V_{2\pi}(\tau)$ has been estimated at the physical point and in the continuum and infinite volume limits in Ref.~\cite{Giusti:2018mdh}\footnote{There, LQCD simulations of the light-quark vector correlator have been performed in isosymmetric QCD (isoQCD). The results at finite lattice volumes have been fitted using the so-called L\"uscher-Lellouch-Meyer model~\cite{Luscher:1985dn, Luscher:1986pf, Luscher:1990ux, Luscher:1991cf, Lellouch:2000pv, Meyer:2011um, Francis:2013fzp} adopting the Gounaris-Sakurai parameterization\,\cite{Gounaris:1968mw} for the pion form factor $F_\pi^V$.}. 
Thus, using the lattice-based correlator $V_{2\pi}(\tau)$ we have evaluated the quantity $4M_\pi^2 \chi_T(\overline{Q}_0^2)$ using Eq.\,(\ref{eq:chiT}). The $\overline{Q}_0^2$ dependence obtained in this way is shown in Fig.~\ref{fig:chiT} by the blue dots. 
Alternatively, we have calculated Eq.\,(\ref{eq:chiT_FV}) adopting for $|F_\pi^V(\omega)|$ the results of the dispersive analysis of the $e^+ e^-$ data available from Ref.\,\cite{Colangelo:2018mtw} up to $\omega = 1$ GeV and putting $|F_\pi^V(\omega)| = 0$ for $\omega > 1$ GeV. We will refer to the results obtained in this way for the susceptibility $4M_\pi^2 \chi_T(\overline{Q}_0^2)$ as the data-driven ones, represented in Fig.~\ref{fig:chiT} by the black line. 

Since the pion form factor of Ref.\,\cite{Colangelo:2018mtw} is provided up to $\omega = 1$ GeV and the integrand in the r.h.s.~of Eq.\,(\ref{eq:chiT_FV}) is positive definite, the data-driven results for $4M_\pi^2 \chi_T(\overline{Q}_0^2)$ represent a lower bound for the transverse susceptibility.
Reassuringly, the lattice-based results turn out to be slightly higher than the data-driven ones. The agreement is remarkable with differences not exceeding $\sim 10 \%$ up to $\overline{Q}_0^2 \simeq 1$ GeV$^2$.

The agreement shown in Fig.\ref{fig:chiT} can be improved by adding to the data-driven pion form factor a simple power-law tail for $\omega > 1$ GeV of the form  $|F_\pi^V(\omega)| = |F_\pi^V(1\,GeV)| \cdot$ $(1 \, GeV / \omega)^4$, which reproduces within the errors the results of Ref.\,\cite{Colangelo:2018mtw} in the $\omega^2$-range from $\simeq 0.9$ GeV$^2$ up to $1$ GeV$^2$. 
In this case we have found that the differences w.r.t.~the lattice-based results for the susceptibility $4M_\pi^2 \chi_T(\overline{Q}_0^2)$ do not exceed $\simeq 4 \%$ (i.e., less than $\simeq 2$ standard deviations).

It should be kept in mind, however, that the parameterization of the pion form factor $F_\pi^V(\omega)$ adopted in Ref.\,\cite{Colangelo:2018mtw}, while fulfilling the requirements of unitarity and analyticity, includes the contributions from $2\pi$, $3\pi$ and inelastic channels, estimated conservatively up to $\omega = 1$ GeV. Any extension to the $\omega$-region above $1$ GeV requires at least the inclusion of the contributions arising from higher resonances like $\rho(1450)$ and $\rho(1700)$ (see the corresponding note in the PDG review\,\cite{ParticleDataGroup:2022pth}), which is still to be settled\,\cite{Colangelo:2018mtw}.
At the same time, also the application of the approach of Ref.~\cite{Giusti:2018mdh} for estimating the $2\pi$ correlator $V_{2\pi}(\tau)$ from lattice isoQCD simulations is limited to Euclidean time distances $\tau \gtrsim 1$ fm, which qualitatively corresponds to energies below $\approx 1$ GeV (see Eq.\,(\ref{eq:V2pi})). 

According to Eq.\,(\ref{eq:chiT_FV}) the impact of the high-energy tail of  $F_\pi^V(\omega)$ increases as $\overline{Q}_0^2$ increases and, therefore, we consider trustable the results obtained for both the lattice-based and the data-driven transverse susceptibility only up to $\overline{Q}_0^2 \approx1$ GeV$^2$ (as adopted in Fig.\,\ref{fig:chiT}).

In what follows, we will make use of the lattice-based results for the transverse vector susceptibility $4M_\pi^2 \chi_T(\overline{Q}_0^2)$ up to $\overline{Q}_0^2 \simeq 1$ GeV$^2$. The main reason is that we want to analyze the spacelike data within the DM method without using information coming from $e^+ e^-$ data. In this way we will compare our results coming only from the spacelike sector with the corresponding ones obtained in Ref.\,\cite{Colangelo:2018mtw} from timelike data (see later Section\,\ref{sec:CERN+JLABpi}).

At $\overline{Q}_0^2 = 0$ we get 
\be
    \label{eq:chiT_0}
    4M_\pi^2 \chi_T(0) = 0.00574 \, (10) ~ . ~
\ee
This value can be compared with the upper bound provided by the quantity $4M_\pi^2 \, \Pi_1^{(I = 1)}$, where $\Pi_1^{(I = 1)}$ is the isovector contribution to the slope of the HVP polarization function evaluated at vanishing four-momentum transfer. The isovector HVP slope, which contains contributions  also from intermediate states other than the $2\pi$ states, has been calculated (in isoQCD) by several lattice collaborations, namely BMW\,\cite{Borsanyi:2016lpl}, RBC\,\cite{RBC:2018dos} and FHM\,\cite{FermilabLattice:2019ugu}, obtaining, respectively, $4M_\pi^2 \, \Pi_1^{(I = 1)} = 0.00607\,(19)$, $0.00624\,(17)$, $0.00611\,(9)$.

\section{Analysis of the electroproduction data}
\label{sec:JLABpi}

\begin{figure}[htb!]
\begin{center}
\includegraphics[scale=0.50]{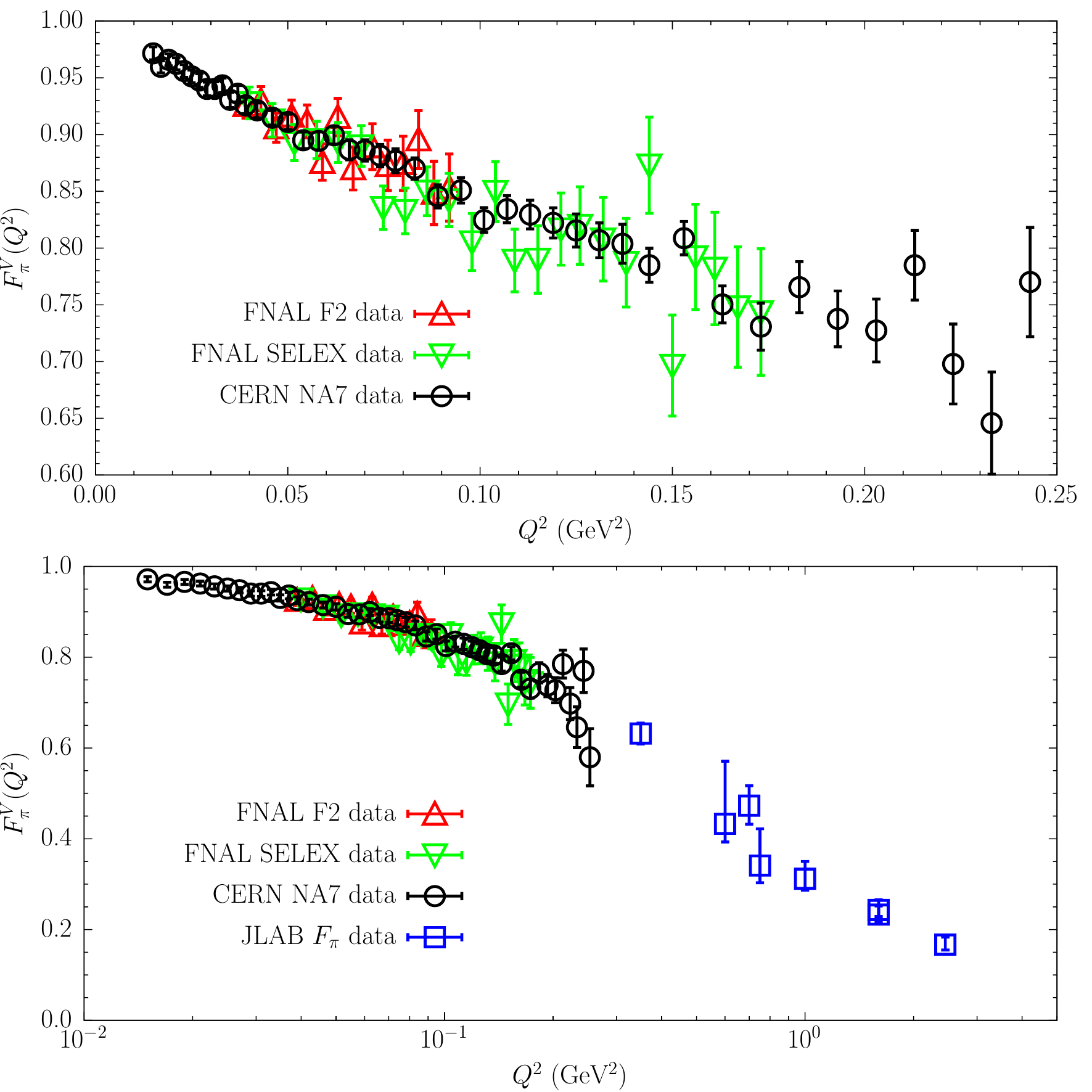}
\end{center}
\vspace{-0.75cm}
\caption{\it \small Experimental data on the em pion form factor $F_\pi^V(Q^2)$ obtained at FNAL F2\,\cite{Dally:1982zk} (red upper triangles), CERN SPS~\cite{NA7:1986vav} (black circles), FNAL SELEX\,\cite{SELEX:2001fbx} (green lower triangles) and from the reanalysis of the electroproduction data performed by the JLAB-$\pi$ Collaboration in Ref.~\cite{JeffersonLab:2008jve} (blue squares).}
\label{fig:FVpion}
\end{figure}

Presently the experimental data on the em pion form factor at spacelike momenta can be divided into two groups.
For values of $Q^2 \lesssim 0.25$ GeV$^2$ the pion form factor $F_\pi^V(Q^2)$ has been determined by measuring the scattering of high-energy (on-shell) pions off atomic electrons at FNAL F2\,\cite{Dally:1982zk}, CERN SPS~\cite{NA7:1986vav} and FNAL SELEX\,\cite{SELEX:2001fbx}. The data are shown in the upper panel of Fig.\,\ref{fig:FVpion}.

At higher values of $Q^2$ the pion form factor is extracted from cross section measurements of the reaction $^1H(e, e^\prime \pi^+)n$, i.e.~from pion electroproduction off the proton, which implies initial off-shell pions.
In such a case the separation of the longitudinal and transverse response functions as well as the extrapolation of the observed scattering from virtual pions to the one corresponding to on-shell pions have to be carefully considered for estimating the systematic uncertainties.
Using the electroproduction technique the pion form factor $F_V^\pi(Q^2)$ has been determined for $Q^2 \gtrsim 0.35$ GeV$^2$ in various experiments at CEA/Cornell~\cite{Bebek:1977pe}, DESY~\cite{Ackermann:1977rp, Brauel:1979zk} and JLAB~\cite{JeffersonLabFpi:2000nlc, JeffersonLabFpi-2:2006ysh, JeffersonLabFpi:2007vir, JeffersonLab:2008gyl}.

A careful analysis of the systematic uncertainties of all the electroproduction data was carried out by the JLAB-$\pi$ Collaboration in Ref.~\cite{JeffersonLab:2008jve}. Particular attention was paid to estimating the uncertainty due to the extrapolation of the electroproduction data to the pion pole. Within the errors no inconsistency is visible with the pion form factor obtained by the dispersive analysis of the $e^+ e^-$ data of Ref.\,\cite{Colangelo:2018mtw} (see later on Fig.\,\ref{fig:comparison_CHS19}) as well as with available results from lattice (iso)QCD simulations (see the review in Ref.\,\cite{Brandt:2013ffb}).

The results of the JLAB-$\pi$ collaboration are collected in Table~\ref{tab:JLABpi} together with the absolute normalization value $F_0 = 1$ at $Q_0^2 = 0$ (i.e.~$z_0 = 0$) corresponding to the charge conservation. The electroproduction data are shown in the lower panel of Fig.~\ref{fig:FVpion} together with the CERN ones. Note that: ~ i) few experimental data are plagued by large and asymmetric errors, which come from the systematic uncertainty due to a long extrapolation to the pion pole\,\cite{JeffersonLab:2008jve}; ~ ii) at $Q^2 = 1.6$ GeV$^2$ Ref.\,\cite{JeffersonLab:2008jve} quotes two results: $0.233_{-17}^{+19}$  and $0.243_{-14}^{+23}$. Both are shown in the lower panel of Fig.~\ref{fig:FVpion}, while in Table\,\ref{tab:JLABpi} only their average $0.238_{-17}^{+21}$ is considered.
\begin{table}[htb!]
\renewcommand{\arraystretch}{1.2}	 
\begin{center}	
\begin{tabular}{||c|c|c||c||c||c|c||}
\hline
$i$ & $Q_i^2$ (GeV$^2)$ & $z_i$ & $F_i$ & $F_i$ (sym.) & $d_i$ & $\phi_i(\overline{Q}_0^2 = 0)$ \\
\hline \hline
0 & 0.0     & $0.0$     & $1.0$                         & $1.0$            & $-7.02 \cdot 10^{1}$ & $0.0144$ \\ \hline \hline
1 & 0.35 & $0.402$ & $0.632_{-23}^{+23}$ & $0.632~(23)$ & $+2.88 \cdot 10^{4}$ & $0.0219$ \\ \hline
2 & 0.60 & $0.494$ & $0.433_{-40}^{+138}$ & $0.482~(89)$ & $-1.26 \cdot 10^{6}$ & $0.0229$ \\ \hline
3 & 0.70 & $0.519$ & $0.473_{-41}^{+44}$ & $0.475~(43)$ & $+5.48 \cdot 10^{6}$ & $0.0230$ \\ \hline
4 & 0.75 & $0.530$ & $0.341_{-38}^{+81}$ & $0.363~(60)$ & $-4.73 \cdot 10^{6}$ & $0.0231$ \\ \hline
5 & 1.00 & $0.576$ & $0.312_{-25}^{+38}$ & $0.319~(32)$ & $+5.34 \cdot 10^{5}$ & $0.0233$ \\ \hline 
6 & 1.60 & $0.645$ & $0.238_{-17}^{+21}$ & $0.240~(19)$ & $-5.66 \cdot 10^{4}$  & $0.0232$ \\ \hline
7 & 2.45 & $0.701$ & $0.167_{-12}^{+16}$ & $0.169~(14)$ & $+6.45 \cdot 10^{3}$ & $0.0228$ \\ \hline
\end{tabular}
\end{center}
\renewcommand{\arraystretch}{1.0}
\vspace{-0.5cm}
\caption{\it \small Experimental data on the em pion form factor $F_\pi^V(Q^2)$ obtained using the electroproduction technique analysed by the JLAB-$\pi$ Collaboration in Ref.~\cite{JeffersonLab:2008jve} together with  the absolute normalization value $F_0 = 1$ at $Q_0^2 = 0$ (i.e.~$z_0 = 0$) due to charge conservation. The fifth column represents the data after symmetrization of the errors. The sixth column contains the values of the kinematical coefficients $d_i$ (see Eq.~(\ref{eq:di})). The last column shows the values of the kinematical factors $\phi_i$ evaluated at $\overline{Q}_0^2 = 0$ (see Eq.~(\ref{eq:phi})).}
\label{tab:JLABpi}
\end{table}

We now apply the DM approach to the set of $N + 1 = 8$ data collected in Table\,\ref{tab:JLABpi} assigning a very small, but non-vanishing error to the data point $F_0 = 1$ at $Q^2 = 0$, namely $\sigma_0 = 10^{-16}$. We also symmetrize the errors obtaining the set of data points shown in the fifth column of Table~\ref{tab:JLABpi}.
Since no information is available on the covariance matrix of the electroproduction data, the form factor values are considered to be uncorrelated, namely the covariance matrix $C$ is given by
\be
    C_{ij} = \sigma_i^2 \delta_{ij} ~ ,
    \label{eq:datacov}
\ee
where $\sigma_i^2$ is the variance of $F_i$ with $i = 0, 1, ... N$.

We start by choosing $\overline{Q}_0^2 = 0$ and postponing to Section\,\ref{sec:Q02} the discussion about the impact of a generic choice $\overline{Q}_0^2 > 0$. 
We assume a gaussian distribution for the non-perturbative transverse susceptibility $4M_\pi^2\chi_T(\overline{Q}_0^2 = 0) = 0.00574 \, (10)$  (see Eq.\,(\ref{eq:chiT_0})). This distribution is taken to be uncorrelated with those of the form factor points collected in Table\,\ref{tab:JLABpi}.

A sample of $10^5$ uncorrelated events normally distributed is generated using as input the mean values $F \equiv \{ F_i \}$ and uncertainties $\sigma \equiv \{ \sigma_i \}$ with $i = 0, 1, ... N$.
For each event we calculate the susceptibility $\chi_{DM}(\overline{Q}_0^2 = 0)$ given by Eq.~(\ref{eq:chiDM}).
It turns out that the calculated values of $\chi_{DM}(\overline{Q}_0^2 = 0)$ range from a minimum equal to $\sim 2.6$ up to a maximum given by $\sim 1.4 \cdot 10^9$ and, therefore, none of the $10^5$ generated events satisfies the unitary filter (\ref{eq:filter}). The same happens also when we increase the size of the sample up to $10^6$.

The main reason for the above finding can be traced back to the values of the kinematical coefficients $d_i$, given by Eq.~(\ref{eq:di}). These coefficients depend only on the series of values $z_i$ and their numerical values are shown in the sixth column of Table~\ref{tab:JLABpi}. They turn out to be quite large in absolute value and to have alternating signs.
It is therefore very unlikely to generate an event with uncorrelated values of the form factor points leading to a value of $\chi_{DM}(\overline{Q}_0^2 = 0)$ as small as $4M_\pi^2 \chi_T(\overline{Q}_0^2 = 0)$.
A very delicate compensation among the contributions of the various data points to Eq.~(\ref{eq:chiDM}) is required and this naturally implies specific correlations among the form factor points.
In principle, one may increase the size of the sample until some of the events satisfy the unitary filter, but a brute-force increase of the size of the sample may become impracticable for large values of the number of data points $N$ (see also later on Section~\ref{sec:CERN+JLABpi}).

As already pointed out in Section\,\ref{sec:DM}, the unitary filter~(\ref{eq:filter}) acts as a constraint and it allows to select a subset of the initial events $F$ and $C \equiv \{ C_{ij} \}$, made only by unitary events. 
Such a subset corresponds to new values $\overline{F}$ for the form factor points and to a new covariance matrix $\overline{C}$ representing the {\it unitary} form factor points (and correlations) on which any further analysis fulfilling unitarity must be based.
Thus, we have to find an efficient way to determine the unitary values  $\overline{F}$ and $\overline{C}$. 
In the next Section we illustrate a simple procedure able to achieve this goal and applicable for any value of $N$.

\section{Unitary sampling procedure}
\label{sec:sampling}

The gaussian multivariate distribution used in the previous Section is based on the probability density function (PDF) given by 
\be
    PDF(f_i) \propto \mbox{exp}\left[ - \frac{1}{2} \sum_{i,j=0}^N (f_i - F_i) C_{ij}^{-1} (f_j - F_j) \right] ~ , ~
    \label{eq:PDF}
\ee
where $\{ F_i, \}$ and $\{ C_{ij} \}$ are respectively the mean values and the covariance matrix used as inputs.
As well known, the PDF~(\ref{eq:PDF}) favors the relative likelihood of small values of the quadratic form $\sum_{i,j=0}^N (f_i - F_i) C_{ij}^{-1} (f_j - F_j)$, which however may correspond to large values of the susceptibility~(\ref{eq:chiDM}), as shown in the previous Section.

We now modify the above PDF in order to allow the susceptibility~(\ref{eq:chiDM}) to be small enough to fulfill the unitary constraint~(\ref{eq:filter}).
We consider the following new PDF:
\bea
    \label{eq:PDF_DM}
    PDF_{DM}(f_i) & \propto &  PDF(f_i) \cdot \mbox{exp}\left[ - \frac{s}{4M_\pi^2 \chi_T(\overline{Q}_0^2)} \chi_{DM}(\overline{Q}_0^2) \right] ~ \\[2mm]
                            & \propto & \mbox{exp}\left[ - \frac{1}{2} \sum_{i,j=0}^N (f_i - F_i) C_{ij}^{-1} (f_j - F_j) - \frac{s}{4M_\pi^2 \chi_T(\overline{Q}_0^2)} 
                                               \sum_{i,j=0}^N f_i D_{ij}^{-1}(\overline{Q}_0^2) f_j \right] ~ , ~ \nonumber
\eea
where $s$ is a parameter (expected to determine the number of events satisfying the unitary filter~(\ref{eq:filter})) and the matrix $D^{-1}(\overline{Q}_0^2)$ is defined as
\be
    D_{ij}^{-1}(\overline{Q}_0^2) \equiv \frac{\phi_i(\overline{Q}_0^2) d_i (1 - z_i^2) ~ \phi_j(\overline{Q}_0^2) d_j (1 - z_j^2)}{1 - z_i z_j} ~ .
    \label{eq:chiDM_ij}
\ee
The use of Eq.~(\ref{eq:PDF_DM}) as a PDF allows to increase the relative likelihood of small values of the susceptibility $\chi_{DM}(\overline{Q}_0^2)$ at the expense of decreasing the PDF~(\ref{eq:PDF}).
Introducing the matrix $\widetilde{C}$  defined in compact notation as
\be
    \label{eq:Ctilde}
    \widetilde{C}^{-1} = C^{-1} + \frac{2s}{ 4M_\pi^2 \chi_T(\overline{Q}_0^2)} D^{-1}(\overline{Q}_0^2) ~ , ~
\ee
Eq.~(\ref{eq:PDF_DM}) can be easily rewritten in the form
\be
     PDF_{DM}(f_i) \propto \mbox{exp}\left[ - \frac{1}{2} \sum_{i,j=0}^N (f_i - \widetilde{F}_i) \widetilde{C}_{ij}^{-1} (f_j - \widetilde{F}_j) ~ 
                                          - \frac{1}{2} \sum_{i,j=0}^N (F_i - \widetilde{F}_i) C_{ij}^{-1} F_j \right] ~ , ~
    \label{eq:PDF_DM_alt}
\ee
where the new vector of mean values $\widetilde{F}$ is related to the starting one $F$ by
\be
    \label{eq:Ftilde}
    \widetilde{F} = \widetilde{C} ~ C^{-1} ~ F ~ . ~
\ee
Note that the second exponential in the r.h.s.~of Eq.~(\ref{eq:PDF_DM_alt}) does not depend on $\{ f_i \}$ and therefore it is irrelevant for the relative likelihood of the events, so that the new PDF is simply given by
\be
     PDF_{DM}(f_i) \propto \mbox{exp}\left[ - \frac{1}{2} \sum_{i,j=0}^N (f_i - \widetilde{F}_i) \widetilde{C}_{ij}^{-1} (f_j - \widetilde{F}_j) \right]~ , ~
    \label{eq:PDF_DM_final}
\ee
which represents a multivariate gaussian distribution characterized by the new set of input values $\{ \widetilde{F}_i, \widetilde{C}_{ij} \}$ given by Eqs.~(\ref{eq:Ftilde}) and (\ref{eq:Ctilde}), respectively.

\begin{figure}[htb!]
\begin{center}
\includegraphics[scale=0.50]{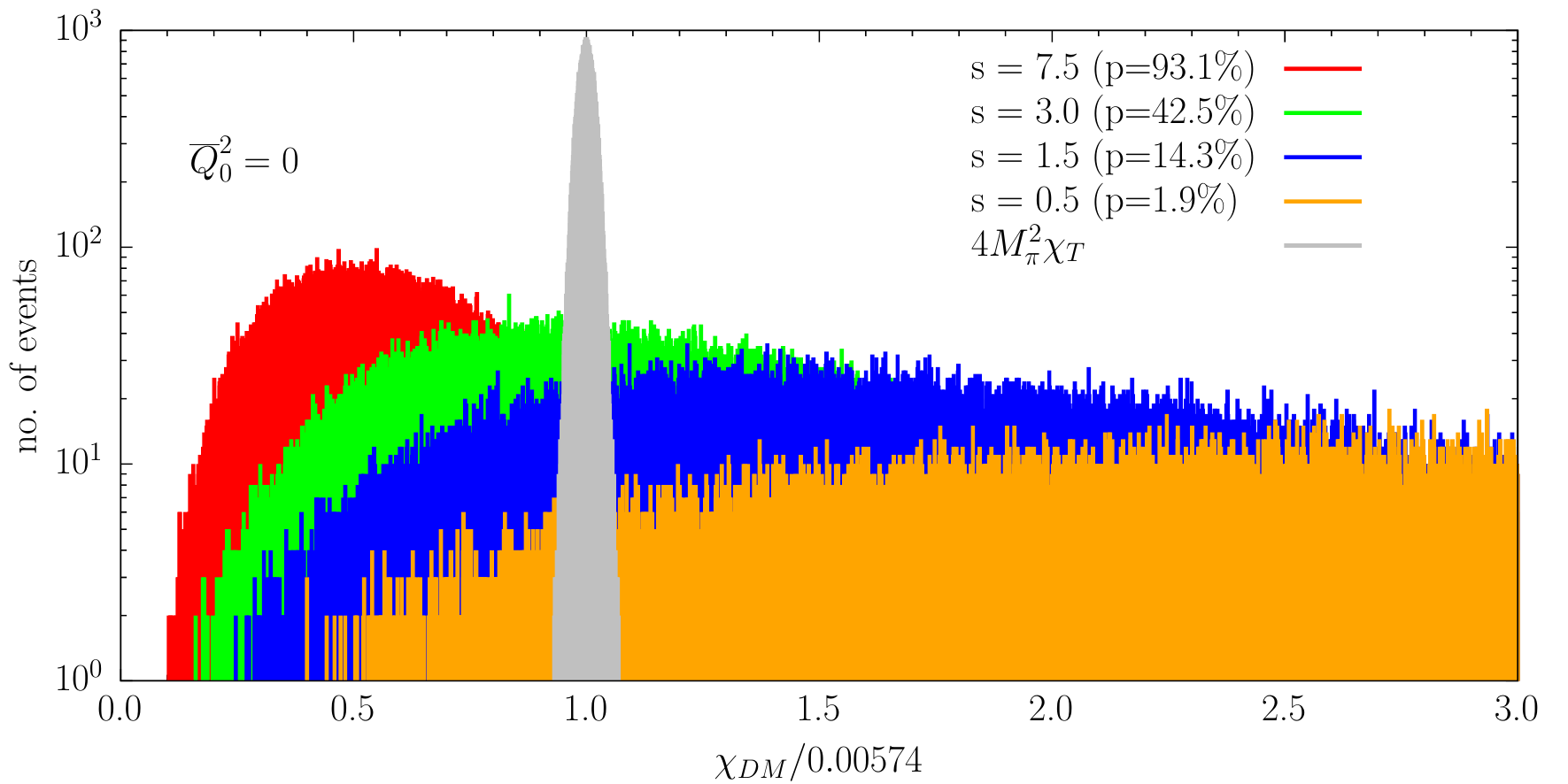}
\end{center}
\vspace{-0.75cm}
\caption{\it \small Histograms of the ratio $\chi _{DM} / 0.00574$ corresponding to the data susceptibility $\chi_{DM}$ (see Eq.~(\ref{eq:chiDM})) obtained using samples of $10^5$ events generated according to the multivariate distribution~(\ref{eq:PDF_DM_final}) for various values of the parameter $s$. The quantity $0.00574$ represents the central value of the non-perturbative transverse susceptibility (\ref{eq:chiT_0}), obtained in Section\,\ref{sec:susceptibility} at $\overline{Q}_0^2 = 0$. The percentage $p$ of events passing the unitary filter\,(\ref{eq:filter}) is given in the inset for each value of $s$. The grey histogram represents the gaussian distribution of the non-perturbative transverse susceptibility ratio $4M_\pi^2 \chi_T (0)/ 0.00574$.}
\label{fig:chiDM}
\end{figure}

In the case of the electroproduction data of Table~\ref{tab:JLABpi} we generate samples with $10^5$ events according to the new PDF~(\ref{eq:PDF_DM_final}) for various values of the parameter $s$ (assuming again $\overline{Q}_0^2 = 0$).
Then, we calculate the corresponding susceptibility $\chi_{DM}(\overline{Q}_0^2 = 0)$ given by Eq.~(\ref{eq:chiDM}). The results are shown in Fig.~\ref{fig:chiDM}, where the (gaussian) distribution corresponding to the non-perturbative transverse result $4M_\pi^2 \chi_T(\overline{Q}_0^2 = 0) = 0.00574 \, (10)$ is also presented.
It can be clearly seen that, as the parameter $s$ increases, the PDF~(\ref{eq:PDF_DM_final}) can be very effective in generating events with $\chi_{DM}(\overline{Q}_0^2 = 0) \leq 4M_\pi^2 \chi_T(\overline{Q}_0^2 = 0)$, i.e.~satisfying the unitary filter\,(\ref{eq:filter}).

Both the vector of mean values $\widetilde{F}$ and the covariance matrix $\widetilde{C}$ depend on the value of the parameter $s$. 
The case $s = 0$ trivially corresponds to $\widetilde{F} = F$ and $\widetilde{C} = C$, while $s > 0$ leads to $\widetilde{F} \neq  F$ and $\widetilde{C} \neq C$.
In order to quantify the deviation of $\widetilde{F}$ from $F$ we introduce the quantity $\Delta$ defined as
\be
   \label{eq:Delta}
    \Delta \equiv \left\{ \frac{1}{N+1} \sum_{i,j=0}^N (\widetilde{F}_i - F_i) C_{ij}^{-1} (\widetilde{F}_j - F_j) \right\}^{1/2} 
                         ~_{\overrightarrow{C_{ij} = \sigma_i^2 \delta_{ij}}} ~ \left\{ \frac{1}{N+1} \sum_{i=0}^N \frac{(\widetilde{F}_i - F_i)^2}{\sigma_i^2} \right\}^{1/2} ~ . ~ 
\ee
The value of $\Delta$ represents the average deviation of the new values $\widetilde{F}$ from the starting ones $F$ measured with respect to the starting covariance.
In other words, $\Delta < 1$ means that on average $\widetilde{F}$ deviates from $F$ by less than one standard deviation.

As a further estimator of the deviation of  $\widetilde{F}$ with respect to $F$, we introduce also the quantity $\eta$ defined as
\be
    \label{eq:eta}
    \eta \equiv \left\{ \frac{1}{N+1} \sum_{i=0}^N \frac{\widetilde{F}_i^2}{F_i^2} \right\}^{1/2} ~ . ~
\ee
The value of $\eta$ can be smaller or larger than unity depending on whether $|\widetilde{F}_i|$ is (on average) smaller or larger than $|F_i|$.

In the same spirit, in order to estimate how the uncertainties of the new mean values, i.e.~$\widetilde{\sigma}_i \equiv \widetilde{C}_{ii}^{1/2}$, deviates (on average) from the starting ones $\sigma_i$ we introduce the quantity $\epsilon$ defined as
\be
    \label{eq:epsilon}
    \epsilon \equiv \left\{ \frac{1}{N+1} \sum_{i=0}^N \frac{\widetilde{C}_{ii}}{C_{ii}} \right\}^{1/2} = 
                            \left\{ \frac{1}{N+1} \sum_{i=0}^N \frac{\widetilde{\sigma}_i^2}{\sigma_i^2} \right\}^{1/2} ~ . ~
\ee

In Table~\ref{tab:FVs} we have collected the mean values $\widetilde{F}$ and the uncertainties $\widetilde{\sigma}$ corresponding to some values of the parameter $s$, which can be compared with the starting values $F$ and $\sigma$.
The values of the quantities $\Delta$ (see Eq.~(\ref{eq:Delta})), $\eta$ (see Eq.~(\ref{eq:eta})), $\epsilon$ (see Eq.~(\ref{eq:epsilon})) and of the percentage $p$ of events passing the unitary filter~(\ref{eq:filter}) are also shown. 
\begin{table}[htb!]
\renewcommand{\arraystretch}{1.2}	 
\begin{center}	
\begin{tabular}{||c|c||c|c|c|c||}
\hline
$Q^2$ (GeV$^2)$ & $F$ (sym.) & $\widetilde{F} ~(s = 0.5)$ & $\widetilde{F} ~(s = 1.5)$ & $\widetilde{F} ~(s = 3.0)$ & $\widetilde{F} ~(s = 7.5)$ \\ 
\hline \hline
0.35 & $0.632~(23)$ & $0.629~(22)$ & $0.624~(21)$  & $0.619~(21)$  & $0.608~(20)$ \\ \hline
0.60 & $0.482~(89)$ & $0.481~(18)$ & $0.482~(16)$  & $0.483~(16)$  & $0.483~(16)$ \\ \hline
0.70 & $0.475~(43)$ & $0.438~(18)$ & $0.440~(16)$  & $0.442~(16)$  & $0.445~(15)$ \\ \hline
0.75 & $0.363~(60)$ & $0.419~(18)$ & $0.422~(17)$  & $0.424~(16)$  & $0.429~(15)$ \\ \hline
1.00 & $0.319~(32)$ & $0.342~(18)$ & $0.346~(17)$  & $0.350~(16)$  & $0.358~(15)$ \\ \hline
1.60 & $0.240~(19)$ & $0.234~(15)$ & $0.237~(15)$  & $0.241~(14)$  & $0.249~(13)$ \\ \hline
2.45 & $0.169~(14)$ & $0.170~(14)$ & $0.168~(14)$  & $0.167~(14)$  & $0.164~(14)$ \\ \hline \hline
$\Delta$    & 0.0 & 0.54 & 0.56 & 0.61 & 0.76 \\ \hline
$\eta$       & 1.0 & 1.02 & 1.02 & 1.02& 1.03 \\ \hline
$\epsilon$ & 1.0 & 0.72 & 0.70 & 0.69 & 0.69 \\ \hline \hline
$p ~ (\%)$ & $< 10^{-4}$ & 1.9 & 14.3 & 42.5 & 93.1 \\ \hline \hline
\end{tabular}
\end{center}
\renewcommand{\arraystretch}{1.0}
\vspace{-0.5cm}
\caption{\it \small Mean values $\widetilde{F}$ and uncertainties $\widetilde{\sigma}$ obtained from Eqs.~(\ref{eq:Ctilde}) and~(\ref{eq:Ftilde}) for various values of the parameter $s$ before the application of the unitary filter\,(\ref{eq:filter}) (assuming $\overline{Q}_0^2 = 0$). The second column show the (symmetrized) electroproduction data from Ref.~\cite{JeffersonLab:2008jve} (see Table~\ref{tab:JLABpi}). The values of the quantities $\Delta$, $\eta$, $\epsilon$ (see Eqs.~(\ref{eq:Delta})-(\ref{eq:epsilon})) and of the percentage $p$ of events passing the unitary filter\,(\ref{eq:filter}) are given in the last four rows.}
\label{tab:FVs}
\end{table}

The values of $p$ and $\Delta$ increase for increasing $s$ as expected, while both $\eta$ and $\epsilon$ are found to be substantially constant.
As the value of the parameter $s$ varies from $\simeq 0.5$ to $\simeq 7.5$, the value of $p$ ranges from $\simeq 2 \%$ to $\simeq 93 \%$ and the one of $\Delta$ from $\simeq 0.5$ to $\simeq 0.8$, while $\eta \simeq 1.02$ and $\epsilon \simeq 0.7$.

\begin{figure}[htb!]
\begin{center}
\includegraphics[scale=0.50]{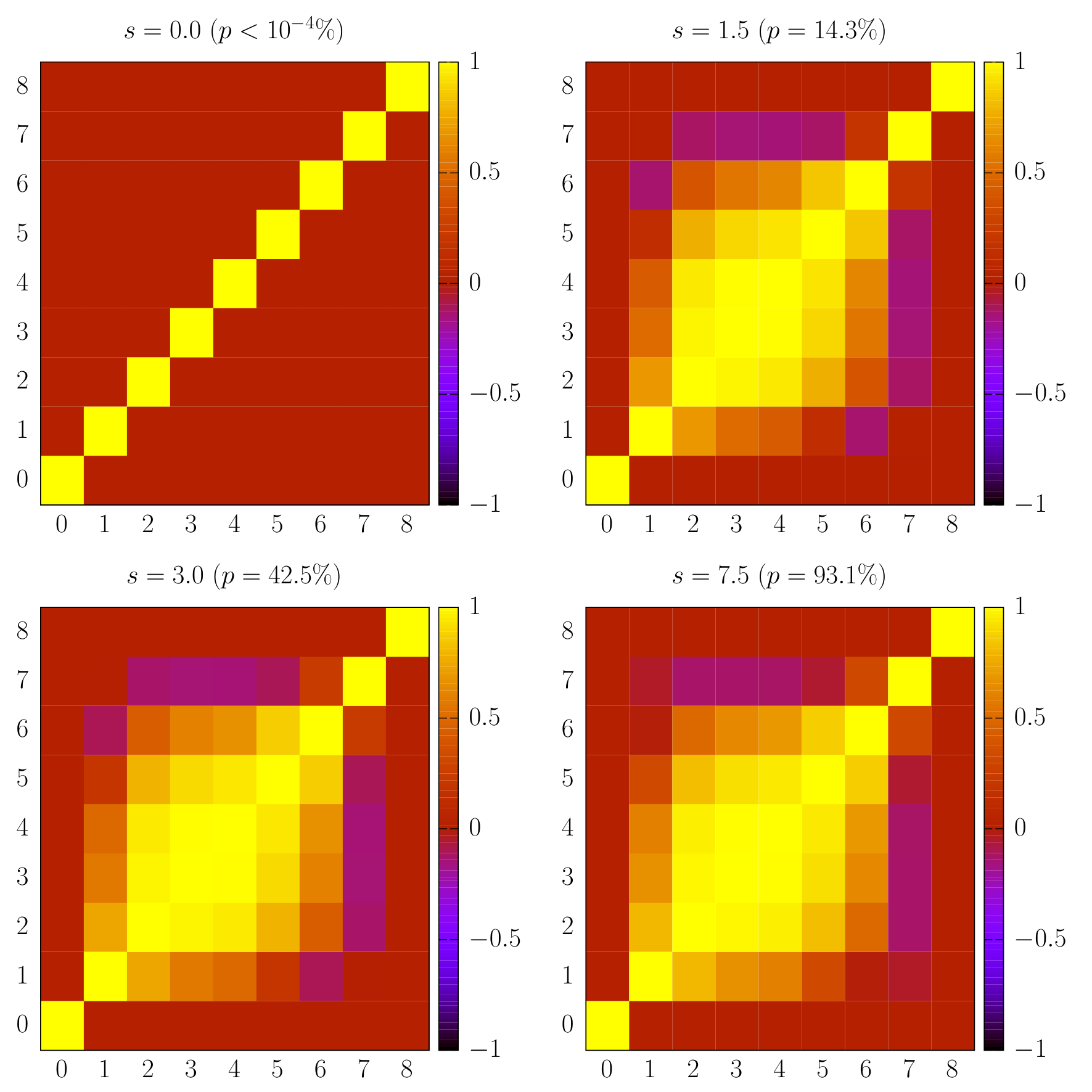}
\end{center}
\vspace{-0.75cm}
\caption{\it \small Heat maps representing the correlation matrix $\widetilde{\rho}_{ij} \equiv \widetilde{C}_{ij} / (\widetilde{\sigma}_i  \widetilde{\sigma}_j)$ corresponding to the unitary sampling\,(\ref{eq:PDF_DM_final}) for various values of the parameter $s$ before the application of the unitary filter\,(\ref{eq:filter}). The case $s = 0$ corresponding to the initial correlation matrix $\rho_{ij} \equiv C_{ij} / (\sigma_i \sigma_j) \to \delta_{ij}$ is also shown. The labels $0, 1, ... 7$ correspond to the form factor points of Table~\ref{tab:JLABpi}, while the label $8$ corresponds to the non-perturbative result for the transverse susceptibility $4M_\pi^2 \chi_T(0)$.}
\label{fig:correlations}
\end{figure}

Not only $\widetilde{F} \neq F$ and $\widetilde{\sigma} \neq \sigma$, but also the correlation matrix of the data corresponding to the new PDF~(\ref{eq:PDF_DM_final}) is different from the starting one, namely
\be
   \label{eq:correlation}
    \widetilde{\rho}_{ij} \equiv \frac{\widetilde{C}_{ij}}{\widetilde{\sigma}_i  \widetilde{\sigma}_j} \neq \rho_{ij} \equiv \frac{C_{ij}}{\sigma_i \sigma_j} ~ . ~
\ee
This point is illustrated through the heat maps of Fig.~\ref{fig:correlations} for various values of the parameter $s$ before the application of the unitary filter\,(\ref{eq:filter}). As expected, the correlations among first neighbors increases for $s > 0$. We observe a slight dependence of $\widetilde{\rho}_{ij}$ on the value of the parameter $s$.

We now apply the unitary filter~(\ref{eq:filter}) and select only the subsets of events satisfying unitarity for each value of the parameter $s$. On such subsets we calculate the mean values, uncertainties and correlations for the form factor and the transverse susceptibility. In order to adopt a compact notation we will denote these quantities by $\overline{F} = \{ \overline{F}_i \}$, $\overline{\sigma} = \{ \overline{\sigma}_i \}$ and $\overline{\rho} = \{\overline{\rho}_{ij} \}$ with $\overline{\rho}_{ij} = \overline{C}_{ij} / (\overline{\sigma}_i \overline{\sigma}_j)$ and $i, j = 0, 1, ... (N+1)$.

The mean values $\overline{F}$ and the uncertainties $\overline{\sigma}$ corresponding to the form factor points are shown in Table~\ref{tab:FVs_1} for some values of the parameter $s$ and compared with the starting values $F$ and $\sigma$.
The quantities $\Delta$, $\eta$  and $\epsilon$ are calculated using $\overline{F}$ and $\overline{\sigma}$ in Eqs.~(\ref{eq:Delta})-(\ref{eq:epsilon}) and shown in Table~\ref{tab:FVs_1}, as well as the size $N_{sample} = p \cdot 10^5$ of the subsets of events passing the unitary filter~(\ref{eq:filter}). 
\begin{table}[htb!]
\renewcommand{\arraystretch}{1.2}	 
\begin{center}	
\begin{tabular}{||c|c||c|c|c|c||}
\hline
$Q^2$ (GeV$^2)$ & $F$ (sym.) & $\overline{F} ~(s = 0.5)$ & $\overline{F} ~(s = 1.5)$ & $\overline{F} ~(s = 3.0)$ & $\overline{F} ~(s = 7.5)$ \\ 
\hline \hline
0.35 & $0.632~(23)$ & $0.615~(19)$ & $0.614~(19)$  & $0.612~(19)$  & $0.607~(20)$ \\ \hline
0.60 & $0.482~(89)$ & $0.483~(16)$ & $0.483~(16)$  & $0.483~(16)$  & $0.483~(16)$ \\ \hline
0.70 & $0.475~(43)$ & $0.443~(16)$ & $0.444~(16)$  & $0.444~(15)$  & $0.446~(15)$ \\ \hline
0.75 & $0.363~(60)$ & $0.426~(16)$ & $0.426~(16)$  & $0.427~(15)$  & $0.429~(15)$ \\ \hline
1.00 & $0.319~(32)$ & $0.353~(15)$ & $0.354~(15)$  & $0.355~(15)$  & $0.359~(15)$ \\ \hline
1.60 & $0.240~(19)$ & $0.244~(13)$ & $0.244~(13)$  & $0.246~(13)$  & $0.249~(13)$ \\ \hline
2.45 & $0.169~(14)$ & $0.166~(14)$ & $0.165~(14)$  & $0.165~(14)$  & $0.163~(14)$ \\ \hline \hline
$\Delta$    & 0.0 & 0.68 & 0.68 & 0.70 & 0.78 \\ \hline
$\eta$       & 1.0 & 1.03 & 1.03 & 1.03& 1.03 \\ \hline
$\epsilon$ & 1.0 & 0.66 & 0.66 & 0.66 & 0.66 \\ \hline \hline
$N_{sample}$ & -- & 1900 & 14300 & 42500 & 93100 \\ \hline \hline
\end{tabular}
\end{center}
\renewcommand{\arraystretch}{1.0}
\vspace{-0.5cm}
\caption{\it \small The same as in Table~\ref{tab:FVs}, but for the mean values $\overline{F}$ and uncertainties $\overline{\sigma}$ obtained using only the subsets of unitary events selected by the filter~(\ref{eq:filter}) for various values of the parameter $s$. The size $N_{sample} = p \cdot 10^5$ of these subsets is shown in the last row.}
\label{tab:FVs_1}
\end{table}
It can be seen that the application of the unitary filter~(\ref{eq:filter}) leads to values $\{ \overline{F}$ and $\overline{\sigma}$, which exhibit a weaker dependence on the parameter $s$ with respect to the values $\widetilde{F}$ and $\widetilde{\sigma}$ obtained before the application of the unitary filter.
As the value of the parameter $s$ varies from $\simeq 0.5$ to $\simeq 7.5$, the value of the quantity  $\Delta$ ranges only from $\simeq 0.7$ to $\simeq 0.8$, while both $\eta \simeq 1.03$ and $\epsilon \simeq 0.7$ do not change significantly.

\begin{figure}[htb!]
\begin{center}
\includegraphics[scale=0.50]{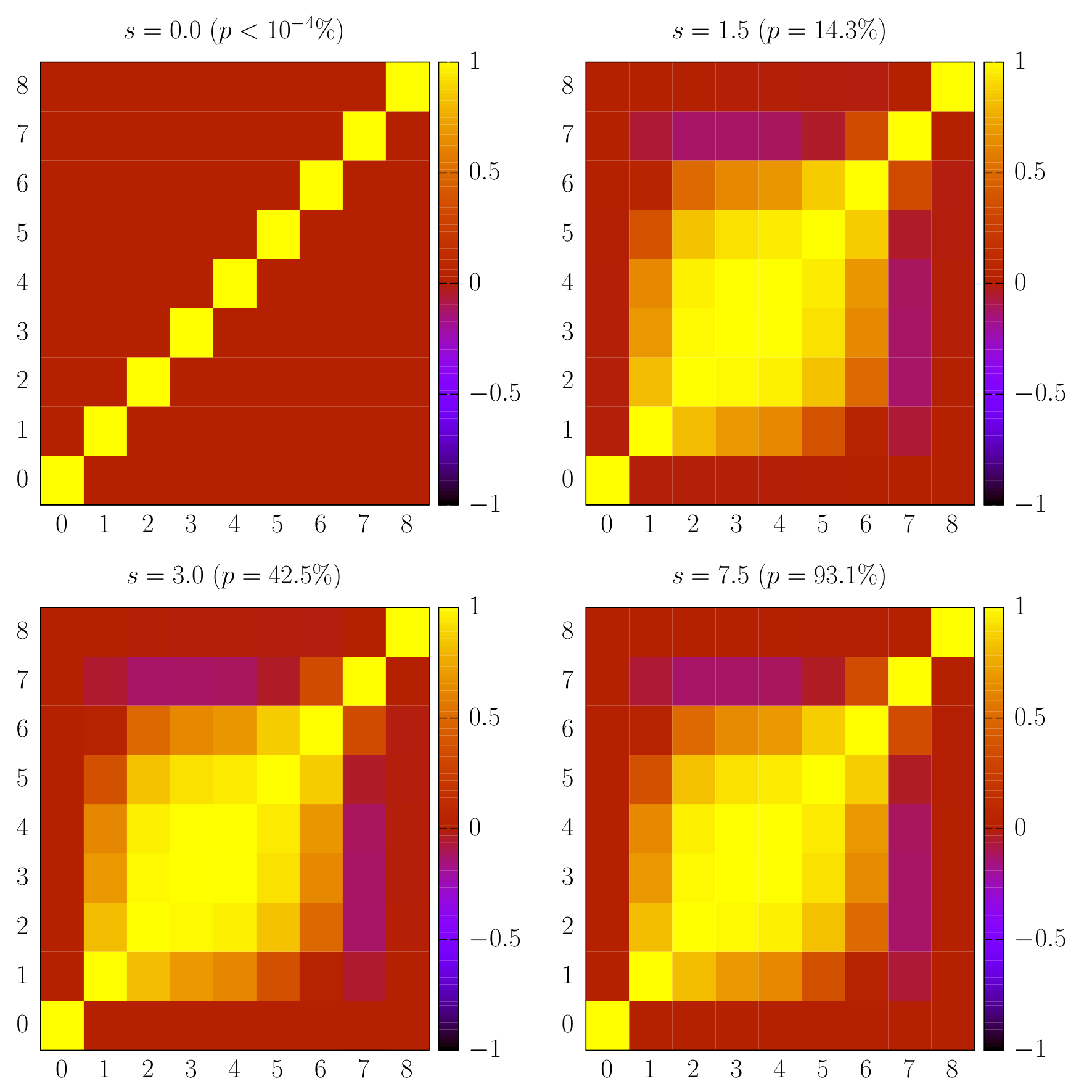}
\end{center}
\vspace{-0.75cm}
\caption{\it \small The same as in Fig.~\ref{fig:correlations}, but for the correlation matrix $\overline{\rho}_{ij} \equiv \overline{C}_{ij} / (\overline{\sigma}_i  \overline{\sigma}_j)$ obtained after the application of the unitary filter~(\ref{eq:filter}) for various values of the parameter $s$ (see Table~\ref{tab:FVs_1}).}
\label{fig:correlations_1}
\end{figure}

Finally, also the correlation matrix $\overline{\rho}$, obtained after the application of the unitary filter and shown in Fig.~\ref{fig:correlations_1}, changes only slightly with respect to the matrix $\widetilde{\rho}$ (see Fig.~\ref{fig:correlations}) obtained before the application of the unitary filter.

\begin{figure}[htb!]
\begin{center}
\includegraphics[scale=0.50]{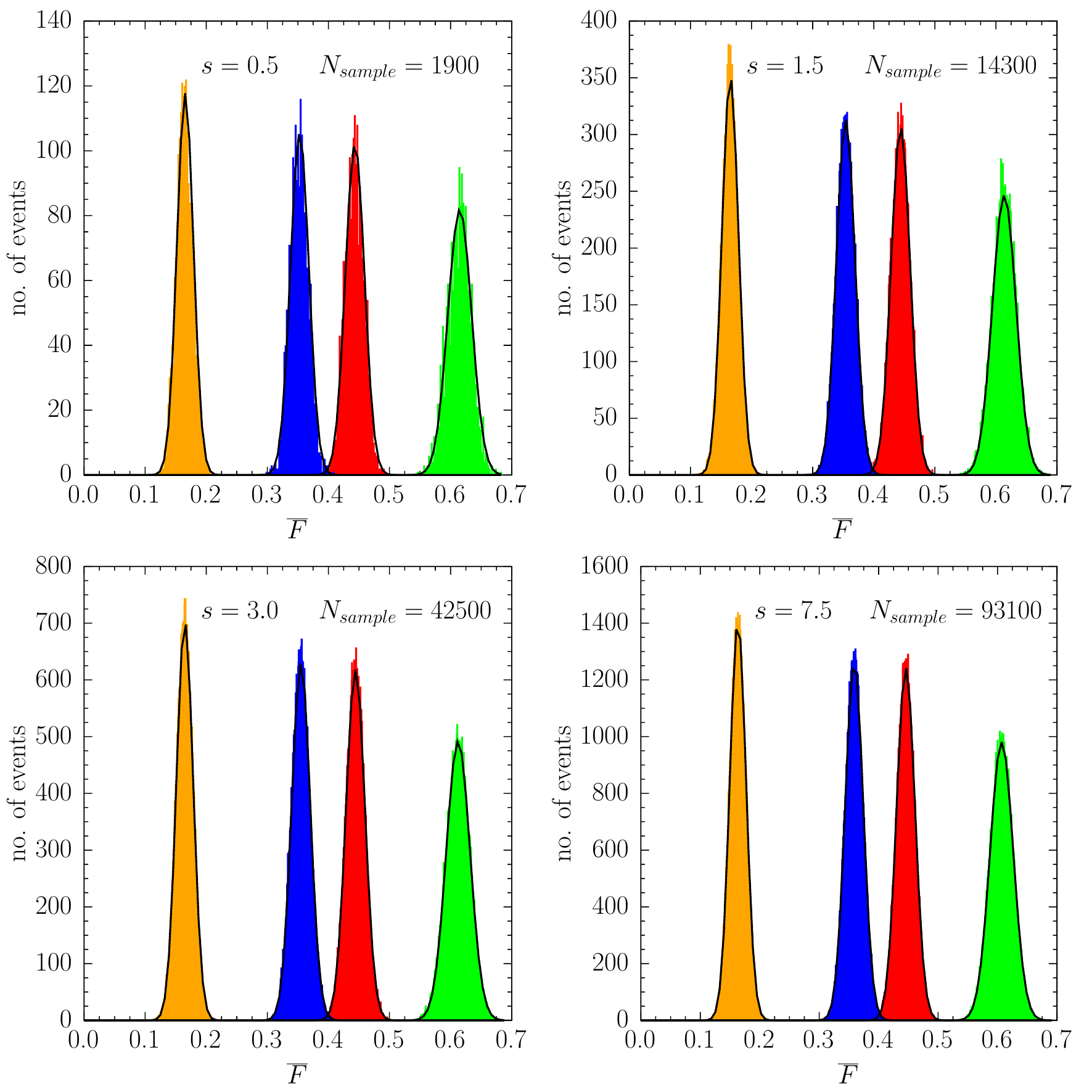}
\end{center}
\vspace{-0.75cm}
\caption{\it \small Distributions of the form factors values $\overline{F}$ obtained after the application of the unitary filter\,(\ref{eq:filter}) for various values of the parameter $s$. The green, red, blue and orange histograms correspond respectively to $Q^2 = 0.35, 0.70 , 1.0, 2.45$ GeV$^2$. The black solid lines represent a gaussian fit of the various histograms.}
\label{fig:ffs}
\end{figure}

The values $\widetilde{F}$ of the form factor data are by construction normally distributed. After the application of the unitary filter~(\ref{eq:filter}) also the distributions of the new values $\overline{F}$ are substantially normal regardless the size of the sample of the unitary events, as shown in Fig.~\ref{fig:ffs}.
Thus, we apply a simple {\it iteration procedure} to get rid off the residual dependence on the value of the parameter $s$. Namely, we apply the unitary filter~(\ref{eq:filter}) to the distributions of the values $\overline{F}$ and recalculate on the new subset of unitary events the mean values, uncertainties and correlations for the form factor and the transverse susceptibility. The percentage of events passing the unitary filter increases drastically and we iterate (few times) the previous steps until the percentage reaches $\simeq 99 \%$. This is done for various values of the parameter $s$ and we find that the final set of values for $\overline{F}$ and for the covariance matrix $\overline{C}$ is almost totally independent on the starting value of $s$ while keeping $\Delta \simeq  0.8$, $\eta \simeq 1.03$ and $\epsilon \simeq 0.7$. We remind that the values obtained for $\Delta$, $\eta$ and $\epsilon$ means that on average the new central values $\overline{F}$ are larger w.r.t.~the original ones $F$ by few percent with changes lower than one standard deviation and that the new uncertainties $\overline{\sigma}$ are on average around $70 \%$ of the original ones $\sigma$.
We stress that $\overline{F} \neq F$ and $ \overline{C} \neq C$ are direct consequences of the application of the unitary filter\,(\ref{eq:filter}).

Using our final set of values for $\overline{F}$ and for the covariance matrix $\overline{C}$ we generate a sample of events for the input values of the form factor, which all satisfy the unitary filter~(\ref{eq:filter}). 
For the $k$-th event, corresponding to form factor values $f_j^{(k)}$ and transverse susceptibility $4M_\pi^2 \chi_T^{(k)}(\overline{Q}_0^2)$, we apply the DM formulae\,(\ref{eq:bounds_pion})-(\ref{eq:gamma_pion}) obtaining
\be
     F_L^{(k)}(Q^2) \leq F_\pi^{V(k)}(Q^2) \leq F_U^{(k)}(Q^2)
     \label{eq:FV_band}
\ee
with
\bea
    \label{eq:FLU}
    F_{L(U)}^{(k)}(Q^2) & = & \frac{1}{\phi(z, \overline{Q}_0^2) d(z)} \sum_{j = 0}^N f_j^{(k)} \phi_j(\overline{Q}_0^2) d_j \frac{1 - z_j^2}{z - z_j} \nonumber \\[2mm]
                                    & \mp & \sqrt{\frac{1}{(1 - z^2) \phi^2(z, \overline{Q}_0^2) d^2(z)} \left[ 4M_\pi^2 \chi_T^{(k)}(\overline{Q}_0^2) - \chi_{DM}^{(k)}(\overline{Q}_0^2) \right] } ~ 
\eea
with $\phi(z, \overline{Q}_0^2) $ given by Eq.~(\ref{eq:phiz}).

After summing over the sample we get the averages $F_{L(U)}(Q^2)$, the standard deviations $\sigma_{L(U)}(Q^2)$ and the correlation coefficient $\rho_{LU}(Q^2)$.
Adopting a uniform distribution between $F_L(Q^2)$ and $F_U(Q^2)$ one finally obtains the following expressions for the em pion form factor $F_\pi^V(Q^2)$ and its variance $\left[ \sigma_\pi^V(Q^2) \right]^2$:
\bea
     \label{eq:average}
     F_\pi^V(Q^2) & = & \frac{1}{2} \left[ F_L(Q^2) + F_U(Q^2) \right] ~ , ~ \\[2mm]
     \label{eq:variance}
     \left[ \sigma_\pi^V(Q^2) \right]^2 & = & \frac{1}{12} \left[ <F_U(Q^2)> - <F_L(Q^2)>\right]^2 \nonumber \\[2mm]
                                                         & + & \frac{1}{3} \left\{ \sigma_L^2(Q^2) + \sigma_U^2(Q^2) +
                                                                   \rho_{LU}(Q^2) \sigma_L(Q^2) \sigma_U(Q^2) \right\} ~ . ~
\eea

\begin{figure}[htb!]
\begin{center}
\includegraphics[scale=0.55]{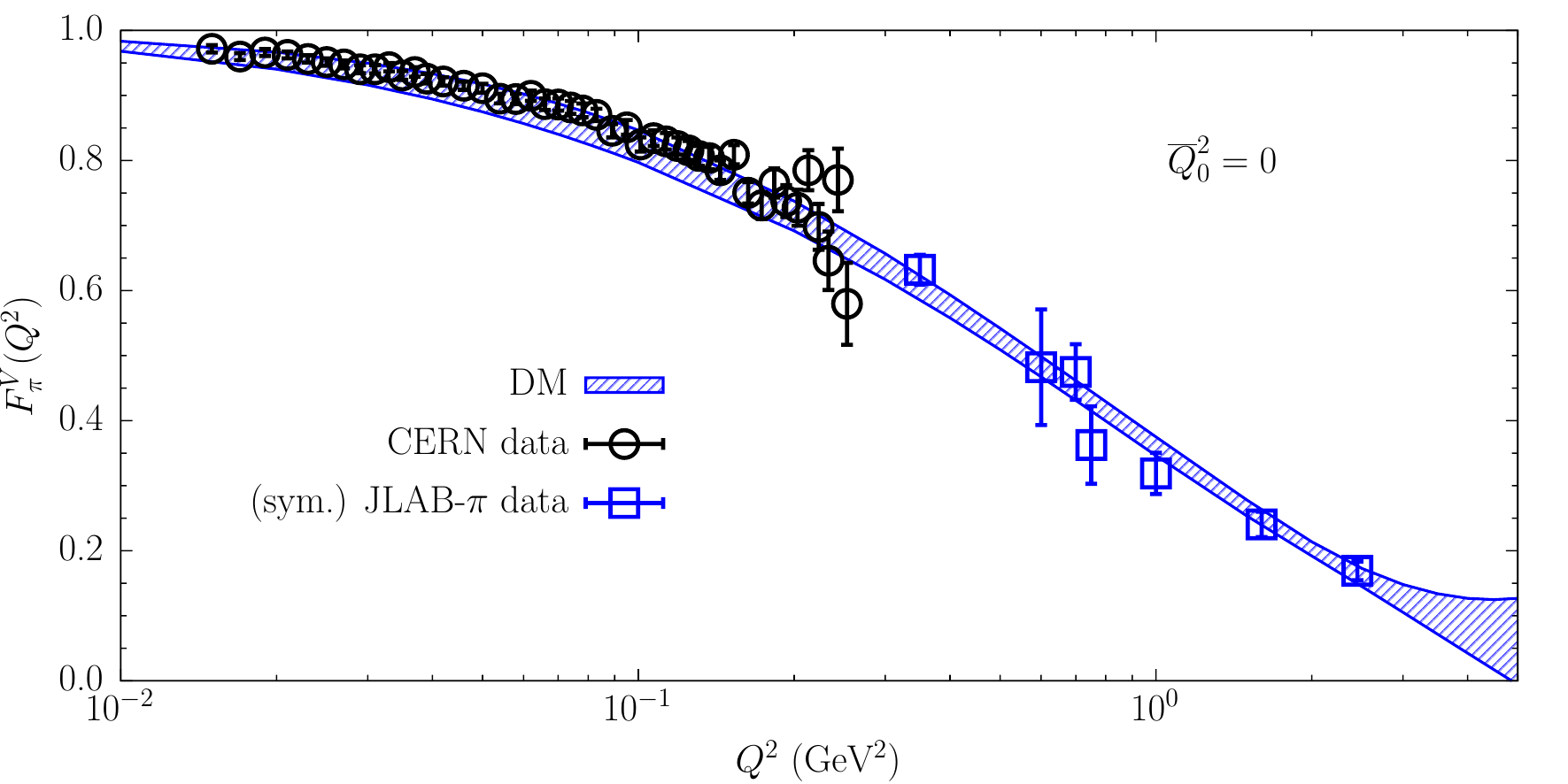}
\end{center}
\vspace{-0.75cm}
\caption{\it \small The DM band (at $1\sigma$ level) for the em pion form factor corresponding to the set of values $\overline{F}$ and the covariance matrix $\overline{C}$ obtained starting from the (symmetrized) electroproduction data analyzed by the JLAB-$\pi$ Collaboration in Ref.~\cite{JeffersonLab:2008jve}, shown as the blue squares. The value $\overline{Q}_0^2 = 0$ is assumed. The experimental data obtained at CERN SPS~\cite{NA7:1986vav} (black circles) are shown just for comparison, but they are not used to construct the DM band (see text).}
\label{fig:bounds}
\end{figure}

The results obtained using Eqs.\,(\ref{eq:average})-(\ref{eq:variance}) starting from the electroproduction data of Ref.~\cite{JeffersonLab:2008jve} are shown in Fig.~\ref{fig:bounds}, where they compared also with the CERN SPS data from Ref.~\cite{NA7:1986vav}. The latter ones are not used to construct the DM band. Nevertheless, the DM predictions at low $Q^2$ based on the electroproduction data at higher $Q^2$ deviates from the CERN data only within $\sim 1\sigma$. 

We close this Section by stressing that the unitary sampling procedure can be easily generalized to any set of hadronic form factors, which must satisfy unitary bounds.

\section{Unitary sampling applied to the CERN and JLAB-$\pi$ experimental data}
\label{sec:CERN+JLABpi}

We now add the direct CERN data~\cite{NA7:1986vav} to the electroproduction JLAB-$\pi$ ones~\cite{JeffersonLab:2008jve} obtaining a total of 52 data points distributed in the $z$-range $[0.043, 0.70]$. We include also the data point $F_0 = 1$ at $Q_0^2 = 0$ (i.e.~$z_0 = 0$) related to the charge conservation, obtaining a total of $N + 1 = 53$ data points.

The CERN data are uncorrelated with the electroproduction JLAB-$\pi$ ones and, therefore, the covariance matrix is block diagonal, namely
\be
    \label{eq:datacov_CERN+JLAB}
    C = \left(
    \begin{tabular}{cc}
    $C^{\rm CERN}$ & $0$ \\[2mm]
    $0$ & $C^{{\rm JLAB-}\pi}$
    \end{tabular}
    \right) ~ , ~
\ee
where, as in the previous Section, the JLAB-$\pi$ covariance matrix $C^{{\rm JLAB-}\pi}$ is diagonal, i.e.\,of the form given in Eq.\,(\ref{eq:datacov}), while for the CERN data it is necessary to include a normalization error $\delta r = 0.45 \%$\,\cite{NA7:1986vav} beyond the tabulated uncertainties $\sigma_i$. We do that by using the following covariance matrix for the CERN data\footnote{Since the DM approach does not make use of any minimization procedure (at variance with the case of explicit $z$-expansions), we should not care about any D'Agostini bias described in Ref.\,\cite{DAgostini:1993arp}.}
\be
    C_{ij}^{\rm CERN} = \sigma_i^2 \delta_{ij} + F_i F_j \delta r^2 ~ , ~
    \label{eq:datacov_CERN}
\ee
where $F_i \pm \sigma_i$ are the tabulated values of the pion form factor in Ref.\,\cite{NA7:1986vav}.

We adopt a sample of $10^4$ events generated according to the PDF~(\ref{eq:PDF}) with the covariance matrix given by Eq.\,(\ref{eq:datacov_CERN+JLAB}). As in Section~\ref{sec:JLABpi} we consider also a gaussian distribution for the non-perturbative transverse susceptibility $4M_\pi^2 \chi_T(\overline{Q}_0^2 = 0) = 0.00574 \, (10)$. This distribution is taken to be uncorrelated with those of the form factor points.
Then, we calculate the values of $\chi_{DM}(\overline{Q}_0^2 = 0)$ (Eq.~(\ref{eq:chiDM})), which range from a minimum equal to $\approx 10^{98}$ up to a maximum equal to $\approx 10^{105}$, i.e.~extremely far from the 2-point bound $4M_\pi^2 \chi_T(\overline{Q}_0^2 = 0) = 5.74 \, (10) \cdot 10^{-3}$.
This is due to the fact that the kinematical coefficients $d_i$ (see Eq.~(\ref{eq:di})) have alternating signs with absolute values ranging from $\sim 10^{14}$ to $\sim 5 \cdot 10^{55}$.
The huge cancellation occurring among the individual contributions to the r.h.s.~of Eq.~(\ref{eq:chiDM}) can be handled using multiple arithmetical precision, achieved by adopting the software package MPFUN from Ref.~\cite{mpfun} and using an adequate number of significant digits for the arithmetic operations.

The use of the PDF~(\ref{eq:PDF_DM_final}) with the modified mean values $\widetilde{F}$ and covariance matrix $\widetilde{C}$, given respectively by Eqs.\,(\ref{eq:Ftilde}) and (\ref{eq:Ctilde}), allows to generate events fulfilling the unitary filter~(\ref{eq:filter}). As the parameter $s$ varies from $\simeq 20$ to $\simeq 45$, the percentage $p$ of unitary events ranges from $\simeq 1\%$ to $\simeq 96 \%$. 
Following the iterative procedure described in the previous Section we obtain a final set of values $\overline{F}$ for the form factor and $\overline{C}$ for the covariance matrix. We stress that both $\overline{F}$ and $\overline{C}$ are almost totally independent on the starting value of $s$, while the quantities $\Delta \simeq  1.0$, $\eta \simeq 1.01$ and $\epsilon \simeq 0.4$ (see Eqs.\,(\ref{eq:Delta})-(\ref{eq:epsilon})) still remain at acceptable values. 

\begin{figure}[htb!]
\begin{center}
\includegraphics[scale=0.50]{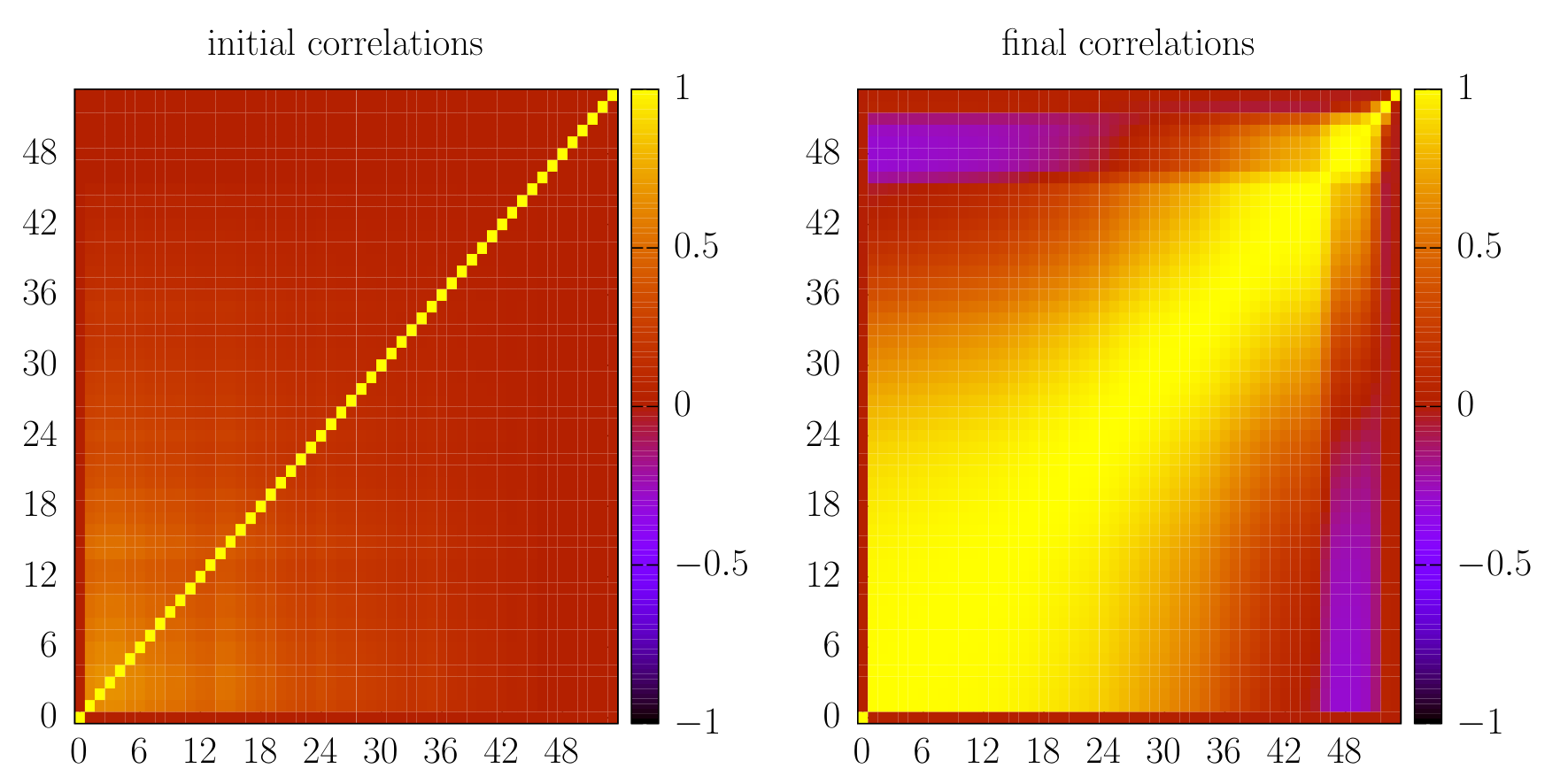}
\end{center}
\vspace{-0.75cm}
\caption{\it \small Heat maps representing the initial correlation matrix of the CERN + JLAB-$\pi$ data (see Eq.\,(\ref{eq:datacov_CERN+JLAB})) and the final correlation matrix $\overline{C}$, obtained using the unitary sampling\,(\ref{eq:PDF_DM_final}) after 10 iterative steps. As in Fig.\,\ref{fig:correlations}  the last label corresponds to the non-perturbative result for the transverse susceptibility $4M_\pi^2 \chi_T(0)$.}
\label{fig:correlations_alldata}
\end{figure}

\begin{figure}[htb!]
\begin{center}
\includegraphics[scale=0.55]{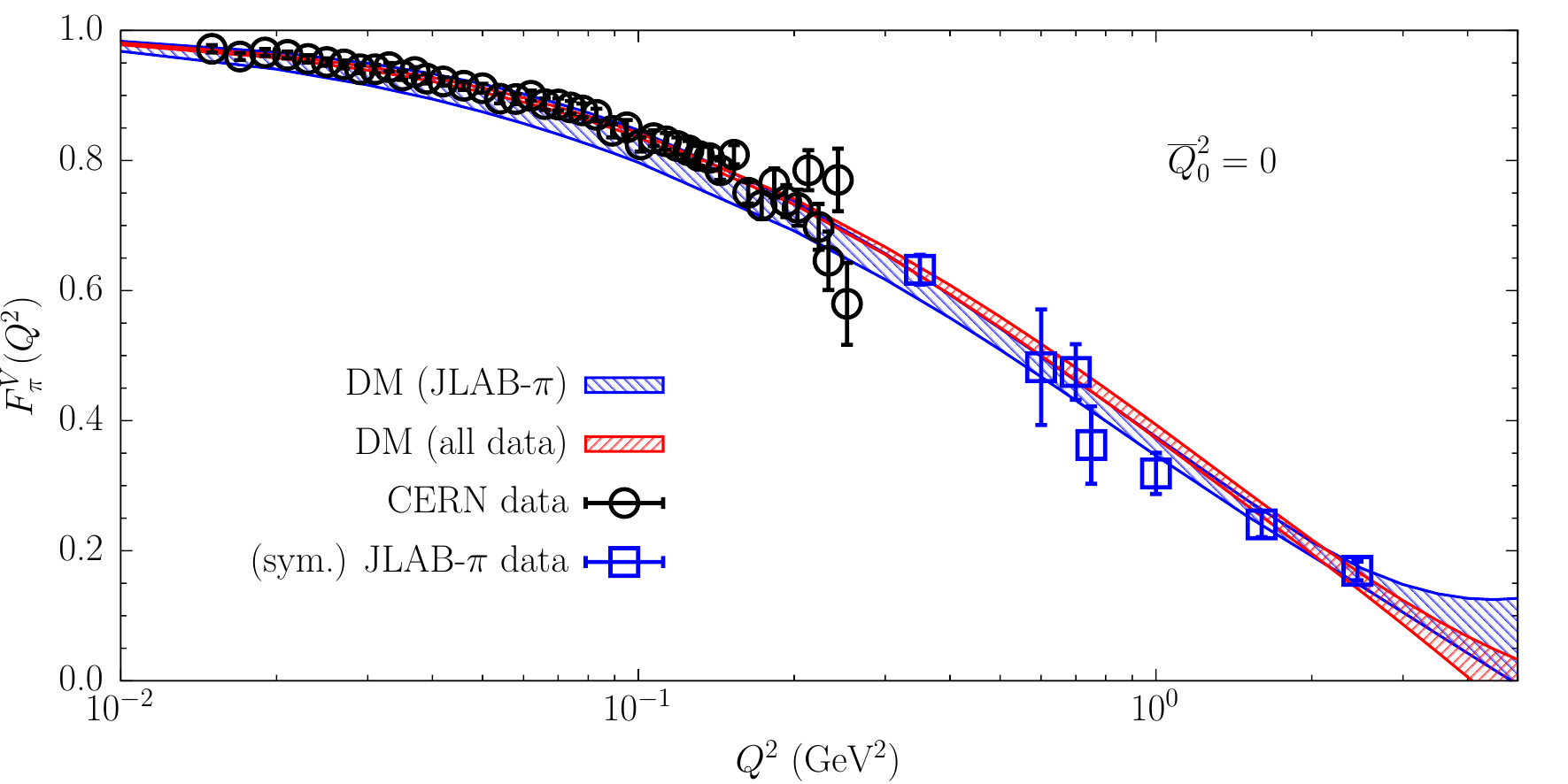}
\end{center}
\vspace{-0.75cm}
\caption{\it \small The red DM band (at $1\sigma$ level) corresponding to the input values $\overline{F}$ and $\overline{C}$ obtained after 10 iterative steps of the unitary sampling procedure applied to both the CERN SPS~\cite{NA7:1986vav} (black circles) and the (symmetrized) electroproduction JLAB-$\pi$~\cite{JeffersonLab:2008jve} (blue squares) experimental data. The blue DM band is the same as in Fig.\,\ref{fig:bounds} and corresponds to the use of only the electroproduction JLAB-$\pi$ data.}
\label{fig:bounds_alldata}
\end{figure}

The unitary form-factor values $\overline{F}$ turn out to be almost normally distributed.
The correlation matrix corresponding to the final covariance matrix $\overline{C}$ is shown in Fig.\,\ref{fig:correlations_alldata} and compared with the initial correlation matrix of the input data $C$ given by Eq.\,(\ref{eq:datacov_CERN+JLAB}).
The corresponding unitary band for the pion form factor, obtained using Eqs.\,(\ref{eq:average})-(\ref{eq:variance}), is shown in Fig.~\ref{fig:bounds_alldata} as the red band.
It can be seen that the inclusion of the CERN data is very effective in producing a more precise band for the pion form factor at all $Q^2$.

Thanks to the DM method we can evaluate the slope of the pion form factor at $Q^2 = 0$ in a way completely independent of any (unitary) parameterization or explicit $z$-expansion.
Using both the CERN and JLAB-$\pi$ data (i.e.~the red band of Fig.\,\ref{fig:bounds_alldata}) we obtain for the pion charge radius $\langle r_\pi \rangle$ the result
\be
    \label{eq:rpi_DM}
    \langle r_\pi \rangle_{DM} = 0. 708 \pm 0.029 ~ \mbox{fm} ~ , ~
\ee
which is definitely larger than the PDG value\,(\ref{eq:rpi_PDG}) by $\approx 1.8\sigma$.
By neglecting the normalization error of the CERN data (i.e.~by putting $\delta r = 0$ in Eq.\,(\ref{eq:datacov_CERN})) we get a more precise result, $\langle r_\pi \rangle_{DM} = 0.695 \pm 0.014$ fm, which differs from the PDG value\,(\ref{eq:rpi_PDG}) by $\approx 2.5\sigma$.

The DM result\,(\ref{eq:rpi_DM}) differs also from the estimate $\langle r_\pi \rangle = 0.663 \pm 0.006$ fm  made in Ref.\,\cite{NA7:1986vav} using the CERN data, but adopting a simple monopole Ansatz for the fitting function. We have checked that using the covariance matrix\,(\ref{eq:datacov_CERN}) a monopole fitting function leads to $\langle r_\pi \rangle = 0.656 \pm 0.008$ fm with a value of $\chi^2/(d.o.f.) \simeq 1.0$. However, when a dipole term is added to the  monopole one, we get  a quite different value of the pion charge radius, namely $\langle r_\pi \rangle = 0.699 \pm 0.024$ fm  (again with a value of $\chi^2/(d.o.f.) \simeq 1.0$), which agrees much better with the DM result\,(\ref{eq:rpi_DM}) for both the mean value and the uncertainty. 
These findings indicate clearly that the estimate of $\langle r_\pi \rangle$ made in Ref.\,\cite{NA7:1986vav} as well as those from Refs.\,\cite{Dally:1982zk, SELEX:2001fbx} are plagued by a significative model dependence, so that they cannot be considered parameterization independent. 

As already pointed out, the DM result\,(\ref{eq:rpi_DM}) differ by $\approx 1.8\sigma$ from the determination of $\langle r_\pi \rangle$ obtained using the abundant and precise timelike $e^+ e^-$ data in Refs.\,\cite{Colangelo:2018mtw, Ananthanarayan:2017efc}, while exhibiting a much larger uncertainty. In order to clarify any possible significance of the above difference a significant improvement of the precision of the experimental data in the spacelike region is called for.

%A further comment concerns the result $\langle r_\pi \rangle = 0.640 \pm 0.007$ fm obtained in Ref.\,\cite{Cui:2021aee}. There, the pion charge radius was extracted from (a subset of) the CERN data\,\cite{NA7:1986vav} using a method based on interpolation via continued fractions augmented by statistical sampling. The attractive feature of this method is that it avoids any assumption on the form of the function used for the representation of the input data and the subsequent extrapolation to $Q^2 = 0$ (as in the case of the DM approach). However, the authors of Ref.\,\cite{Cui:2021aee} did not use the absolute normalization $F_\pi^V(Q^2 = 0) = 1$ (with vanishing uncertainty) and did not discuss the normalization error of the CERN data. Moreover, it is not guaranteed that their approach is consistent with unitarity, because they reproduce quite precisely the input data $F$ that are classified as non-unitary by the DM method.

\begin{figure}[htb!]
\begin{center}
\includegraphics[scale=0.50]{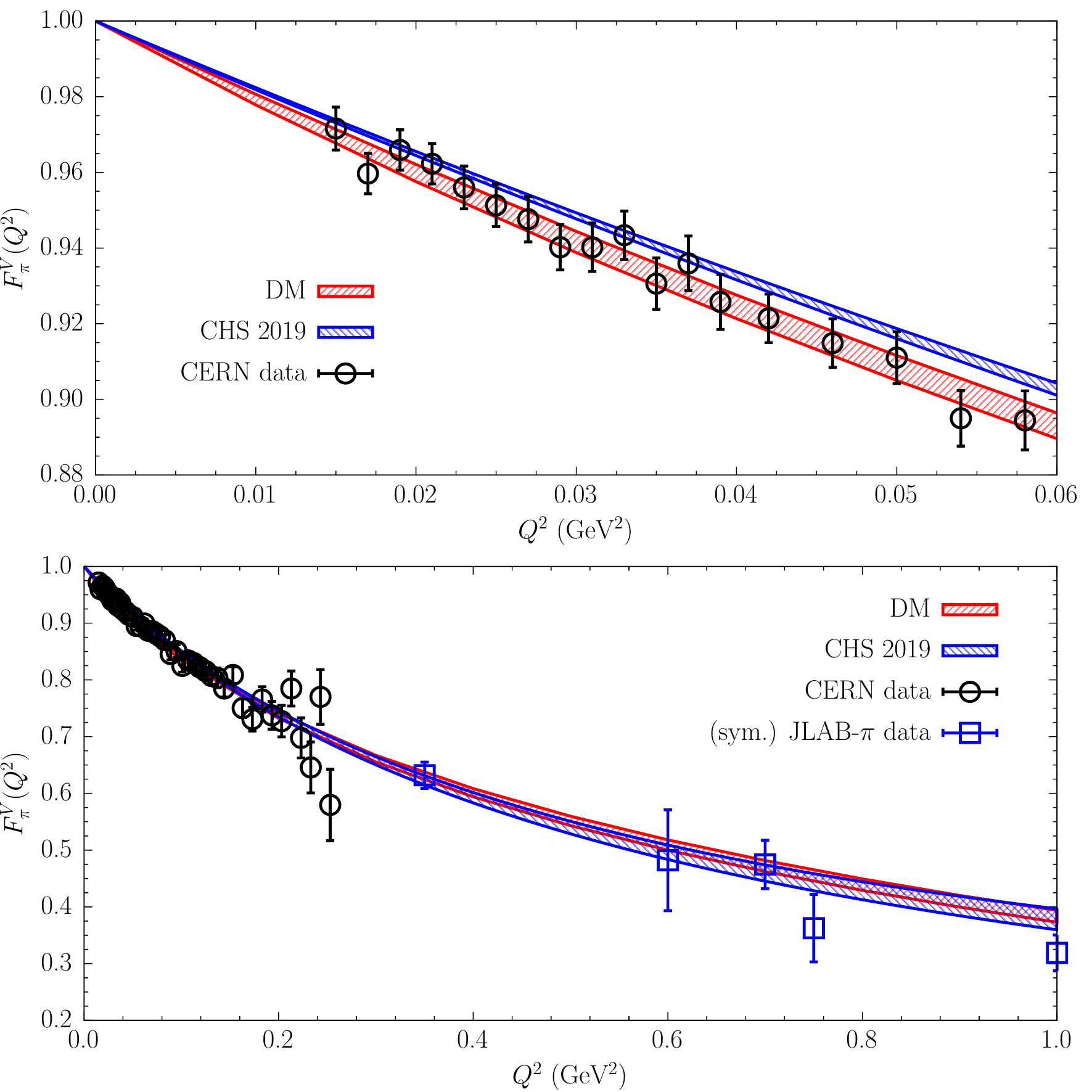}
\end{center}
\vspace{-0.75cm}
\caption{\it \small The red DM band of Fig.\,\ref{fig:bounds_alldata} compared with the results of Ref.\,\cite{Colangelo:2018mtw} (labelled $\mbox{CHS~2019}$), based on a unitary analysis of both timelike $e^+ e^-$ and spacelike CERN data~\cite{NA7:1986vav} (black circles). In the lower panel also the (symmetrized) electroproduction JLAB-$\pi$~\cite{JeffersonLab:2008jve} (blue squares) experimental data are shown.}
\label{fig:comparison_CHS19}
\end{figure}

In Fig.\,\ref{fig:comparison_CHS19} our DM band is compared with the results of Ref.\,\cite{Colangelo:2018mtw}, based on a unitary analysis of both timelike $e^+ e^-$ and spacelike CERN data.
A good overall agreement is observed up to $Q^2 \simeq 1$ (GeV/c)$^2$ (see lower panel), while a zoom in the low-$Q^2$ region (upper panel) shows that:
\begin{itemize}
\item the use of the very precise and dense timelike $e^+ e^-$ data leads to the accurate result for the pion charge radius obtained in Ref.\,\cite{Colangelo:2018mtw}, namely $\langle r_\pi \rangle = 0.655 \pm 0.003$ fm;
\item the DM band is in better agreement with the spacelike CERN data w.r.t.~to the results of the dispersive analysis of Ref.\,\cite{Colangelo:2018mtw}. In this respect, since the pion charge radius is correlated with the $2\pi$ contribution to the muon HVP term\,\cite{Colangelo:2020lcg}, it would be interesting to analyze the possible impact of the new $e^+ e^- \to \pi^+ \pi^-$ experimental data from the CMD-3 Collaboration\,\cite{CMD-3:2023alj} on such correlated quantities\,\cite{Colangelo:2023rqr}.
\end{itemize}

We close this Section by observing that the addition of the spacelike data of the F2\,\cite{Dally:1982zk} and the (less precise) SELEX\,\cite{SELEX:2001fbx} experiments at FNAL do not change significantly the results shown in Fig.\,\ref{fig:bounds_alldata} and in Eq.\,(\ref{eq:rpi_DM}).
In particular, using as input the F2\,\cite{Dally:1982zk}, CERN\,\cite{NA7:1986vav} and JLAB-$\pi$\,\cite{JeffersonLab:2008jve} data sets (for a total of 66 data points including the absolute normalization at $Q^2 = 0$) we get
 \be
    \label{eq:rpi_DM_alldata}
    \langle r_\pi \rangle_{DM} = 0. 703 \pm 0.027 ~ \mbox{fm} ~ , ~
\ee
which differs from the PDG value\,(\ref{eq:rpi_PDG}) by $\approx 1.6\sigma$. Had we chosen the vector susceptibility $4 M_\pi^2 \chi_T(\overline{Q}_0^2 = 0)$ to be equal to the value $0.00550\,(4)$, which corresponds to the evaluation of Eq.\,(\ref{eq:chiT_FV}) using the dispersive pion form factor $|F_\pi^V(\omega)|$ from Ref.\,\cite{Colangelo:2018mtw} up to $\omega = 1$ GeV, the value of the pion charge radius $\langle r_\pi \rangle_{DM}$ would result to be $0.702 \pm 0.026$ fm, i.e.~very close to Eq.\,(\ref{eq:rpi_DM_alldata}).

\section{Unitary BGL fit}
\label{sec:BGLfit}

In this Section we perform a BGL analysis of the spacelike data after constructing a truncated $z$-expansion in which unitarity is built-in.

As known, in the BGL approach\,\cite{Boyd:1997kz} the product of the pion form factor $F_\pi^V(Q^2)$ times the kinematical function $\phi$ is analytic inside the unit circle $|z| = 1$ and, therefore, it can be expanded as
\be
    \label{eq:BGL}
    F_\pi^V(Q^2) = \frac{\sqrt{4M_\pi^2 \chi_T(\overline{Q}_0^2)}}{ \phi(z, \overline{Q}_0^2)} \sum_{k = 0}^\infty a_k(\overline{Q}_0^2) z^k ~ , ~
\ee
where $\phi(z, \overline{Q}_0^2)$ is given by Eq.\,(\ref{eq:phiz}) and the coefficients $a_k(\overline{Q}_0^2)$ (which are real and depend implicitly also on the choice of the auxiliary quantity $t_0$ appearing in the definition of the conformal variable\,(\ref{eq:conformal})) are constrained by the unitary bound
\be
    \label{eq:unitarity_BGL}
    \sum_{k = 0}^\infty a_k^2(\overline{Q}_0^2) \leq 1 ~ . ~
\ee
In order to lighten the notation, we will indicate hereafter the coefficients $a_k(\overline{Q}_0^2)$ simply as $a_k$. 

The $z$-expansion\,(\ref{eq:BGL}), truncated at some order $N_{\rm BGL}$, can be used as a fitting Ansatz to describe the spacelike data for the pion form factor, namely
\be
    \label{eq:BGL_truncated}
    F_\pi^{\rm BGL}(Q^2) = \frac{\sqrt{4M_\pi^2 \chi_T(\overline{Q}_0^2)}}{ \phi(z, \overline{Q}_0^2)} \sum_{k = 0}^{N_{\rm BGL}} a_k z^k ~ 
\ee
with the unitary bound given by
\be
    \label{eq:unitarity_BGL_truncated}
    \sum_{k = 0}^{N_{\rm BGL}} a_k^2 \leq 1 ~ . ~
\ee
The truncation introduces unavoidably a model dependence. In Ref.\,\cite{Boyd:1997kz} it was proposed to look  at the truncation error $\delta F_\pi^{\rm BGL}(Q^2)$, defined as
\be
    \label{eq:truncation_error}
    \delta F_\pi^{\rm BGL}(Q^2) = \frac{\sqrt{4M_\pi^2 \chi_T(\overline{Q}_0^2)}}{ \phi(z, \overline{Q}_0^2)} \sum_{k = N_{\rm BGL} + 1}^\infty a_k z^k ~ , ~
\ee
which has un upper bound given by
\be
     \label{eq:truncation_bound}
     |\delta F_\pi^{\rm BGL}(Q^2)| \leq \frac{\sqrt{4M_\pi^2 \chi_T(\overline{Q}_0^2)}}{ |\phi(z, \overline{Q}_0^2)|} ~ \sqrt{1 - \sum_{k = 0}^{N_{\rm BGL}} a_k^2} ~ 
                                             \frac{|z|^{N_{\rm BGL} + 1}}{\sqrt{1 - z^2}} ~ . ~
\ee
%Note that the bound\,(\ref{eq:truncation_bound}) requires that the truncated BGL fit\,(\ref{eq:BGL_truncated}) must satisfy the unitary bound\,(\ref{eq:unitarity_BGL_truncated}).
The weak point of the bound\,(\ref{eq:truncation_bound}) on the truncation error\,(\ref{eq:truncation_error}) is that it may represent an upper limit on the difference between the true function\,(\ref{eq:BGL}) and the truncated fit\,(\ref{eq:BGL_truncated}) {\em if and only if} the  coefficients $a_k$ with $k = 0, 1, ... N_{\rm BGL}$ of the true function coincides exactly with those of the truncated fit. However, generally speaking, this is not guaranteed and, therefore, the truncation bound\,(\ref{eq:truncation_bound}) is of limited use, particularly when the bound\,(\ref{eq:unitarity_BGL_truncated}) is almost saturated.

We now want to apply the truncated BGL fit\,(\ref{eq:BGL_truncated}) to the description of the spacelike pion data with the unitary bound\,(\ref{eq:unitarity_BGL_truncated}) built-in. 
This can be achieved through a simple procedure based on a {\em hyperspherical} transformation (see, e.g., Ref.\,\cite{hyperspherical}) described in Appendix\,\ref{sec:appB}. 
We generate a sample of $10^3$ events $F$ according to the PDF~(\ref{eq:PDF}) using the direct CERN\,\cite{NA7:1986vav} and electroproduction JLAB-$\pi$\,\cite{JeffersonLab:2008jve} data with the covariance matrix $C$ given by Eq.\,(\ref{eq:datacov_CERN+JLAB}).  At the same time, as in Sections~\ref{sec:JLABpi} and \ref{sec:CERN+JLABpi}, we consider also a gaussian distribution for the non-perturbative transverse susceptibility $4M_\pi^2 \chi_T(\overline{Q}_0^2 = 0) = 0.00574 \, (10)$. This distribution is taken to be uncorrelated with those of the form factor points.

For each event we fit the input data using the BGL Ansatz\,(\ref{eq:BGL_truncated}) corresponding to a given truncation order $N_{\rm BGL}$ with the unitarity bound\,(\ref{eq:unitarity_BGL_truncated}) built-in through the hyperspherical procedure of Appendix\,\ref{sec:appB}. Then, we minimize the reduced $\chi_r^2$-variable given by
\be
    \label{eq:chi2_correlated}
    \chi_r^2 \equiv \frac{1}{N - N_{\rm BGL}} \sum_{i,j = 0}^N \left( F_i^{\rm BGL} - F_i) C_{ij}^{-1} (F_j^{\rm BGL} - F_j \right) ~ , ~
\ee 
obtaining the best unitary BGL fit for a given truncation order $N_{\rm BGL}$.
We have considered values of $N_{\rm BGL}$ between 2 and 10 and the corresponding results for the BGL parameters $a_k$ are shown in Table\,\ref{tab:ak} for $N_{\rm BGL} = 2, 4, 6, 8, 10$.
Note that the value of the parameter $a_0$ is constrained by the absolute normalization $F_\pi^V(Q^2 = 0) = 1$ and by the value of the transverse susceptibility $4M_\pi^2 \chi_T(\overline{Q}_0^2 = 0) = 0.00574 \, (10)$, leading to $a_0 = 0.190 \pm 0.002$ for any value of $N_{\rm BGL}$. 
\begin{table}[htb!]
\renewcommand{\arraystretch}{1.2}
\begin{center}
\begin{tabular}{||c||c|c|c|c|c||}
\hline
           & $N_{\rm BGL} = 2$ & $N_{\rm BGL} = 4$ & $N_{\rm BGL} = 6$ & $N_{\rm BGL} = 8$ & $N_{\rm BGL} = 10$ \\ \hline \hline
$a_0$ & $+0.190 \pm 0.002$ & $+0.190 \pm 0.002$ & $+0.190 \pm 0.002$ & $+0.190 \pm 0.002$ & $+0.190 \pm 0.002$ \\ \hline
$a_1$ & $+0.203 \pm 0.007$ & $+0.159 \pm 0.014$ & $+0.153 \pm 0.015$ & $+0.154 \pm 0.016$ & $+0.155 \pm 0.016$ \\ \hline
$a_2$ & $-0.58 \pm 0.01$      & $-0.25 \pm 0.07$      & $-0.24 \pm 0.09$      & $-0.26 \pm 0.101$    & $-0.27 \pm 0.11$ \\ \hline
$a_3$ &                                 & $-0.75 \pm 0.08$       & $-0.53 \pm 0.22$      & $-0.42 \pm 0.27$       & $-0.40 \pm 0.28$ \\ \hline
$a_4$ &                                 & $+0.52 \pm 0.07$      & $-0.24 \pm 0.20$      & $-0.27 \pm 0.21$       & $-0.25 \pm 0.21$ \\ \hline
$a_5$ &                                 &                                   & $+0.25 \pm 0.12$     & $+0.05 \pm 0.20$       & $-0.00 \pm 0.21$ \\ \hline
$a_6$ &                                 &                                   & $+0.57 \pm 0.17$     & $+0.27 \pm 0.13$       & $+0.24 \pm 0.17$ \\ \hline
$a_7$ &                                 &                                   &                                  & $+0.35 \pm 0.16$       & $+0.30 \pm 0.14$ \\ \hline
$a_8$ &                                 &                                   &                                  & $+0.35 \pm 0.21$       & $+0.25 \pm 0.16$ \\ \hline
$a_9$ &                                 &                                   &                                  &                                    & $+0.21 \pm 0.18$ \\ \hline 
$a_{10}$  &                            &                                   &                                  &                                    & $+0.18 \pm 0.19$ \\ \hline \hline
$(r_0^2)_{min}$               & 0.36 & 0.25 & 0.34 & 0.32 & 0.29 \\ \hline
$(r_0^2)_{max}$              & 0.47 & 1.00 & 1.00 & 1.00 & 1.00 \\ \hline
$\langle r_0^2 \rangle$    & 0.41 & 0.98 & 0.98 & 0.96 & 0.96 \\ \hline \hline
$\langle \chi_r^2 \rangle$ & $1.13 \pm 0.35$ & $0.98 \pm 0.35$ & $0.98 \pm 0.35$ & $1.04 \pm 0.36$ &  $1.13 \pm 0.38$ \\ \hline \hline
\end{tabular}
\end{center}
\renewcommand{\arraystretch}{1.0}
\vspace{-0.5cm}
\caption{\it \small Values of the parameters $a_k$ of the unitary BGL fit\,(\ref{eq:BGL_truncated}) for various values of the truncation order $N_{\rm BGL}$. The minimum $(r_0^2)_{min}$, maximum $(r_0^2)_{max}$ and the average $\langle r_0^2 \rangle$ values of the parameter $r_0^2$ representing the unitary bound (see Eq.\,(\ref{eq:hyperradius}) for its definition) are shown together with the average value of the reduced $\chi_r^2$-variable\,(\ref{eq:chi2_correlated}) corresponding to a sample of $10^3$ events generated using the direct CERN\,\cite{NA7:1986vav} and electroproduction JLAB-$\pi$\,\cite{JeffersonLab:2008jve} data with the covariance matrix given by Eq.\,(\ref{eq:datacov_CERN+JLAB}).}
\label{tab:ak}
\end{table}

The following comments are in order.

\begin{figure}[htb!]
\begin{center}
\includegraphics[scale=0.55]{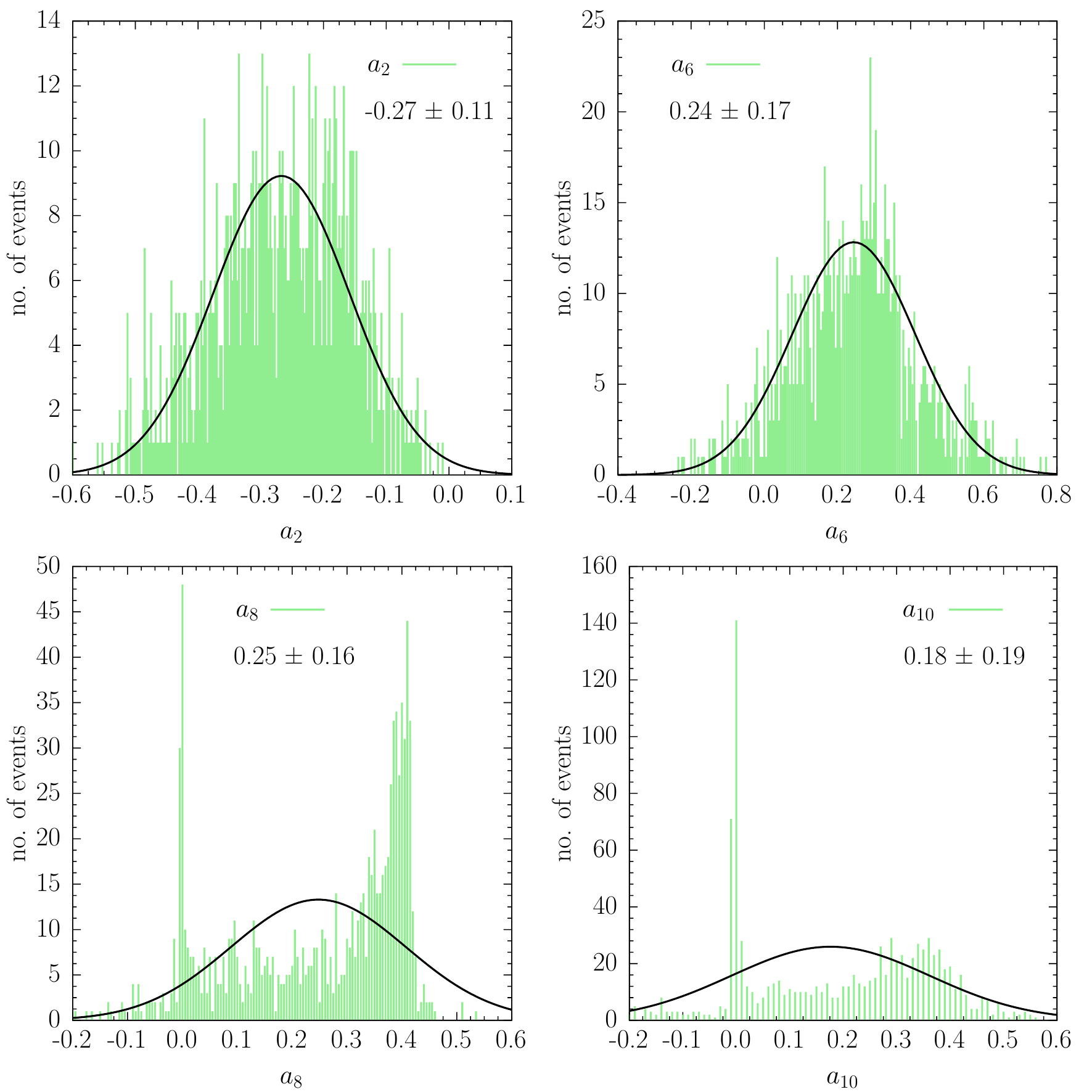}
\end{center}
\vspace{-0.75cm}
\caption{\it \small Distributions of the parameter $a_2$, $a_6$, $a_8$ and $a_{10}$ of the unitary BGL fit\,(\ref{eq:BGL_truncated}) with $N_{BGL} = 10$ shown in Table\,\ref{tab:ak}. The black solid lines represent a gaussian fit of the various histograms.}
\label{fig:parameters}
\end{figure}

\begin{figure}[htb!]
\begin{center}
\includegraphics[scale=0.50]{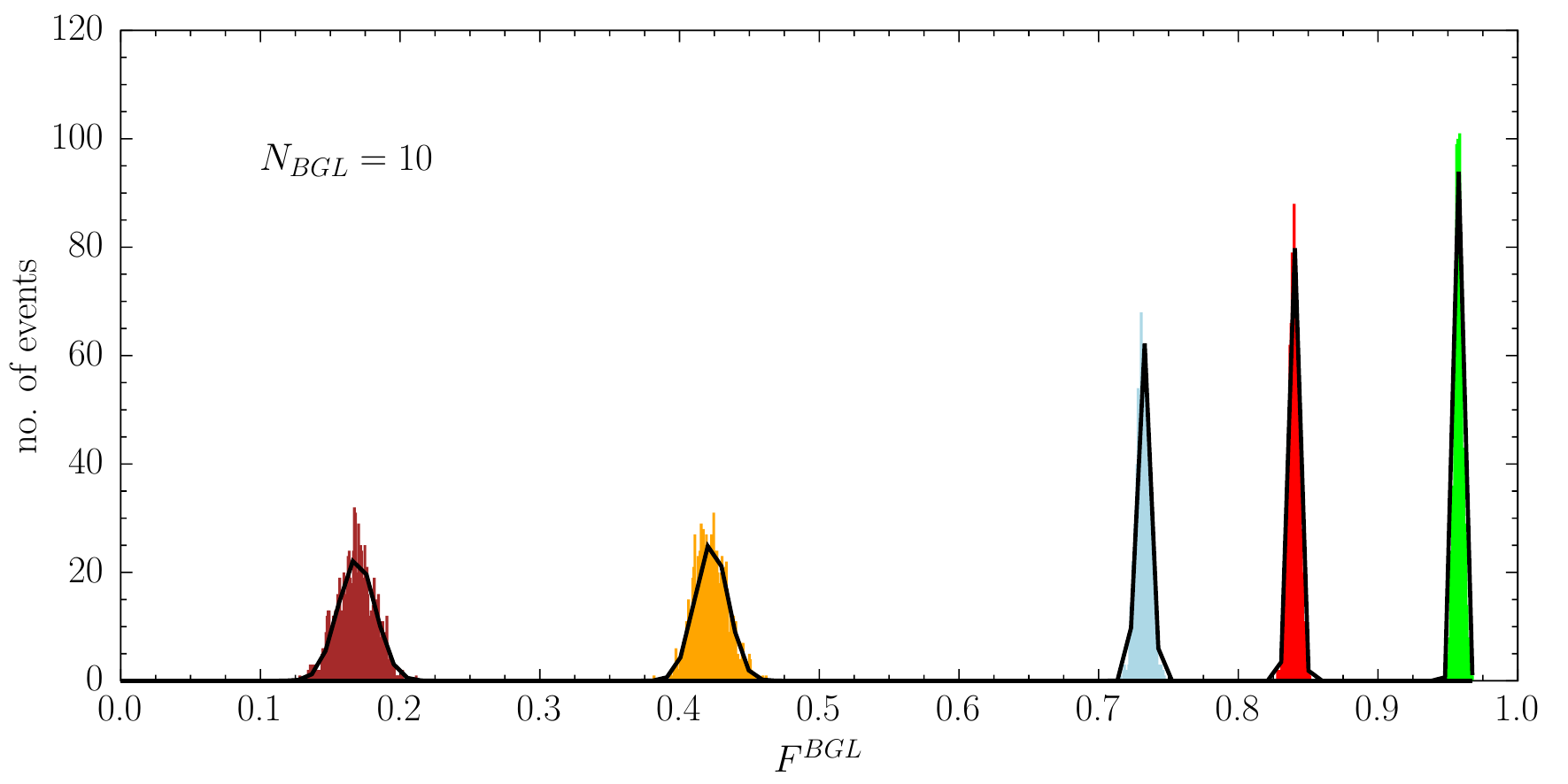}
\end{center}
\vspace{-0.75cm}
\caption{\it \small Distributions of the form factors values $F^{BGL}$ of the unitary BGL fit\,(\ref{eq:BGL_truncated}) with $N_{BGL} = 10$. The green, red, cyan, orange and brown histograms correspond respectively to $Q^2 = 0.021, 0.101, 0.203,  0.75, 2.45$ GeV$^2$. The black solid lines represent a gaussian fit of the various histograms.}
\label{fig:ffs_BGL}
\end{figure}

\begin{figure}[htb!]
\begin{center}
\includegraphics[scale=0.55]{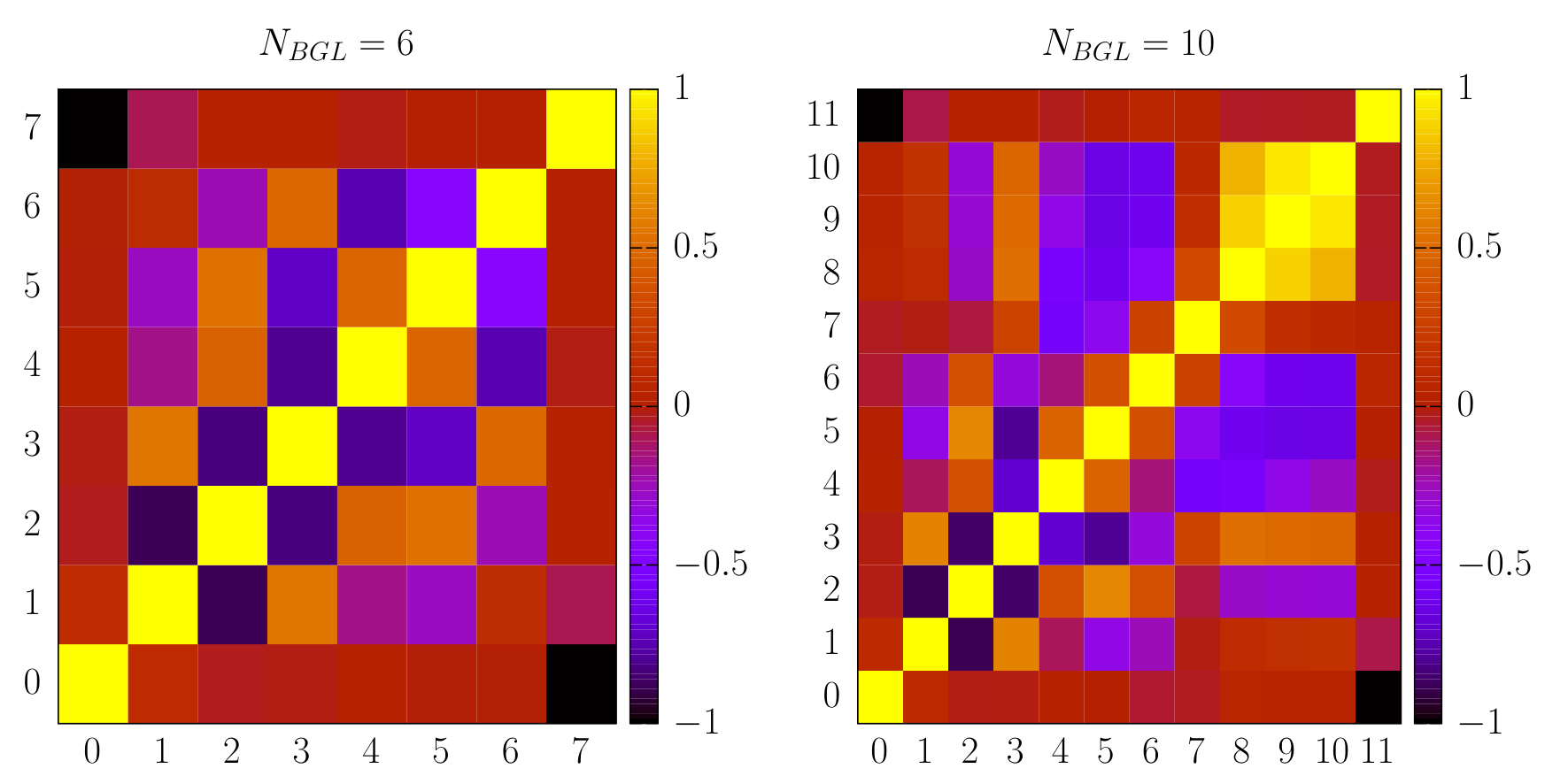}
\end{center}
\vspace{-0.75cm}
\caption{\it \small Heat maps representing the correlation matrix for the parameter $a_k$ of the unitary BGL fit\,(\ref{eq:BGL_truncated}) with $N_{BGL} = 6$ (left panel) and $N_{BGL} = 10$ (right panel). The last label ($7$ in the left panel and $11$ in the right one) corresponds to the transverse susceptibility $4M_\pi^2 \chi_T(\overline{Q}_0^2 = 0)$. Note that this quantity is anticorrelated with $a_0$ because of the absolute normalization condition $F_\pi^V(Q^2 = 0) = 1$.}
\label{fig:correlations_parameters}
\end{figure}

\begin{itemize}

\item As $N_{BGL}$ increases, the mean values and uncertainties of the coefficients $a_k$ with $k \lesssim 6$ tend to remain stable.

\item The coefficients $a_k$ for high-order monomials ($k \gtrsim 6$) have large uncertainties (up to $\approx 100 \%$).

\item For $N_{BGL} \geq 4$ the unitary bound\,(\ref{eq:unitarity_BGL_truncated}) is almost saturated, while unitarity is strictly fulfilled (see in Table\,\ref {tab:ak} the maximum and average values of the parameter $r_0^2$, defined in Eq.\,(\ref{eq:hyperradius})) .

\item The values of the reduced $\chi_r^2$-variable\,(\ref{eq:chi2_correlated}) are always consistent with unity.

\item The distribution of the coefficients $a_k$ is approximately normal for low-order monomials ($k \lesssim 6$), while significative deviations from a Gaussian distribution occur in the case of higher-order monomials (see Fig.\,\ref{fig:parameters}) due mainly to the saturation of the unitary bound\,(\ref{eq:unitarity_BGL_truncated}). Nevertheless, the form factors values are almost normally distributed, as shown in Fig.\,\ref{fig:ffs_BGL}.

\item For a given order $N_{BGL}$ of the truncation the parameters $a_k$ are generally anticorrelated (see Fig.\,\ref{fig:correlations_parameters}).

\end{itemize}


\begin{figure}[htb!]
\begin{center}
\includegraphics[scale=0.55]{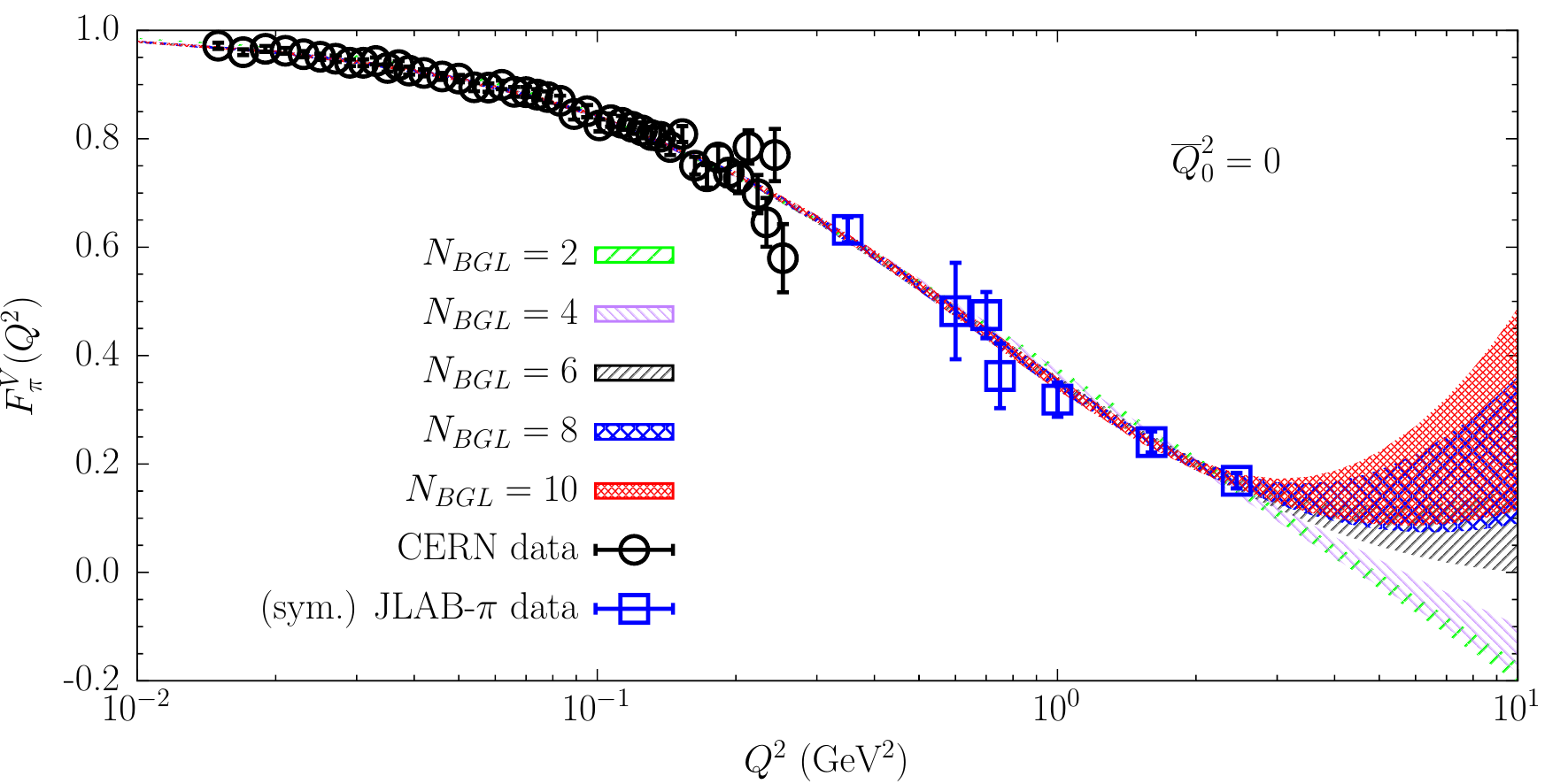}
\end{center}
\vspace{-0.75cm}
\caption{\it \small The unitary bands (at $1\sigma$ level) obtained using the unitary BGL fit\,(\ref{eq:BGL_truncated}) for various values of the truncation order $N_{\rm BGL}$, applied to the CERN SPS~\cite{NA7:1986vav} (black circles) and the (symmetrized) electroproduction JLAB-$\pi$~\cite{JeffersonLab:2008jve} (blue squares) experimental data.}
\label{fig:BGL}
\end{figure}

The bands for the pion form factors corresponding to the unitary BGL fits of Table\,\ref{tab:ak} are shown in Fig.\,\ref{fig:BGL} for various values of the truncation order $N_{\rm BGL}$. 
It can be seen that in the kinematical region covered by the CERN and JLAB-$\pi$ data (i.e.\,for $Q^2 \lesssim 2.5$ GeV$^2$) the results of the unitary BGL fit are stable against the order $N_{\rm BGL}$ of the truncation, while at larger values of $Q^2$ the bands are largely unstable and no extrapolation is possible at least from $N_{BGL} \leq 10$.
This finding clearly indicates the inadequacy of the estimate of the truncation error based on Eq.\,(\ref{eq:truncation_bound}), since the saturation of the unitary bound\,(\ref{eq:unitarity_BGL_truncated}) would imply a negligible truncation error that is not observed at all at large $Q^2$.

Moreover, a closer look to Figs.\,\ref{fig:bounds_alldata} and\,\ref{fig:BGL} reveals that the BGL and DM bands differ by $1 \div 2\sigma$ in the kinematical region of the electroproduction JLAB-$\pi$ data (i.e.~for $0.35 \leq Q^2 ~ (\mbox{GeV}^2) \leq 2.45$). This observation will be discussed in a while.

\begin{figure}[htb!]
\begin{center}
\includegraphics[scale=0.55]{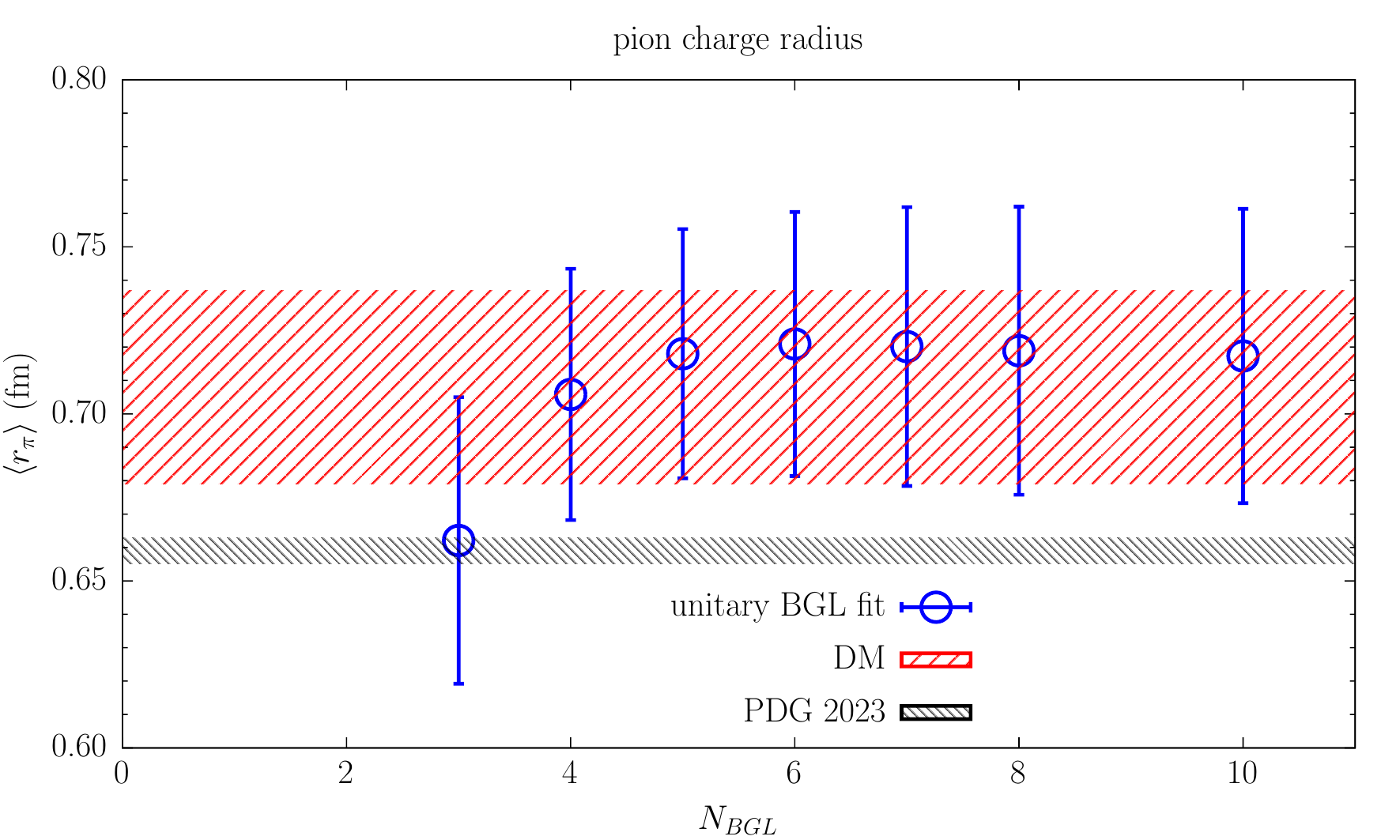}
\end{center}
\vspace{-0.75cm}
\caption{\it \small The pion charge radius $\langle r_\pi \rangle$ corresponding to the unitary BGL fit\,(\ref{eq:BGL_truncated}) for various values of the truncation order $N_{\rm BGL}$, applied to the CERN~\cite{NA7:1986vav} and the (symmetrized) electroproduction JLAB-$\pi$~\cite{JeffersonLab:2008jve} experimental data. The red band correspond to the DM result\,(\ref{eq:rpi_DM}), while the black one to the PDG value\,(\ref{eq:rpi_PDG}) given in Ref.\,\cite{ParticleDataGroup:2022pth}.}
\label{fig:radius}
\end{figure}

The pion charge radius $\langle r_\pi \rangle$ corresponding to the BGL fit\,(\ref{eq:BGL_truncated}) is explicitly given by
\be
     \label{eq:rpi_BGL_a1}
     \langle r_\pi \rangle_{BGL} = \sqrt{\frac{3}{8 M_\pi^2} \left( \frac{3}{2} - \frac{a_1}{a_0} \right)} ~ . ~
\ee
The results for $\langle r_\pi \rangle_{BGL} $ obtained for the unitary BGL fits of Table\,\ref{tab:ak} are shown in Fig.\,\ref{fig:radius} and exhibit a good convergence as a function of the truncation order $N_{BGL}$. We get $\langle r_\pi \rangle_{BGL} = 0.717 \pm 0.044$ fm, which is consistent with the DM result\,(\ref{eq:rpi_DM}) with an uncertainty larger by a factor $\simeq 1.5$.
Including also the spacelike data of the F2 experiment at FNAL\,\cite{Dally:1982zk} we obtain
\be
    \label{eq:rpi_BGL}
    \langle r_\pi \rangle_{BGL} = 0.711 \pm 0.039 ~ \mbox{fm} ~ , ~
\ee
which differs from the PDG value\,(\ref{eq:rpi_PDG}) by $\simeq 1.3\sigma$.
 We mention that, if we adopt for the vector susceptibility $4 M_\pi^2 \chi_T(\overline{Q}_0^2 = 0)$ the value $0.00550\,(4)$, which corresponds to the evaluation of Eq.\,(\ref{eq:chiT_FV}) using the dispersive pion form factor $|F_\pi^V(\omega)|$ from Ref.\,\cite{Colangelo:2018mtw} up to $\omega = 1$ GeV, the pion charge radius $\langle r_\pi \rangle_{BGL}$ remains basically unchanged with respect to Eq.\,(\ref{eq:rpi_BGL}).

The instability of the unitary BGL fits of the pion form factor observed in Fig.\,\ref{fig:BGL} at $Q^2 \gtrsim 2.5$ GeV$^2$ and the larger uncertainty of the BGL result\,(\ref{eq:rpi_BGL}) w.r.t.~the corresponding DM result\,(\ref{eq:rpi_DM}) are connected to non-unitary effects present in the {\em fitted} CERN and JLAB-$\pi$ experimental data.
According to the DM method these data, corresponding to the input sets $F$ and $C$ for the pion form factor values and their covariance matrix, do not fulfill the unitary bound\,(\ref{eq:bounds}). The unitary sampling procedure, described in Section\,\ref{sec:sampling}, has allowed us to get in Section\,\ref{sec:CERN+JLABpi} a new set of input data $\overline{F}$ and $\overline{C}$ fulfilling the unitary bound\,(\ref{eq:bounds}). 

\begin{figure}[htb!]
\begin{center}
\includegraphics[scale=0.55]{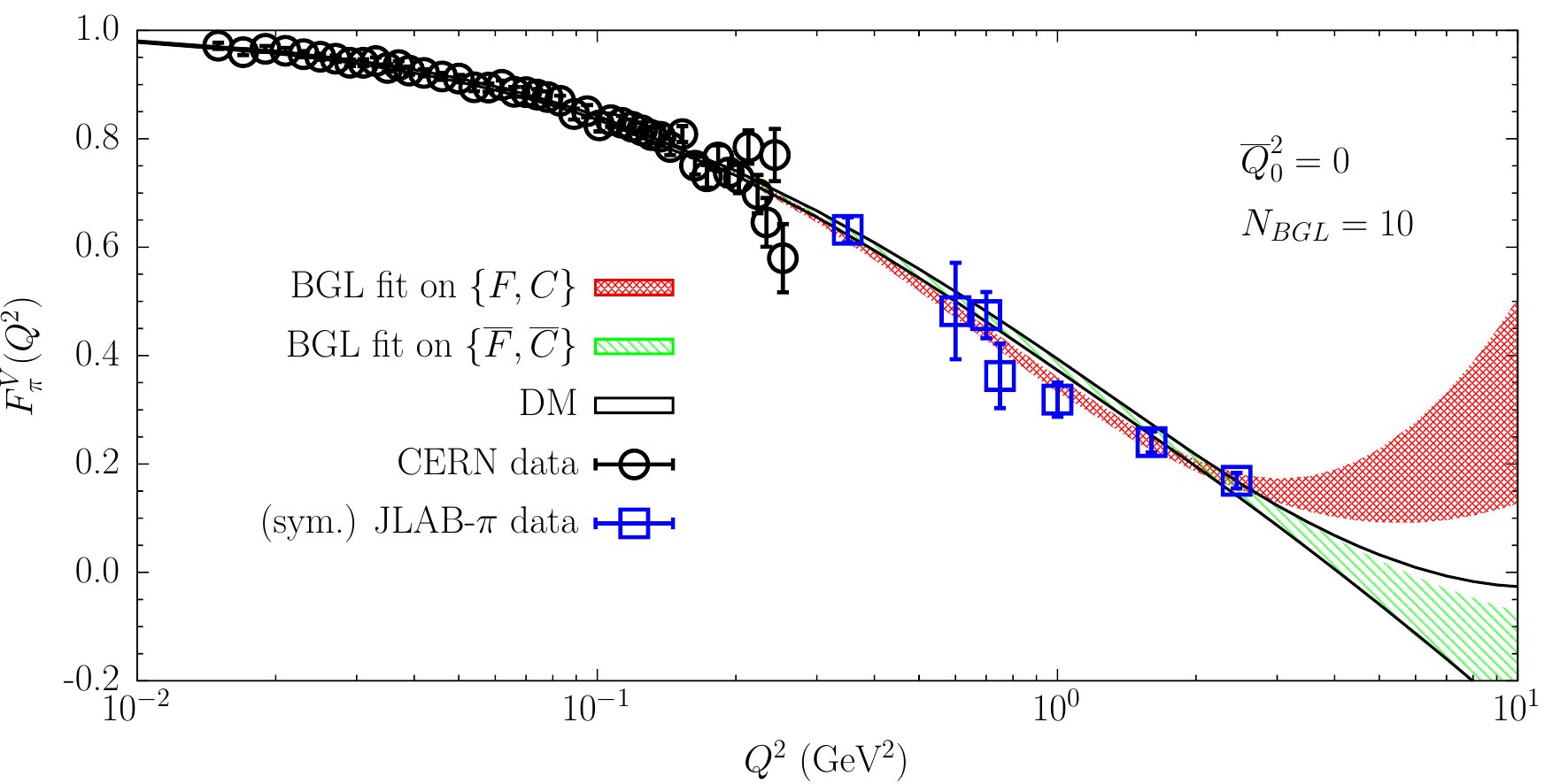}
\end{center}
\vspace{-0.75cm}
\caption{\it \small The unitary band (at $1\sigma$ level) obtained using the unitary BGL fit\,(\ref{eq:BGL_truncated}) for various values of the truncation order $N_{\rm BGL}$, applied to the CERN SPS~\cite{NA7:1986vav} (black circles) and the (symmetrized) electroproduction JLAB-$\pi$~\cite{JeffersonLab:2008jve} (blue squares) experimental data.}
\label{fig:BGL_DM}
\end{figure}

Therefore, we apply the unitary BGL fit\,(\ref{eq:BGL_truncated}) with $N_{BGL} = 10$ to the unitary set of input data $\overline{F}$ and $\overline{C}$  defined by the DM method, i.e.~we replace in Eq.\,(\ref{eq:chi2_correlated}) the input data set $F$ and $C$ with the DM set $\overline{F}$ and $\overline{C}$. 
We stress that only in this way unitarity is fulfilled both by the fitting function (by construction) and by the fitted data (by the unitary sampling procedure). 
The corresponding band for the pion form factor is shown in Fig.\,\ref{fig:BGL_DM} as the green band. 
The comparison with the bands from Figs.\,\ref{fig:bounds_alldata} and\,\ref{fig:BGL} indicates very clearly that non-unitary effects are significantly present when the set $F$ and $C$ of input data are adopted regardless the fact that the BGL fit satisfies unitarity by construction. Such non-unitary effects produces not only a large instability of the BGL fits in the kinematical region not covered by the experimental data, but they can also affect significantly the fitting results in the electroproduction region.
Finally, the pion charge radius corresponding to the green band of Fig.\,\ref{fig:BGL_DM} is $\langle r_\pi \rangle = 0.707 \pm 0.029$ fm; both the central value and the uncertainty are now in nice agreement with the DM result\,(\ref{eq:rpi_DM}).

\section{Unitary BCL approach}
\label{sec:BCLfit}

An alternative $z$-expansion is the so-called BCL one, originally proposed in Ref.\,\cite{Bourrely:2008za} to address the momentum dependence of the hadronic form factors describing the semileptonic $B \to \pi \ell \nu_\ell$ decays.

In the case of the em pion form factor the BCL expansion is a direct $z$-expansion and its truncated form read as
\be
    \label{eq:BCL_truncated}
    F_\pi^{\rm BCL}(Q^2) = \sum_{k = 0}^{N_{\rm BCL}} b_k z^k ~ . ~
\ee

An interesting feature of the BCL approach is the inclusion of the analytic constraint at the annihilation threshold $z = -1$. Indeed, angular momentum conservation requires that in the timelike region $\mbox{Im}[F_\pi^V(\omega)] \propto (\omega^2 - 4 M_\pi^2)^{3/2}$. In turn this implies that the real part of pion form factor should have a vanishing first derivative at the annihilation threshold, namely $dF_\pi^V(z) / dz = 0$ at $z = -1$. Such a constraint can be easily implemented in the truncated BCL approach (at variance with the BGL one) by adding a further monomial term $b_{N_{BCL}+1} z^{N_{BCL}+1}$ to Eq.\,(\ref{eq:BCL_truncated}), namely
\be
    \label{eq:BCL_constrained}
    F_\pi^{\rm BCL}(Q^2) = \sum_{k = 0}^{N_{\rm BCL}} b_k \left[ z^k + (-)^{k - N_{\rm BCL}} ~ \frac{k}{N_{BCL}+1} z^{N_{\rm BCL} + 1} \right] ~ , ~
\ee
where $b_{N_{\rm BCL} + 1} = \sum_{k = 0}^{N_{\rm BCL}} k \, b_k (-)^{k - N_{\rm BCL}} / (N_{\rm BCL} + 1)$.

The unitary constraint for the coefficients $b_k$ is more involved w.r.t.~the case of the BGL fit. It makes the coefficients $b_k$ dependent on $\overline{Q}_0^2$  and reads as
\be
    \label{eq:unitarity_BCL}
    \sum_{j, k = 0}^{N_{\rm BCL} + 1} b_j(\overline{Q}_0^2) ~ B_{jk}(\overline{Q}_0^2) ~ b_k(\overline{Q}_0^2) \leq 1 ~ , ~
\ee
where the matrix $B_{jk}(\overline{Q}_0^2)$ is calculable in terms of the kinematical function $\phi(z, \overline{Q}_0^2)$ and of the transverse susceptibility $4 M_\pi^2 \chi_T(\overline{Q}_0^2)$ (see Ref.\,\cite{Bourrely:2008za}).
In Appendix\,\ref{sec:appC} we extend the procedure used in the case of BGL fits to include the unitary condition\,(\ref{eq:unitarity_BCL}) in a $\chi^2$-minimization fitting approach based on the BCL fit\,(\ref{eq:BCL_constrained}). We calculate also the matrix $B_{jk}(\overline{Q}_0^2)$ for $\overline{Q}_0^2 = 0$ and $N_{\rm BCL}$ up to 10.

We have applied the truncated BCL fit\,(\ref{eq:BCL_constrained}) to the description of the spacelike pion data with the unitary bound\,(\ref{eq:unitarity_BCL}) built-in. 
As in the case of the unitary BGL fits performed in the previous Section, we have used a sample of $10^3$ events $F$ generated according to the PDF~(\ref{eq:PDF}) using the direct CERN\,\cite{NA7:1986vav} and electroproduction JLAB-$\pi$\,\cite{JeffersonLab:2008jve} data with the covariance matrix $C$ given by Eq.\,(\ref{eq:datacov_CERN+JLAB}).  A gaussian distribution for the non-perturbative transverse susceptibility $4M_\pi^2 \chi_T(\overline{Q}_0^2 = 0) = 0.00574 \, (10)$, uncorrelated with those of the form factor points, has been assumed.
We have minimized the reduced $\chi_r^2$-variable given by
\be
    \label{eq:chi2_correlated_BCL}
    \chi_r^2 \equiv \frac{1}{N - N_{\rm BCL}} \sum_{i,j = 0}^N \left( F_i^{\rm BCL} - F_i) C_{ij}^{-1} (F_j^{\rm BCL} - F_j \right) ~ , ~
\ee 
obtaining the best unitary BCL fit for a given truncation order $N_{\rm BCL}$. Note that the value of the parameter $b_0$ is constrained by the absolute normalization $F_\pi^V(Q^2 = 0) = 1$ to be equal to $b_0 = 1$ for any value of $N_{\rm BCL}$.

We have considered values of the truncation order $N_{\rm BCL}$ between 2 and 10, obtaining results similar to those of the BGL fits shown in Fig.\,\ref{fig:BGL} with a slightly better precision.
In the kinematical region covered by the CERN and JLAB-$\pi$ data (i.e.\, for $Q^2 \lesssim 2.5$ GeV$^2$) the results of the unitary BCL fit are stable against the order of the truncation $N_{\rm BCL}$, while at larger values of $Q^2$ the bands are unstable and no extrapolation is possible at least from $N_{\rm BCL} \leq 10$.  
We find that for $N_{\rm BCL} \geq 4$ the unitary bound\,(\ref{eq:unitarity_BCL}) is almost saturated.
As in the case of the BGL fits performed in the previous Section, this finding implies the inadequacy of truncation errors based only on higher order terms in the BCL fit\,(\ref{eq:BCL_constrained}), since the saturation of the unitary bound would imply a negligible truncation error that is not observed at all at large $Q^2$.

For the pion charge radius the results corresponding to the BCL fits exhibit a good convergence as a function of the truncation order $N_{\rm BCL}$, obtaining $\langle r_\pi \rangle_{BCL} = 0.713 \pm 0.031$ fm, which agrees very well with the DM result\,(\ref{eq:rpi_DM}). Including also the spacelike data of the F2 experiment at FNAL\,\cite{Dally:1982zk} we get
\be
    \label{eq:rpi_BCL}
    \langle r_\pi \rangle_{BCL} = 0.709 \pm 0.028 ~ \mbox{fm} ~ , ~
\ee
which differs from the PDG value\,(\ref{eq:rpi_PDG}) by $\simeq 1.8\sigma$.
As already observed in Sections\,\ref{sec:CERN+JLABpi} and \ref{sec:BGLfit} in the cases of the DM and BGL approaches, respectively, if we adopt for the vector susceptibility $4 M_\pi^2 \chi_T(\overline{Q}_0^2 = 0)$ the value $0.00550\,(4)$, which corresponds to the evaluation of Eq.\,(\ref{eq:chiT_FV}) using the dispersive pion form factor $|F_\pi^V(\omega)|$ from Ref.\,\cite{Colangelo:2018mtw} up to $\omega = 1$ GeV, the pion charge radius $\langle r_\pi \rangle_{BCL}$ remains basically unchanged with respect to Eq.\,(\ref{eq:rpi_BCL}).

\section{Impact of $\overline{Q}_0^2 > 0$}
\label{sec:Q02}

In this Section we address the issue of the impact of the value of the auxiliary quantity $\overline{Q}_0^2$, at which the transverse susceptibility $4M_\pi^2 \chi_T$ is evaluated, on the unitary filter\,(\ref{eq:filter}) and on the corresponding DM band for the pion form factor.

The $\overline{Q}_0^2$-dependence of the l.h.s.~of the inequality\,(\ref{eq:filter}) is shown in Fig.\,\ref{fig:chiT} for $\overline{Q}_0^2 \leq 1$ GeV$^2$. The transverse susceptibility $4M_\pi^2 \chi_T(\overline{Q}_0^2)$ decreases as $\overline{Q}_0^2$ increases and such a drop is mainly governed by the mass of the dominant $\rho(775)$-meson resonance in $F_\pi^V(\omega)$.

Instead, the $\overline{Q}_0^2$-dependence of the r.h.s.~of the inequality\,(\ref{eq:filter}) is due to the kinematical function $\phi(z, \overline{Q}_0^2)$, defined in Eq.\,(\ref{eq:phiz}) where $\overline{z}_0$ is given by Eq.\,(\ref{eq:z0bar}).
By definition the function $\phi(z, \overline{Q}_0^2)$ does not know anything about meson resonances.
Let us factorize out the term $(1 - \overline{z}_0) = 4M_\pi / \overline{Q}_0 + {\cal O}(1 / \overline{Q}_0^2)$ by introducing the quantities
\bea
     \label{eq:chiT_bar}
    4 M_\pi^2  \overline{\chi}_T(\overline{Q}_0^2) & \equiv &\frac{ 4 M_\pi^2 \chi_T(\overline{Q}_0^2)}{(1 - \overline{z}_0)^6} ~ , ~ \\[2mm]
     \label{eq:phiz_bar}
     \overline{\phi}(z, \overline{Q}_0^2) & \equiv & \frac{\phi(z, \overline{Q}_0^2)}{(1 - \overline{z}_0)^3} = 
         \frac{1}{\sqrt{1536 \pi}} (1 + z)^2 \frac{\sqrt{1 - z} }{(1 - \overline{z}_0 z)^3} ~ . ~
\eea
In this way, the DM band for the pion form factors can be obtained from Eqs.\,(\ref{eq:bounds_pion})-(\ref{eq:gamma_pion}) simply by replacing $\chi_T$ and $\phi$ with $\overline{\chi}_T$ and $\overline{\phi}$, namely
\bea
      \label{eq:bounds_pion_bar}
      && \overline{\beta}(z) - \sqrt{\overline{\gamma}(z)} \leq F_\pi^V(z) \leq \overline{\beta}(z) + \sqrt{\overline{\gamma}(z)} ~ , ~ \\[2mm]
      \label{eq:beta_pion_bar}
      && \overline{\beta}(z) = \frac{1}{\overline{\phi}(z, \overline{Q}_0^2) d(z)} 
            \sum_{i = 0}^N \overline{\phi}_i F_i d_i \frac{1 - z_i^2}{z - z_i} ~ , ~ \\[2mm]
      \label{eq:gamma_pion_bar}
      && \overline{\gamma}(z) = \frac{1}{(1 - z^2) \overline{\phi}^2(z, \overline{Q}_0^2) d^2(z)} \left[ 4 M_\pi^2 \overline{\chi}_T(\overline{Q}_0^2) - 
                                \overline{\chi}_\text{DM}(\overline{Q}_0^2) \right] ~ , ~
\eea
and the unitary filter\,(\ref{eq:filter}) becomes
\be
    \label{eq:filter_bar}
     4 M_\pi^2 \overline{\chi}_T(\overline{Q}_0^2) \geq \overline{\chi}_{DM}(\overline{Q}_0^2) \equiv 
         \sum_{i, j = 0}^N F_i F_j \frac{\overline{\phi}_i(\overline{Q}_0^2) d_i (1 - z_i^2) ~ 
                                                            \overline{\phi}_j(\overline{Q}_0^2) d_j (1 - z_j^2)}{1 - z_i z_j} ~ . ~
\ee

\begin{figure}[htb!]
\begin{center}
\includegraphics[scale=0.60]{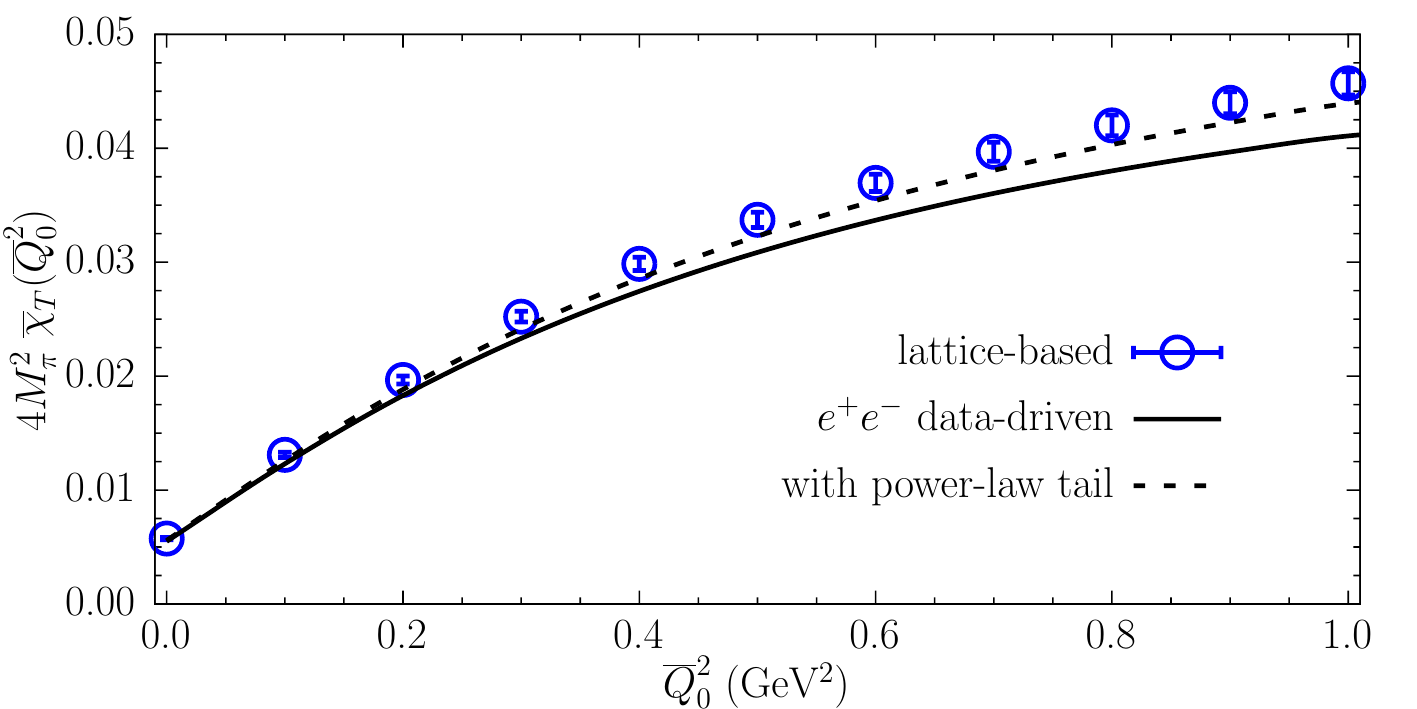}
\end{center}
\vspace{-0.5cm}
\caption{\it \small Blue dots: transverse vector susceptibility $4M_\pi^2  \overline{\chi}_T(\overline{Q}_0^2)$ evaluated adopting the $2\pi$ correlator $V_{2\pi}(\tau)$ obtained in Ref.~\cite{Giusti:2018mdh} using LQCD simulations (see Section\,\ref{sec:susceptibility}). Black line: the susceptibility $4M_\pi^2 \overline{\chi}_T(\overline{Q}_0^2)$ obtained using for $|F_\pi^V(\omega)|$ the results of the dispersive analysis of the $e^+ e^-$ data available from Ref.\,\cite{Colangelo:2018mtw} up to $\omega = 1$ GeV and putting $|F_\pi^V(\omega)| = 0$ for $\omega > 1$ GeV. The dashed line corresponds to add to the data-driven pion form factor a power-law tail for $\omega > 1$ GeV of the form  $|F_\pi^V(\omega)| = |F_\pi^V(1\,GeV)| \cdot$ $(1 \, GeV / \omega)^4$ (see Section\,\ref{sec:susceptibility}). The range of $\overline{Q}_0^2$ is limited to $\overline{Q}_0^2 \lesssim 1$ GeV$^2$ for the reasons explained in Section\,\ref{sec:susceptibility}.}
\label{fig:chiT_Q0}
\end{figure}

\begin{figure}[htb!]
\begin{center}
\includegraphics[scale=0.50]{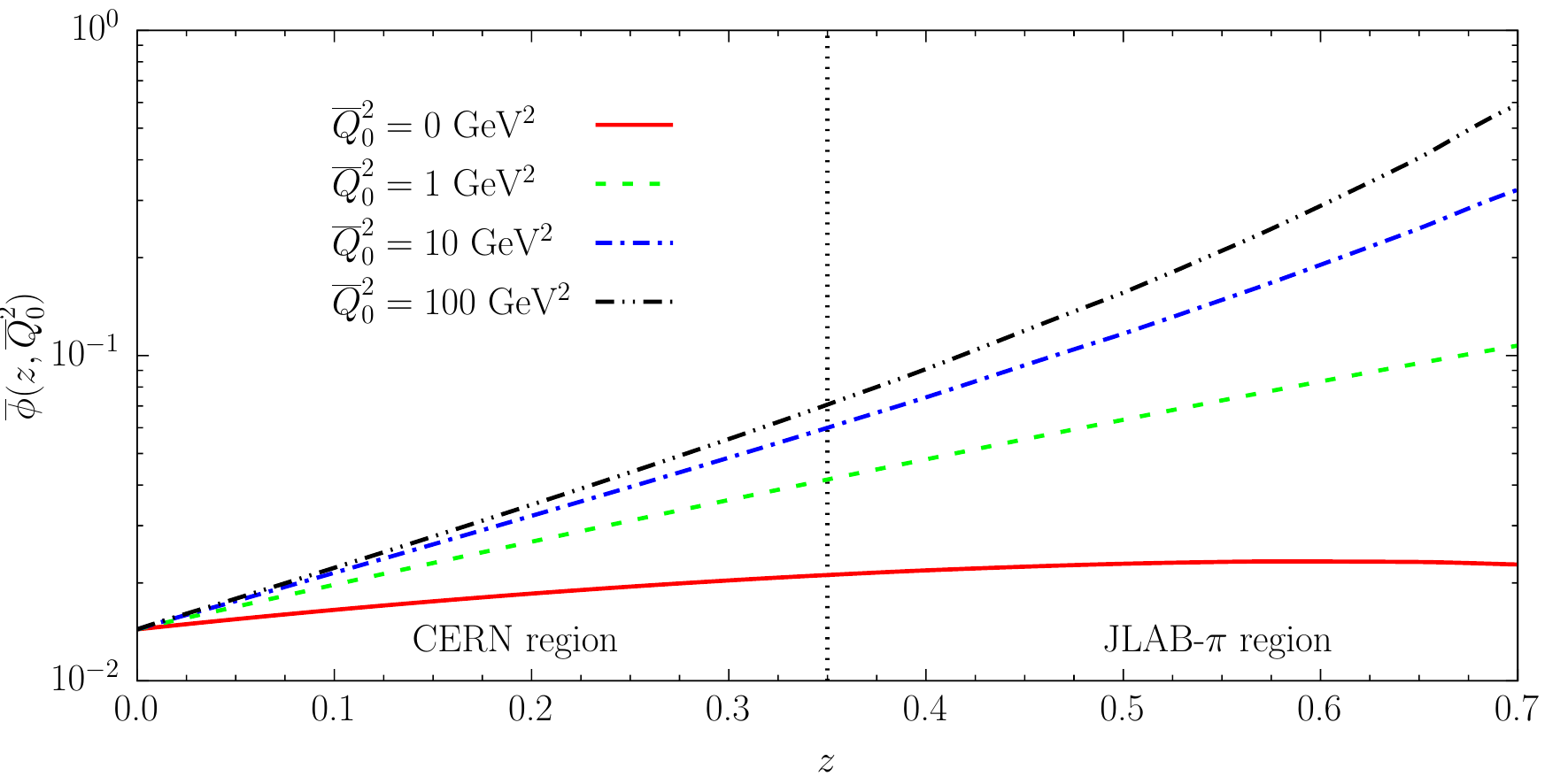}
\end{center}
\vspace{-0.5cm}
\caption{\it \small The kinematical function $\overline{\phi}(z, \overline{Q}_0^2)$, given by Eq.\,(\ref{eq:phiz_bar}), versus the conformal variable $z$ at various values of $\overline{Q}_0^2$ specified in the inset. The vertical dotted line separates the $Q^2$-region of the CERN data $(0.04 \lesssim z \lesssim 0.35)$ from the one of the electroproduction JLAB-$\pi$ data $(0.4 \lesssim z \lesssim 0.7)$.}
\label{fig:phi_Q0}
\end{figure}

The $\overline{Q}_0^2$-dependencies of $4M_\pi^2\overline{\chi}_T(\overline{Q}_0^2)$ and $\overline{\phi}(z, \overline{Q}_0^2)$ are shown in Figs.~\ref{fig:chiT_Q0} and \ref{fig:phi_Q0}, respectively.
The following comments are in order.
\begin{itemize}

\item The susceptibility $4M_\pi^2 \overline{\chi}_T(\overline{Q}_0^2)$ increases as $\overline{Q}_0^2$ increases and goes for $\overline{Q}_0^2 \to \infty$ to the limiting value
\be
    \label{eq:chiT_bar_asympt}
    4M_\pi^2 \overline{\chi}_T(\infty) = \frac{1}{1536 \pi^2} \int_{2M_\pi}^\infty \frac{d\omega}{2M_\pi} \, \left( \frac{\omega}{2 M_\pi} \right)^3 
                                                            \left( 1 - \frac{4M_\pi^2}{\omega^2} \right)^{3/2} \, |F_\pi^V(\omega)|^2 ~ ,
\ee
which is very sensitive to the high-energy tail of $ |F_\pi^V(\omega)|$.

\item The kinematical function $\overline{\phi}(z, \overline{Q}_0^2)$ is approximately flat at $\overline{Q}_0^2 = 0$, while it increases sizably at high values of $z$ as $\overline{Q}_0^2$ increases. Thus, as $\overline{Q}_0^2$ increases, the quantity $\overline{\chi}_{DM}(\overline{Q}_0^2)$ increases, so that the r.h.s.~of the DM filter\,(\ref{eq:filter_bar}) becomes more sensitive to the input data in the large $Q^2$-region.

\end{itemize}

Thus, from the above findings we can conclude that at $\overline{Q}_0^2 = 0$ the unitary filter\,(\ref{eq:filter_bar}) is dominated by the CERN data, which are more precise and dense w.r.t.~to the JLAB-$\pi$ data, while as $\overline{Q}_0^2$ increases the impact of the electroproduction data increases. 

\begin{figure}[htb!]
\begin{center}
\includegraphics[scale=0.55]{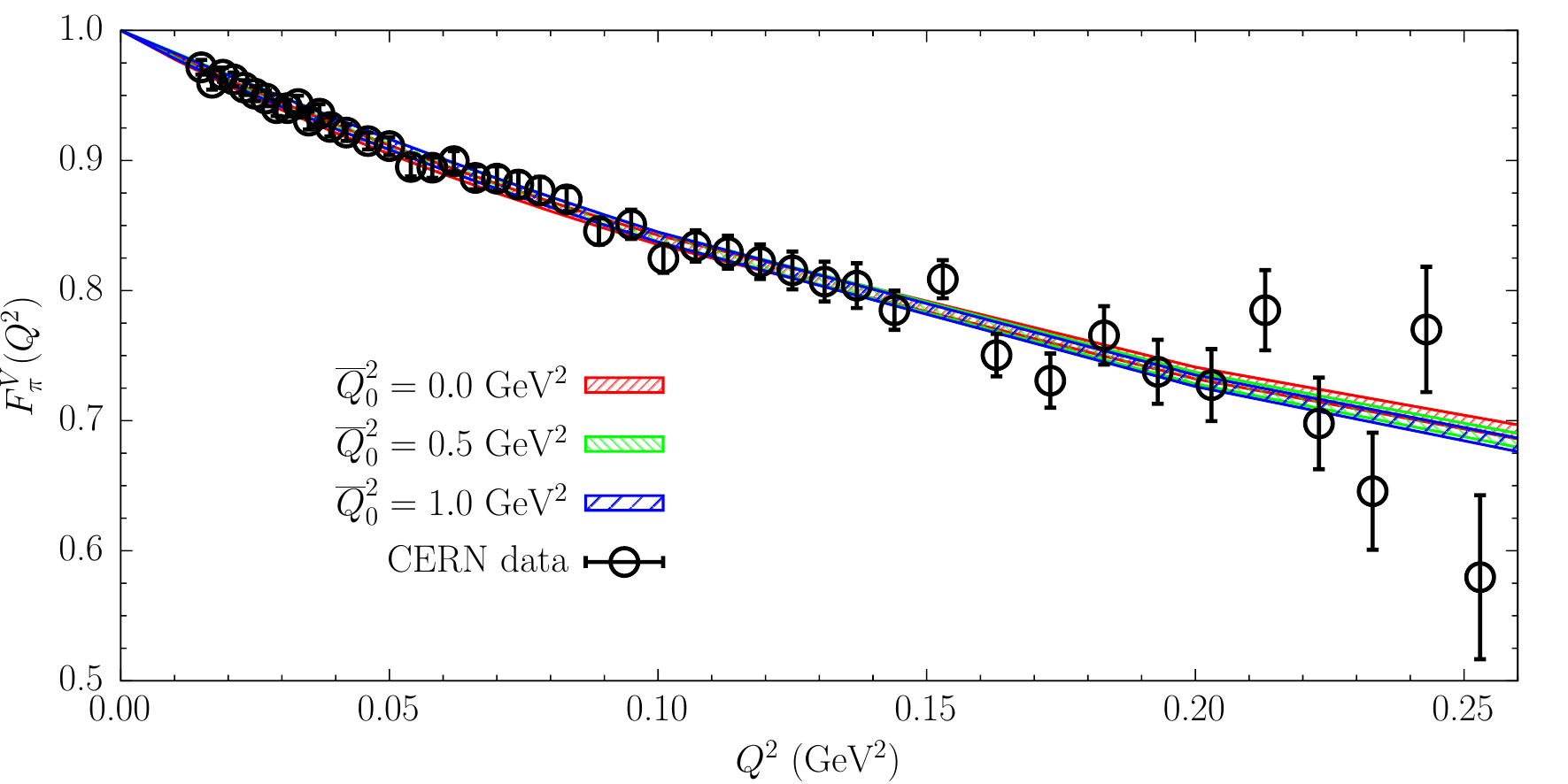} \\
\includegraphics[scale=0.55]{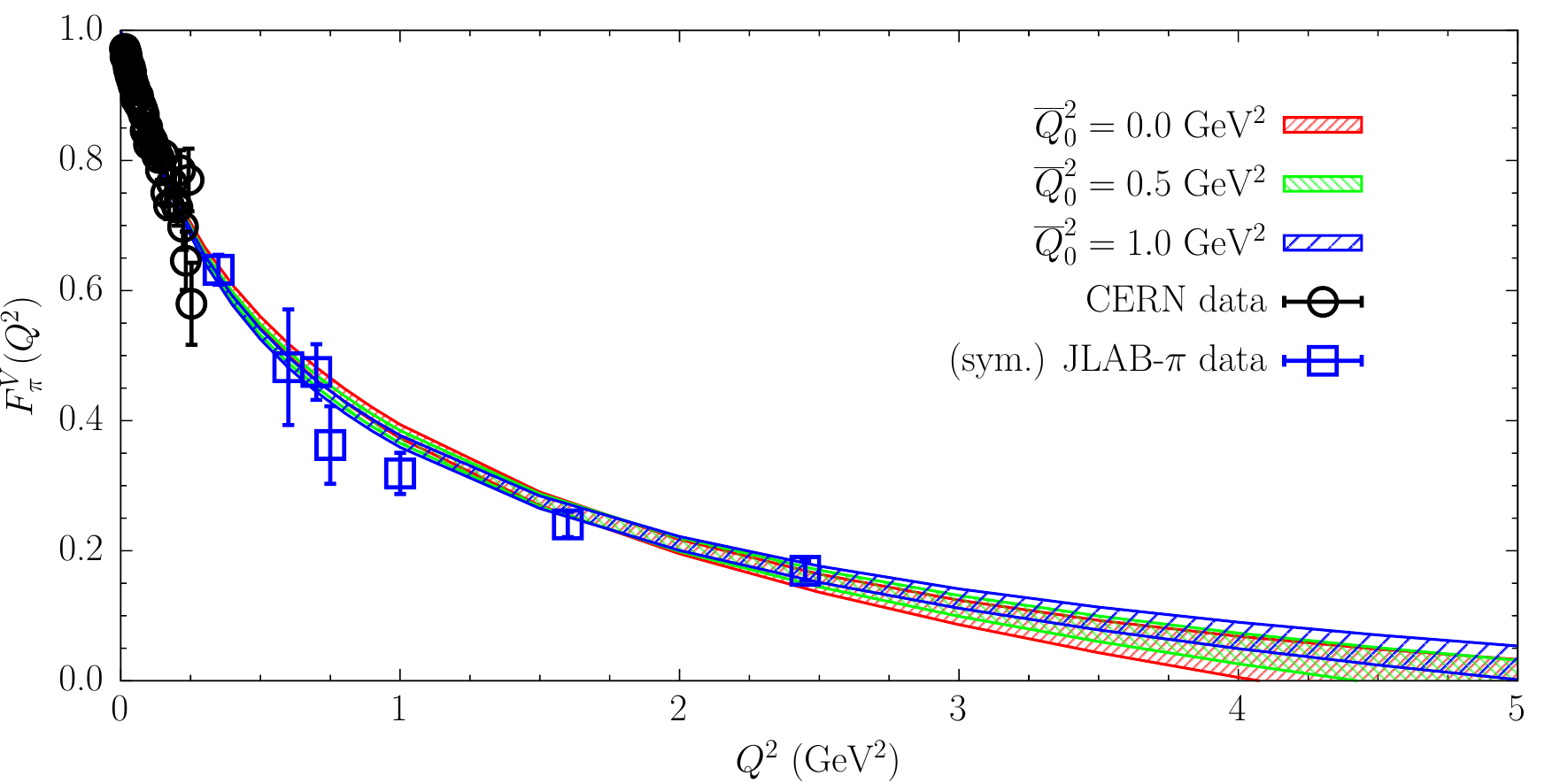}
\end{center}
\vspace{-0.75cm}
\caption{\it \small The DM bands (at $1\sigma$ level) obtained at three different values of $\overline{Q}_0^2 = 0, 0.5, 1$ GeV$^2$ using the unitary sampling procedure (with 10 iterative steps) applied to both the CERN SPS~\cite{NA7:1986vav} (black circles) and the (symmetrized) electroproduction JLAB-$\pi$~\cite{JeffersonLab:2008jve} (blue squares) experimental data.}
\label{fig:bounds_Q0}
\end{figure}

As $\overline{Q}_0^2$ increases, both sides of the DM filter\,(\ref{eq:filter_bar}) increase. Whether this filter may lead to more precise form factor bands for $\overline{Q}_0^2 > 0$, can be established only by a direct numerical investigation. This has been done for three different values of $\overline{Q}_0^2$ not exceeding the limiting value $\overline{Q}_0^2 = 1$ GeV$^2$, as discussed in Section\,\ref{sec:susceptibility}. The corresponding results are shown in Fig.\,\ref{fig:bounds_Q0}.

At low values of $Q^2$ ($\lesssim 0.25$ GeV$^2$) the sensitivity to the choice of $\overline{Q}_0^2$ is quite limited and the more precise determination of the pion charge radius $\langle r_\pi \rangle$ is the one obtained at $\overline{Q}_0^2 = 0$, presented in Section\,\ref{sec:CERN+JLABpi}. In the $Q^2$-region of few GeV$^2$ the DM band becomes more precise as $\overline{Q}_0^2$ increases. In particular, the DM band at $\overline{Q}_0^2 = 0$ may be consistent with negative values of the pion form factor for $Q^2 \approx 4 - 5$ GeV$^2$. This tendency is less pronounced as $\overline{Q}_0^2$ increases and it is expected to disappear for $\overline{Q}_0^2 > 1$ GeV$^2$. Note that it is reasonable to exclude zeros in the pion form factor, particularly in the spacelike region, where in quantum mechanics the pion form factor is the Fourier transform of a charge distribution proportional to the square of the pion wave function\footnote{For an interesting discussion about the absence of zeros of $F_\pi^V(\omega)$ and new dispersion relations applicable to the logarithm of $|F_\pi^V(\omega)|$ see Ref.\,\cite{Leutwyler:2002hm}.}.

The main conclusion of this Section is that the choice of the value of $\overline{Q}_0^2$ can have an impact on the DM predictions in the region of $Q^2 \approx$ few GeV$^2$ (and beyond), where the optimized choice is expected to be $\overline{Q}_0^2 \approx$  few GeV$^2$. On the contrary, for the pion charge radius $\langle r_\pi \rangle$, which represents an important quantity investigated in this work, the optimized choice is given by $\overline{Q}_0^2 = 0$, as properly considered in Section\,\ref{sec:CERN+JLABpi}.

Thus, it would be valuable to obtain in the next future a reliable determination of the susceptibility $4M_\pi^2 \overline{\chi}_T(\overline{Q}_0^2)$ for $\overline{Q}_0^2 \gtrsim 1$ GeV$^2$ either coming from LQCD simulations or driven by analyses of timelike $e^+ e^-$ data. 

We close this Section by recalling that the precision of the unitary DM band for $F_\pi^V(Q^2)$ depends also on the quantity and the $Q^2$-range of the input data. In this respect, both the addition of experimental results for $Q^2$ up to $\approx 8.5$ GeV$^2$, planned at JLAB\,\cite{Dudek:2012vr, JLAB-E12-06-101}, or even up to $\approx 30$ GeV$^2$, expected at future facilities like Electron-Ion Colliders\,\cite{Accardi:2012qut, Anderle:2021wcy}, and the inclusion of precise LQCD determinations of $F_\pi^V(Q^2)$ at low and high $Q^2$ would be very valuable.

\section{The onset of perturbative QCD}
\label{sec:pQCD}

The behavior of the pion form factor at large spacelike momentum transfer is predicted by the pQCD hard-scattering mechanism\,\cite{Lepage:1979zb, Efremov:1979qk, Chernyak:1977as, Farrar:1979aw} to be
\be
    \label{eq:pQCD}
    F_\pi^V(Q^2) ~_{\overrightarrow{Q^2 \to \infty}} ~ 8 \pi f_\pi^2 \frac{\alpha_s(Q^2)}{Q^2} \left[ 1 + G(Q^2) \right]
\ee
where $f_\pi \simeq 130$ MeV is the pion decay constant,
\be
    \label{eq:alphas}
    \alpha_s(Q^2) = \frac{4\pi}{(11 - 2N_f/3) ~ \mbox{ln}\left(Q^2 / \Lambda_{QCD}^2\right)}
\ee
is the running strong coupling at leading order (with $N_f$ being the number of active flavors and $\Lambda_{QCD}$ the QCD scale) and $G(Q^2)$ describes the corrections due to the pre-asymptotic structure of the scale-dependent pion distribution amplitude, which is function of the light-front fraction of the pion’s total momentum carried by a valence quark.

The question at which energy scale the asymptotic behaviour\,(\ref{eq:pQCD}) sets in has long been debated in literature and the answer is not trivial because of the presence of nonperturbative effects at intermediate values of $Q^2$ (see, e.g., the recent review in Ref.\,\cite{Horn:2016rip}).

As already noted in Ref.\,\cite{DiCarlo:2021dzg}, at finite values of $\overline{Q}_0^2$ the DM unitary width $\gamma(z)$ in Eq.\,(\ref{eq:gamma_pion}) (or equivalently $\overline{\gamma}(z)$ in Eq.\,(\ref{eq:gamma_pion_bar})) is proportional to $(1 - z)^{-2}$ for $z \to 1$ and, therefore, it diverges proportionally to $Q^2$. This implies that the DM method looses its predicting power at the end-point $z = 1$ (i.e.~$Q^2 \to \infty$)\footnote{A similar situation occurs at the end-point $z = -1$ corresponding to the annihilation threshold $q^2 = -Q^2 = 4 M_\pi^2$.}.

However, the situation changes when $\overline{Q}_0^2$ becomes sufficiently large. Indeed, from Eq.\,(\ref{eq:phiz_bar}) for $\overline{Q}_0^2 >> Q^2$ one has that the function $\overline{\phi}(z, \overline{Q}_0^2) \propto (1 - z)^{-5/2} \propto Q^{5/2}$, so that $\sqrt{\overline{\gamma}(z)} \propto (1 - z)^2 \propto 1 / Q^2$. While the central value $\overline{\beta}(z)$ drops down as $1 / Q^{5/2}$, i.e.~faster than the pQCD prediction\,(\ref{eq:pQCD}), the unitary width $\sqrt{\overline{\gamma}(z)}$ does not. This means that at a certain value of $Q^2 = Q_{pQCD}^2$ the pQCD prediction\,(\ref{eq:pQCD}) may start to be within the unitary bounds $\overline{\beta}(z) \pm \sqrt{\overline{\gamma}(z)}$ at $1\sigma$ level (for $\overline{Q}_0^2 >> Q_{pQCD}^2$). Such a value $Q_{pQCD}^2$ provides an estimate of the energy scale at which the asymptotic behaviour\,(\ref{eq:pQCD}) sets in and it is based only on unitarity and spacelike experimental data.

Note that in principle one should calculate the DM unitary bands for the pion form factor for increasing, but finite values of $\overline{Q}_0^2$ and then extrapolate such DM bands to the limit $\overline{Q}_0^2 \to \infty$. In this way the analytic property of the kinematical function $\overline{\phi}(z, \overline{Q}_0^2)$ is kept at each step of the calculation. In practice we have checked that at large $Q^2$ the DM band extrapolated to $\overline{Q}_0^2 \to \infty$ can be obtained directly by considering {\it ab initio} the kinematical function $\overline{\phi}(z, \overline{Q}_0^2)$ in the limit $\overline{Q}_0^2 \to \infty$. 

\begin{figure}[htb!]
\begin{center}
\includegraphics[scale=0.55]{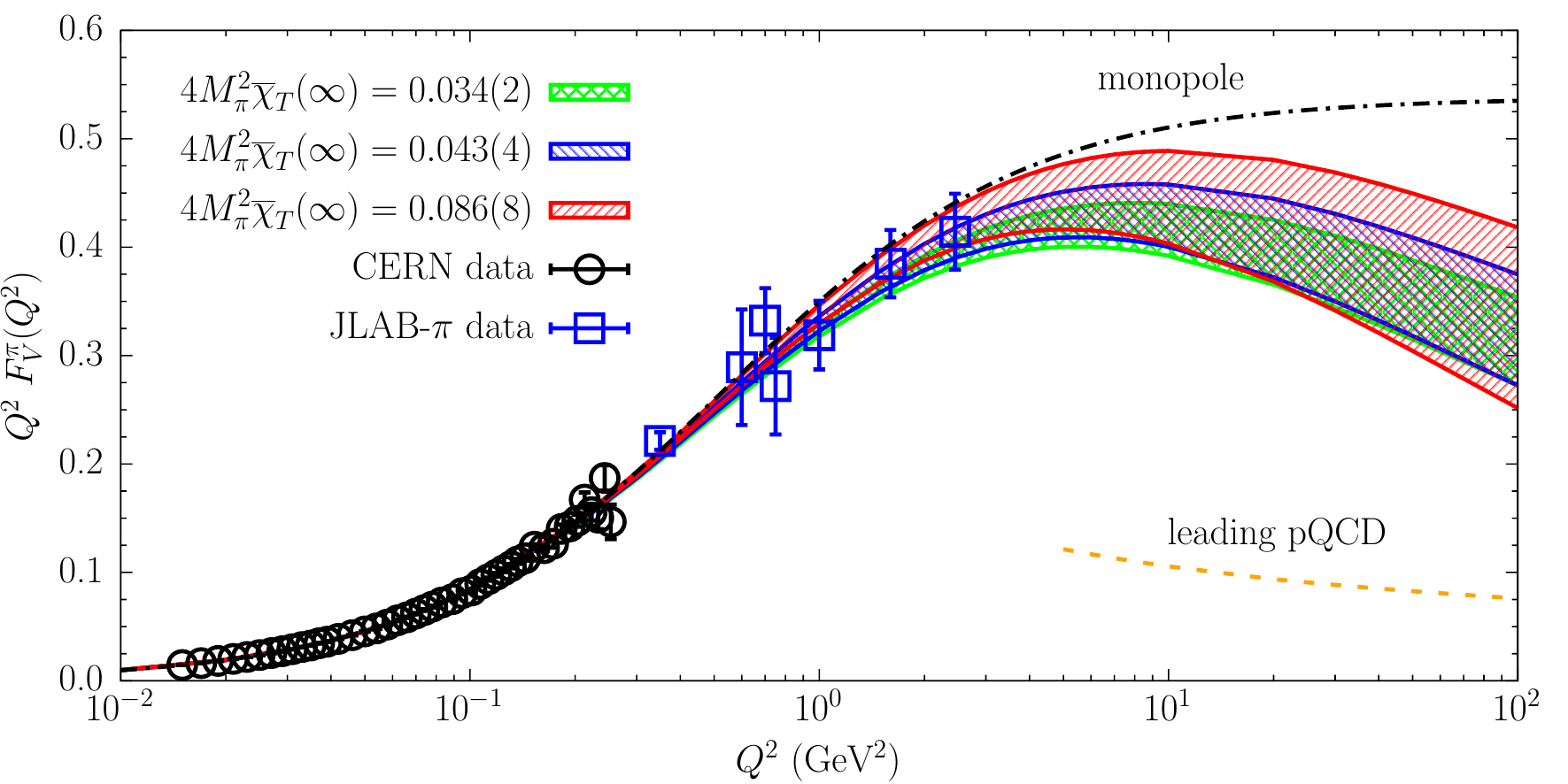}
\end{center}
\vspace{-0.75cm}
\caption{\it \small The pion form factor $Q^2 F_\pi^V(Q^2)$ predicted by the DM method adopting the three estimates\,(\ref{eq:chiT_bar_CHS19})-(\ref{eq:chiT_bar_final}) for the transverse susceptibility $4M_\pi^2 \overline{\chi}_T(\infty)$ in the limit $\overline{Q}_0^2 \to \infty$. The input data are the CERN~\cite{NA7:1986vav} (black circles) and the (symmetrized) electroproduction JLAB-$\pi$~\cite{JeffersonLab:2008jve} (blue squares) experimental data. The orange line represents the leading pQCD prediction\,(\ref{eq:pQCD}) with $G(Q^2) = 0$ and $\Lambda_{QCD} \simeq 300$ MeV\,\cite{FlavourLatticeAveragingGroupFLAG:2021npn}. The black line corresponds to the monopole shape $Q^2 / (1 + \langle r_\pi^2 \rangle_{PDG} \, Q^2 / 6)$ reproducing the PDG central value\,(\ref{eq:rpi_PDG}) for the pion charge radius.}
\label{fig:bounds_pQCD}
\end{figure}

Since a precise estimate of the transverse susceptibility $4M_\pi^2 \overline{\chi}_T(\overline{Q}_0^2)$ is not available for $\overline{Q}_0^2 > 1$ GeV$^2$, we limit ourselves to investigate the sensitivity of the DM unitary band to a range of possible values for the quantity $4M_\pi^2 \overline{\chi}_T(\infty)$ given by Eq.\,(\ref{eq:chiT_bar_asympt}). 
The first estimate is calculated using for $|F_\pi^V(\omega)|$ the results of Ref.\,\cite{Colangelo:2018mtw} and cutting the integral in the r.h.s.~of Eq.\,(\ref{eq:chiT_bar_asympt}) at $\omega = 1$ GeV.  Due to the positivity of the integrand function in Eq.\,(\ref{eq:chiT_bar_asympt}), such a value represents a {\it lower bound} to the transverse susceptibility, namely
\be
    \label{eq:chiT_bar_CHS19}
     4M_\pi^2 \overline{\chi}_T(\infty) \geq 0.034 \pm 0.002 ~ . ~
\ee
By considering for $\omega > 1$ GeV a power-law tail of the form $|F_\pi^V(\omega)| = |F_\pi^V(1\,GeV)| \cdot$ $(1 \, GeV / \omega)^4$ an additional contribution equal to $0.009 \pm 0.003$ is obtained, leading to
 \be
     \label{eq:chiT_bar_CHS19+tail}
     4M_\pi^2 \overline{\chi}_T(\infty) = 0.043 \pm 0.004 ~ , ~
\ee
 which corresponds to an increase of about $30 \%$. 
 Since the susceptibility $4M_\pi^2 \overline{\chi}_T(\infty)$ is very sensitive to the high-energy tail of the em pion form factor, we consider for our sensitivity study a third higher value by applying to the result\,(\ref{eq:chiT_bar_CHS19+tail}) a conservative factor equal to $2$, obtaining\footnote{By assuming conservatively for $\omega > 1$ GeV a tail of $|F_\pi^V(\omega)|$ proportional to the leading behaviour expected in pQCD for timelike momenta\,\cite{Farrar:1979aw}, i.e.~$\omega^{-2} \mbox{log}^{-1}(\omega^2 / \Lambda_{QCD}^2)$ with $\Lambda_{QCD} \simeq 300$ MeV\,\cite{FlavourLatticeAveragingGroupFLAG:2021npn}, one gets an additional contribution to Eq.\,(\ref{eq:chiT_bar_CHS19}) equal to $\simeq 0.04$, yielding a value of $4M_\pi^2 \overline{\chi}_T(\infty)$ not exceeding the one given in Eq.\,(\ref{eq:chiT_bar_final}).}
\be
    \label{eq:chiT_bar_final}
    4M_\pi^2 \overline{\chi}_T(\infty) = 0.086 \pm 0.008 ~ . ~
\ee
The DM bands for the quantity $Q^2 F_\pi^V(Q^2)$ corresponding to the three choices\,(\ref{eq:chiT_bar_CHS19})-(\ref{eq:chiT_bar_final}) are shown in Fig.\,\ref{fig:bounds_pQCD}.
The lower unitary bound (at $1\sigma$ level) turns out to be almost insensitive to the chosen value of $4M_\pi^2 \overline{\chi}_T(\infty)$ and remains significantly much larger than the leading pQCD prediction\,(\ref{eq:pQCD}) (with $G(Q^2) = 0$) at least up to $Q^2 \sim 100$ GeV$^2$. Therefore, the pre-asymptotic structure of the pion distribution amplitude is expected to produce significant effects on the pion form factor up to quite large values of $Q^2$. This is in qualitative agreement with the findings of several estimates available in the literature based both on models\,\cite{Chang:2013nia, Ydrefors:2021dwa} and on LQCD simulations\,\cite{Gao:2022vyh, LatticeParton:2022zqc, Holligan:2023rex}.

We point out that the DM bands shown in Fig.\,\ref{fig:bounds_pQCD}, which we recall are based only on unitarity and spacelike experimental data, may provide important information on the scale dependence of the pion distribution amplitude.

\begin{figure}[htb!]
\begin{center}
\includegraphics[scale=0.55]{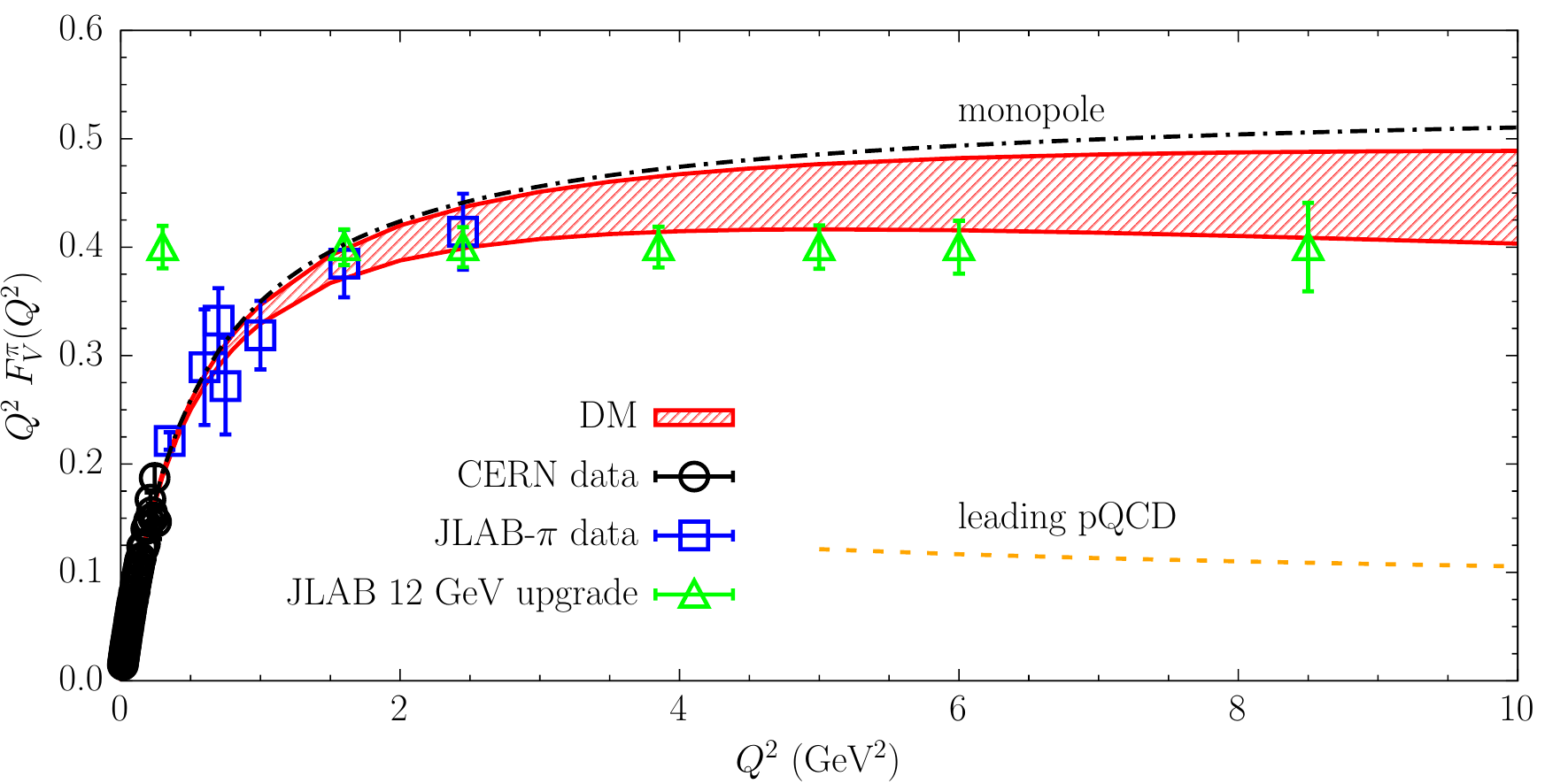}
\end{center}
\vspace{-0.75cm}
\caption{\it \small The pion form factor $Q^2 F_\pi^V(Q^2)$ predicted by the DM method in the $Q^2$-range up to 10 GeV$^2$ adopting the estimate\,(\ref{eq:chiT_bar_final}) for the transverse susceptibility $4M_\pi^2 \overline{\chi}_T(\infty)$ at $\overline{Q}_0^2 \to \infty$. The input data are the CERN~\cite{NA7:1986vav} (black circles) and the (symmetrized) electroproduction JLAB-$\pi$~\cite{JeffersonLab:2008jve} (blue squares) experimental data. The green triangles (fixed at an arbitrary value of $0.4$) correspond to the projected precisions (statistical+systematic) expected in the experimental proposal\,\cite{JLAB-E12-06-101} for the 12 GeV JLAB upgrade. The orange and black lines are the same as in Fig.\,\ref{fig:bounds_pQCD}.}
\label{fig:bounds_pQCD_10GeV}
\end{figure}

As noted in Section\,\ref{sec:Q02}, the precision of the unitary DM band for $F_\pi^V(Q^2)$ at large $Q^2$ can be improved by adding new experimental results, like those planned at JLAB\,\cite{Dudek:2012vr, JLAB-E12-06-101} for $Q^2$ up to $\approx 8.5$ GeV$^2$. The projected precision of the forthcoming JLAB experimental data is shown in Fig.\,\ref{fig:bounds_pQCD_10GeV} by the green triangles and compared with the DM unitary band corresponding to the transverse susceptibility\,(\ref{eq:chiT_bar_final}).

\section{Conclusions}
\label{sec:conclusions}

The experimental data on the em form factor of charged pions available at spacelike momenta have been analyzed using the DM approach\,\cite{DiCarlo:2021dzg}, which describes the momentum dependence of hadronic form factors without introducing any explicit parameterization and includes properly the constraint coming from unitarity and analyticity. The latter one is given by a transverse vector susceptibility, which has been evaluated nonperturbatively from the results of lattice QCD simulations of suitable two-point correlation functions contributing to the Hadronic Vacuum Polarization term of the muon.  

We have elucidated in detail the role played by the DM unitary filter\,(\ref{eq:filter}), which allows to select the subset of input data that can be reproduced exactly by a unitary $z$-expansion.
Since the unitary bound turns out to be extremely selective as the number of data points increases, we have develop a unitary sampling procedure, which allows to generate in an efficient way a distribution of values for the pion form factor satisfying unitarity for any value of the number of data points. Such a procedure can be generalized straightforwardly to any set of hadronic form factors, which must satisfy unitary bounds.

We have applied the unitary sampling method to the spacelike data from both the CERN SPS experiment\,\cite{NA7:1986vav} and the JLAB-$\pi$ Collaboration\,\cite{JeffersonLab:2008jve} for a total of more than 50 data points.
The pion charge radius has been determined in a completely model-independent way and consistently with the unitary bound. This is at variance with the results obtained in the experimental works\,\cite{Dally:1982zk, NA7:1986vav, SELEX:2001fbx}, where the spacelike data have been fitted assuming a simple monopole Ansatz, which introduces a non-negligible model dependence. 

The DM  result is $\langle r_\pi \rangle_{DM} = 0.703 \pm 0.027$ fm, which differs by $\simeq 1.6$ standard deviations from the latest PDG\,\cite{ParticleDataGroup:2022pth} value $\langle r_\pi \rangle_{PDG} = 0.659 \pm 0.004$ fm, dominated by the very precise results of dispersive analyses of timelike data\,\cite{Colangelo:2018mtw, Ananthanarayan:2017efc} coming from measurements of the cross section  of the $e^+ e^- \to \pi^+ \pi^-$ process.
In order to clarify any possible significance of such a difference it is crucial to improve significantly the precision of the experimental data in the spacelike region.
 
 We have analyzed the spacelike data using also traditional $z$-expansions, like the BGL\,\cite{Boyd:1997kz} or BCL\,\cite{Bourrely:2008za} fitting functions, using a new procedure that easily incorporates {\it ab initio} the non-perturbative unitary bound, obtaining $\langle r_\pi \rangle_{BGL} = 0.711 \pm 0.039$ fm and $\langle r_\pi \rangle_{BCL} = 0.709 \pm 0.028$ fm in nice agreement with the DM result. 
 
A detailed comparison among the BGL/BCL fitting procedures and the DM method has been carried out in a wide range of values of $Q^2$, showing that unitarity is fulfilled properly only by the DM approach even if the BGL/BCL fitting functions are constructed to be unitary. This is due to the fact that unitarity must be imposed both on the fitting function and on the input data.
Even if an explicit $z$-expansion is constructed to satisfy unitarity, the fitting procedure is usually applied to all the input data regardless whether they satisfy unitarity or not (i.e., regardless whether the input data can be exactly reproduced by a unitary $z$-expansion). Fitting non-unitary data might introduce distortions in a unitary $z$-expansion, as we have found explicitly in the case of the unitary BGL/BCL fitting procedures. It will be interesting to investigate such a potential problem in the case of the hadronic form factors describing semileptonic $B$-meson weak decays.
 
We have addressed also the issue of the onset of pQCD at large spacelike momenta. Using the DM method we have performed a sensitivity study of the pion form factor at large spacelike momenta based only on experimental spacelike data and unitarity. Although the leading pQCD behaviour is found to set in only at very large momenta, our DM bands may provide information about the pre-asymptotic effects related to the scale dependence of the pion distribution amplitude.

We stress that the DM approach is equally well suited to be applied to experimental and/or theoretical data. The DM analyses of LQCD data as well as of timelike data for the em pion form factor are in progress.
An extension of the present work to the case of the spacelike data on the em form factors of the nucleon is also in progress.

\section*{Acknowledgments}

We warmly thank G.~Martinelli for many helpful discussions and for his continuous support. We are deeply indebted with G.~Colangelo, M.~Hoferichter and P.~Stoffer for having provided us the numerical results of the dispersive analysis of Ref.\,\cite{Colangelo:2018mtw} for the pion form factor $F_\pi^V(\omega)$ both in the spacelike sector down to $\omega^2 = -1$ GeV$^2$ and in the timelike one up to $\omega^2 = 1$ GeV$^2$.
S.S.~is supported by the Italian Ministry of Research (MIUR) under grant PRIN 20172LNEEZ.
The work of L.V.~is supported by Agence Nationale de la Recherche (ANR) under contract n. 202650 (ANR-19-CE31-0016, GammaRare).

\appendix

\section{Independence of the DM approach on the auxiliary variable $t_0$}
\label{sec:appA}

In this Appendix we provide some details about the independence of the DM approach by any specific choice of the auxiliary variable $t_0$ introduced in the definition of the conformal variable $z$, namely (see Eq.\,(\ref{eq:conformal}))
\be
\label{eq:conformal_z}
z \equiv z(t, t_0) \equiv \frac{\sqrt{t_+ - t} - \sqrt{t_+ - t_0}}{\sqrt{t_+ - t} + \sqrt{t_+ - t_0}} ~ . ~
\ee
It is straightforward to show that the quantities
\bea
      \frac{1 - z z_i}{z - z_i} & = & \frac{\sqrt{t_+ - t} + \sqrt{t - t_i}}{\sqrt{t_+ - t} - \sqrt{t - t_i}} ~ , ~ \nonumber \\[2mm]
      \label{eq:factors}
      \frac{\sqrt{1 - z^2} \sqrt{1 - z_i^2}}{z - z_i} & = & 2 \frac{\sqrt{t_+ - t} ^{1/4} \sqrt{t - t_i}^{1/4}}{\sqrt{t_+ - t} - \sqrt{t - t_i}} ~ , ~ \\[2mm]
      \frac{\sqrt{1 - z_i^2} \sqrt{1 - z_j^2}}{1 - z_i z_j} & = & 2 \frac{\sqrt{t_+ - t_i} ^{1/4} \sqrt{t - t_j}^{1/4}}{\sqrt{t_+ - t_i} + \sqrt{t - t_j}} ~ \nonumber
\eea
are independent on $t_0$. The first of the above equations implies that the coefficients $d(z)$ and $d_i$, given respectively by Eqs.\,(\ref{eq:dz_final}) and (\ref{eq:di_final}), do not depend upon the choice of $t_0$.  

On the contrary, the kinematical function $\phi(z)$ depends on $t_0$ through a simple factor given by (see, e.g., Ref.\,\cite{Bharucha:2010im})
\be
    \frac{(t_+ - t)^{1/4}}{ (t_+ - t_0)^{1/4}} \left[ \sqrt{t_+ - t} - \sqrt{t_+ - t_0} \right] ~ , ~
\ee
which can be rewritten as $2 \sqrt{t_+ - t} / \sqrt{1 - z^2}$. Therefore, for any kinematical function $\phi$ the product $\phi(z) \sqrt{1 - z^2}$ is independent on $t_0$.  This property guarantees that the bound\,(\ref{eq:JQ2z}) is independent on $t_0$ as it should be\footnote{The change of the integration variable from $z$ to $\widetilde{z} \equiv z(t, \widetilde{t}_0)$ is given by the Jacobian $d z/ d \widetilde{z} = (1 - z^2) / (1 - \widetilde{z}^2)$. On the circle one has $z = e^{i \alpha}$ and $1 - z^2 = |1 - z^2| e^{i(\alpha - \pi / 2)}$, which implies $( \widetilde{z} / z) (1 - z^2) / (1 - \widetilde{z}^2) = |(1 - z^2) / (1 - \widetilde{z}^2)|$, so that the  bound\,(\ref{eq:JQ2z}) does not depend on the value of  $t_0$.} and, together with Eq.\,(\ref{eq:factors}), that the quantities $\beta(z)$, $\gamma(z)$ and $\chi_{DM}$, given respectively by Eqs.\,(\ref{eq:beta_final}), (\ref{eq:gamma_final}) and (\ref{eq:chi0_final}), are independent on $t_0$.

\section{Inclusion of the unitary constraint in truncated BGL fits}
\label{sec:appB}

In this Appendix we describe a simple procedure that allows to span the space of the values of the BGL coefficients $a_k$, appearing in the truncated BGL fit
\be
    \label{eq:BGL_app}
    F_\pi^{\rm BGL}(Q^2) = \frac{\sqrt{4M_\pi^2 \chi_T(\overline{Q}_0^2)}}{ \phi(z, \overline{Q}_0^2)} \sum_{k = 0}^{N_{\rm BGL}} a_k z^k ~ 
\ee
and satisfying the (truncated) unitary constraint
\be
    \label{eq:unitarity_BGL_app}
    \sum_{k = 0}^{N_{\rm BGL}} a_k^2 \leq 1 ~ . ~
\ee

Let us introduce $N_{\rm BGL} + 1$ parameters $r_k$ with $k = 0, 1, ... N_{\rm BGL}$, whose values can vary by construction in the range $[0, 1]$. Then, we define $N_{\rm BGL}$ hyperangles $\theta_k$ (with $k = 1, 2, ... N_{\rm BGL}$) as
\bea
    \label{eq:hyperangles}
    \theta_k & = & \pi ~ r_k \qquad \qquad \mbox{for ~}  k = 1, 2, ... N_{\rm BGL} - 1  ~ , ~ \nonumber \\[2mm]
    \theta_{N_{\rm BGL}} & = & 2 \pi ~ r_{N_{\rm BGL}} ~ , ~
\eea
so that the hyperangles $\theta_k$ vary in the range $[0, \pi]$ for $ k = 1, 2, ... N_{\rm BGL} - 1$, while the angle $ \theta_{N_{\rm BGL}} $ is in the range $[0, 2\pi]$.
Then, the coefficients $a_k$ are related to the hyperradius $r_0$ and to the hyperangles $\theta_k$ ($k = 1, 2, ... N_{\rm BGL}$) by
\bea
     \label{eq:ak}
     a_0 & = & r_0  ~ \mbox{cos}\theta_1 ~ , ~ \nonumber \\[2mm]
     a_k & = & r_0 \left[ \prod_{j = 1}^k \mbox{sin}\theta_j \right] \mbox{cos}\theta_{k+1} \qquad \qquad \mbox{for ~} k = 1, 2, ... N_{\rm BGL} - 1 ~ , ~ \\
     a_{N_{\rm BGL}} & = & r_0 \left[ \prod_{j = 1}^{N_{\rm BGL} - 1} \mbox{sin}\theta_j \right] \mbox{sin}\theta_{N_{\rm BGL}} ~ . ~ \nonumber
\eea
It is straightforward to check that 
\be
    \label{eq:hyperradius}
    \sum_{k = 0}^{N_{\rm BGL}} a_k^2 = r_0^2 ~ , ~
\ee
so that the parameter space $0 \leq r_k \leq 1$ maps the one of the parameters $a_k$ fulfilling the unitary bound\,(\ref{eq:unitarity_BGL_truncated}).
The relation between the coefficients $\{ a_k \}$ and $\{ r_0, \theta_k \}$ can be inverted obtaining
\bea
    \label{eq:rk_ak}
    r_0 & = & \sqrt{\sum_{j = 0}^{N_{\rm BGL}} a_j^2} ~ , ~ \nonumber \\[2mm]
    \theta_k & = & \mbox{Arccos}\frac{a_{k-1}}{\sqrt{\sum_{j = k - 1}^{N_{\rm BGL}} a_j^2} } \qquad \qquad \mbox{for ~} k = 1, 2, ... N_{\rm BGL} - 1 ~ , ~  \\[2mm]
    \theta_{N_{\rm BGL}} & = & \mbox{Arccos}\frac{a_{N_{\rm BGL} - 1}}{\sqrt{a_{N_{\rm BGL} -1}^2 + a_{N_{\rm BGL}}^2}} \qquad \qquad \mbox{for ~} a_{N_{\rm BGL}} \geq 0  ~ , ~ \nonumber \\[2mm]
                         & = & 2 \pi - \mbox{Arccos}\frac{a_{N_{\rm BGL} - 1}}{\sqrt{a_{N_{\rm BGL} -1}^2 + a_{N_{\rm BGL}}^2}} \qquad \qquad \mbox{for ~} a_{N_{\rm BGL}} < 0 ~ . ~ \nonumber
\eea

We stress that the procedure described in this Appendix for constructing a unitary BGL fitting function can be applied to a generic hadronic form factor which should fulfill a unitary constraint.

\section{Inclusion of the unitary constraint in truncated BCL fits}
\label{sec:appC}

In this Appendix we describe briefly the procedure that allows to span the space of the values of the BCL coefficients $b_k$, appearing in the truncated BCL fit
\be
    \label{eq:BCL_app}
    F_\pi^{\rm BCL}(Q^2) = \sum_{k = 0}^{N_{\rm BCL}} b_k(\overline{Q}_0^2) ~ z^k ~ 
\ee
and satisfying the (truncated) unitary constraint
\be
    \label{eq:unitarity_BCL_app}
    \sum_{j, k = 0}^{N_{\rm BCL}} b_j(\overline{Q}_0^2) ~ B_{jk}(\overline{Q}_0^2) ~ b_k(\overline{Q}_0^2) \leq 1 ~ . ~
\ee

According to Ref.\,\cite{Bourrely:2008za}, in the case of the em pion form factor, the matrix $B_{jk}(\overline{Q}_0^2)$ is given by
\be
    \label{eq:Bjk_app}
    B_{jk}(\overline{Q}_0^2) = \overline{B}_{|j - k|}(\overline{Q}_0^2) \equiv \sum_{n = 0}^\infty \eta_n(\overline{Q}_0^2) ~ \eta_{n + |j -k|}(\overline{Q}_0^2)
\ee
with $\eta_n(\overline{Q}_0^2)$ being the coefficients of the $z$-expansion of the kinematical function $\phi(z, \overline{Q}_0^2)$ divided by $\sqrt{4 M_\pi^2 \chi_T( \overline{Q}_0^2)}$, namely
\be
    \label{eq:eta_app}
    \frac{\phi(z, \overline{Q}_0^2)}{\sqrt{4 M_\pi^2 \chi_T( \overline{Q}_0^2)}} = \sum_{n = 0}^\infty \eta_n(\overline{Q}_0^2) z^n ~ . ~
\ee

Since the l.h.s.~of Eq.(\ref{eq:eta_app}) represents an analytic, bounded function inside the unit disc $|z| \leq 1$, the coefficients $\overline{B}_{|j - k|}(\overline{Q}_0^2)$ can be evaluated numerically by truncating the sum over $n$ in Eq.\,(\ref{eq:Bjk_app}) up to a finite order. We have performed such a calculation at $ \overline{Q}_0^2 = 0$ using $n \leq 300$ and $4 M_\pi^2 \chi_T(0) = 0.00574$. 
The corresponding results for the first 12 coefficients $\overline{B}_{|j -k|}(0)$ are
\bea
    \label{eq:Bm}
    \left\{ \overline{B}(0) \right\} & = & \left\{ +0.1470510, +0.0630217, -0.0396807, -0.0388320, -0.0148702,  \right. ~ \nonumber \quad  \\[2mm]
                                                && \left. -0.0085370, -0.0056373, -0.0040269, -0.0030298, -0.0023662, \right. ~ \nonumber \quad \\[2mm]
                                                && \left. -0.0019011, -0.0015618 \right\} ~ , ~ 
 \eea
 which allow to construct the matrix $B_{jk}(0)$ for $N_{BCL} \leq 10$. The distribution of values of the transverse susceptibility $4 M_\pi^2 \chi_T(0)$ can be taken exactly into account by dividing all the coefficients  $\overline{B}_{|j -k|}(0)$ given in Eq.\,(\ref{eq:Bm}) by the common factor $4 M_\pi^2 \chi_T(0) / 0.00574$.

Following the strategy described in the case of the unitary BGL fit in Appendix\,\ref{sec:appB} and dropping for sake of simplicity the dependence upon $\overline{Q}_0^2$, we introduce $N_{\rm BCL} + 1$ parameters $r_k$ with $k = 0, 1, ... N_{\rm BCL}$, whose values can vary by construction in the range $[0, 1]$. Then, using the hyperspherical rotation defined by Eqs.\,(\ref{eq:hyperangles})-(\ref{eq:ak}), we transform the set of parameters $r_k$ into a set of (intermediate) coefficients $a_k$ satisfying the constraint
\be
    \label{eq:unitarity_ak}
    \sum_{k = 0}^{N_{\rm BCL}} a_k^2 = r_0^2 \leq 1~ . ~
\ee
The BCL coefficients $b_k$ can be obtained by observing that the matrix $B_{jk}$ is symmetric and positive definite, so that it can be diagonalized having only positive eigenvalues. Thus, we can obtain the $N_{\rm BCL} + 1$ coefficients $b_k$ from the set $\{ a_k \}$ as
\be
    \label{eq:bk}
    b_k = \sum_{k^\prime = 0}^{N_{\rm BCL}} B_{k k^\prime}^{-1/2} a_{k^\prime} ~
\ee
with
\be
    \label{eq:Binv_app}
    B_{k k^\prime}^{-1/2} = \sum_{m = 0}^{N_{\rm BCL}} v_k^m \frac{1}{\sqrt{\lambda_m}}v_{k^\prime}^m ~ , ~
\ee
where $v^m$ is the eigenvector of the matrix $B$ corresponding to the eigenvalue $\lambda_m$.
It follows that
\be
    \label{eq:unitarity_bk}
    \sum_{j, k = 0}^{N_{\rm BCL}} b_j B_{jk} b_k =  \sum_{k = 0}^{N_{\rm BCL}} a_k^2 = r_0^2 \leq 1 ~ . ~
\ee

Charge conservation requires that $F_\pi^V(Q^2 = 0) = 1$, which implies $b_0 = 1$ in Eq.\,(\ref{eq:BCL_app}). Thus, the unitary constraint\,(\ref{eq:unitarity_BCL_app}) should be evaluated putting $b_0 = 1$. In the case the unitary sum exceeds unity, the coefficients $b_k$ with $k = 1, 2, ... N_{\rm BCL}$ can be multiplied by a common factor chosen to ensure that the unitary sum is equal to unity, while $b_0$ is kept equal to unity.

Finally,  as described in Section\,\ref{sec:BCLfit}, the inclusion of the analytical constraint at the annihilation threshold $z = -1$ corresponds to add in Eq.\,(\ref{eq:BCL_app}) a further monomial term $b_{N_{\rm BCL} + 1} z^{N_{\rm BCL} + 1}$ with the value of the coefficient $b_{N_{\rm BCL} + 1}$ fixed by those of the coefficients $b_k$ with $k = 1, 2, ... N_{\rm BCL}$. Such an addition requires to re-evaluate the unitary constraint\,(\ref{eq:unitarity_BCL_app}) with $N_{\rm BCL}$ replaced by $N_{\rm BCL} + 1$. In the case the new unitary sum exceeds unity, the coefficients $b_k$ with $k = 1, 2, ... N_{\rm BCL} + 1$ can be multiplied by a common factor to ensure that the unitary sum is equal to unity, while the analytical constraint remains fulfilled.

The procedure described in this Appendix is not limited to the case of the em pion form factor, but it can be applied to a generic hadronic form factor which should fulfill a unitary constraint.

\bibliography{biblio}

\providecommand{\href}[2]{#2}\begingroup\raggedright\begin{thebibliography}{10}

\bibitem{DiCarlo:2021dzg}
M.~Di~Carlo, G.~Martinelli, M.~Naviglio, F.~Sanfilippo, S.~Simula and
  L.~Vittorio, \emph{{Unitarity bounds for semileptonic decays in lattice
  QCD}}, \href{http://dx.doi.org/10.1103/PhysRevD.104.054502}{\emph{Phys. Rev.
  D} {\bf 104} (2021) 054502}, [\href{http://arxiv.org/abs/2105.02497}{{\tt
  2105.02497}}].

\bibitem{ParticleDataGroup:2022pth}
{\scshape Particle Data Group} collaboration, R.~L. Workman et~al.,
  \emph{{Review of Particle Physics}},
  \href{http://dx.doi.org/10.1093/ptep/ptac097}{\emph{PTEP} {\bf 2022} (2022)
  083C01}.

\bibitem{Aoyama:2020ynm}
T.~Aoyama et~al., \emph{{The anomalous magnetic moment of the muon in the
  Standard Model}},
  \href{http://dx.doi.org/10.1016/j.physrep.2020.07.006}{\emph{Phys. Rept.}
  {\bf 887} (2020) 1--166}, [\href{http://arxiv.org/abs/2006.04822}{{\tt
  2006.04822}}].

\bibitem{Dally:1982zk}
E.~B. Dally et~al., \emph{{Elastic Scattering Measurement of the Negative Pion
  Radius}}, \href{http://dx.doi.org/10.1103/PhysRevLett.48.375}{\emph{Phys.
  Rev. Lett.} {\bf 48} (1982) 375--378}.

\bibitem{NA7:1986vav}
{\scshape NA7} collaboration, S.~R. Amendolia et~al., \emph{{A Measurement of
  the Space - Like Pion Electromagnetic Form-Factor}},
  \href{http://dx.doi.org/10.1016/0550-3213(86)90437-2}{\emph{Nucl. Phys. B}
  {\bf 277} (1986) 168}.

\bibitem{SELEX:2001fbx}
{\scshape SELEX} collaboration, I.~M. Gough~Eschrich et~al., \emph{{Measurement
  of the Sigma- Charge Radius by Sigma- Electron Elastic Scattering}},
  \href{http://dx.doi.org/10.1016/S0370-2693(01)01285-0}{\emph{Phys. Lett. B}
  {\bf 522} (2001) 233--239}, [\href{http://arxiv.org/abs/hep-ex/0106053}{{\tt
  hep-ex/0106053}}].

\bibitem{Bebek:1977pe}
C.~J. Bebek et~al., \emph{{Electroproduction of single pions at low epsilon and
  a measurement of the pion form-factor up to $q^2$ = 10-GeV$^2$}},
  \href{http://dx.doi.org/10.1103/PhysRevD.17.1693}{\emph{Phys. Rev. D} {\bf
  17} (1978) 1693}.

\bibitem{Ackermann:1977rp}
H.~Ackermann, T.~Azemoon, W.~Gabriel, H.~D. Mertiens, H.~D. Reich, G.~Specht
  et~al., \emph{{Determination of the Longitudinal and the Transverse Part in
  pi+ Electroproduction}},
  \href{http://dx.doi.org/10.1016/0550-3213(78)90523-0}{\emph{Nucl. Phys. B}
  {\bf 137} (1978) 294--300}.

\bibitem{Brauel:1979zk}
P.~Brauel, T.~Canzler, D.~Cords, R.~Felst, G.~Grindhammer, M.~Helm et~al.,
  \emph{{Electroproduction of $\pi^+ n$, $\pi^- p$ and $K^+ \Lambda$, $K^+
  \Sigma^0$ Final States Above the Resonance Region}},
  \href{http://dx.doi.org/10.1007/BF01443698}{\emph{Z. Phys. C} {\bf 3} (1979)
  101}.

\bibitem{JeffersonLabFpi:2000nlc}
{\scshape Jefferson Lab F(pi)} collaboration, J.~Volmer et~al.,
  \emph{{Measurement of the Charged Pion Electromagnetic Form-Factor}},
  \href{http://dx.doi.org/10.1103/PhysRevLett.86.1713}{\emph{Phys. Rev. Lett.}
  {\bf 86} (2001) 1713--1716},
  [\href{http://arxiv.org/abs/nucl-ex/0010009}{{\tt nucl-ex/0010009}}].

\bibitem{JeffersonLabFpi-2:2006ysh}
{\scshape Jefferson Lab F(pi)-2} collaboration, T.~Horn et~al.,
  \emph{{Determination of the Charged Pion Form Factor at Q**2 = 1.60 and
  2.45-(GeV/c)**2}},
  \href{http://dx.doi.org/10.1103/PhysRevLett.97.192001}{\emph{Phys. Rev.
  Lett.} {\bf 97} (2006) 192001},
  [\href{http://arxiv.org/abs/nucl-ex/0607005}{{\tt nucl-ex/0607005}}].

\bibitem{JeffersonLabFpi:2007vir}
{\scshape Jefferson Lab F(pi)} collaboration, V.~Tadevosyan et~al.,
  \emph{{Determination of the pion charge form-factor for Q**2 = 0.60-GeV**2 -
  1.60-GeV**2}},
  \href{http://dx.doi.org/10.1103/PhysRevC.75.055205}{\emph{Phys. Rev. C} {\bf
  75} (2007) 055205}, [\href{http://arxiv.org/abs/nucl-ex/0607007}{{\tt
  nucl-ex/0607007}}].

\bibitem{JeffersonLab:2008jve}
{\scshape Jefferson Lab} collaboration, G.~M. Huber et~al., \emph{{Charged pion
  form-factor between Q**2 = 0.60-GeV**2 and 2.45-GeV**2. II. Determination of,
  and results for, the pion form-factor}},
  \href{http://dx.doi.org/10.1103/PhysRevC.78.045203}{\emph{Phys. Rev. C} {\bf
  78} (2008) 045203}, [\href{http://arxiv.org/abs/0809.3052}{{\tt 0809.3052}}].

\bibitem{JeffersonLab:2008gyl}
{\scshape Jefferson Lab} collaboration, H.~P. Blok et~al., \emph{{Charged pion
  form factor between $Q^2$=0.60 and 2.45 GeV$^2$. I. Measurements of the cross
  section for the ${^1}$H($e,e'\pi^+$)$n$ reaction}},
  \href{http://dx.doi.org/10.1103/PhysRevC.78.045202}{\emph{Phys. Rev. C} {\bf
  78} (2008) 045202}, [\href{http://arxiv.org/abs/0809.3161}{{\tt 0809.3161}}].

\bibitem{Ananthanarayan:2016mns}
B.~Ananthanarayan, I.~Caprini, D.~Das and I.~Sentitemsu~Imsong, \emph{{Precise
  determination of the low-energy hadronic contribution to the muon $g-2$ from
  analyticity and unitarity: An improved analysis}},
  \href{http://dx.doi.org/10.1103/PhysRevD.93.116007}{\emph{Phys. Rev. D} {\bf
  93} (2016) 116007}, [\href{http://arxiv.org/abs/1605.00202}{{\tt
  1605.00202}}].

\bibitem{Colangelo:2018mtw}
G.~Colangelo, M.~Hoferichter and P.~Stoffer, \emph{{Two-pion contribution to
  hadronic vacuum polarization}},
  \href{http://dx.doi.org/10.1007/JHEP02(2019)006}{\emph{JHEP} {\bf 02} (2019)
  006}, [\href{http://arxiv.org/abs/1810.00007}{{\tt 1810.00007}}].

\bibitem{Kahane:1964zz}
J.~Kahane, \emph{{Radiative Corrections to pi-e Scattering}},
  \href{http://dx.doi.org/10.1103/PhysRev.135.B975}{\emph{Phys. Rev.} {\bf 135}
  (1964) B975--B1004}.

\bibitem{Adylov:1977kj}
G.~T. Adylov et~al., \emph{{A Measurement of the Electromagnetic Size of the
  Pion from Direct Elastic Pion Scattering Data at 50-GeV/c}},
  \href{http://dx.doi.org/10.1016/0550-3213(77)90056-6}{\emph{Nucl. Phys. B}
  {\bf 128} (1977) 461--505}.

\bibitem{Ananthanarayan:2017efc}
B.~Ananthanarayan, I.~Caprini and D.~Das, \emph{{Electromagnetic charge radius
  of the pion at high precision}},
  \href{http://dx.doi.org/10.1103/PhysRevLett.119.132002}{\emph{Phys. Rev.
  Lett.} {\bf 119} (2017) 132002}, [\href{http://arxiv.org/abs/1706.04020}{{\tt
  1706.04020}}].

\bibitem{Masjuan:2008fv}
P.~Masjuan, S.~Peris and J.~J. Sanz-Cillero, \emph{{Vector Meson Dominance as a
  first step in a systematic approximation: The Pion vector form-factor}},
  \href{http://dx.doi.org/10.1103/PhysRevD.78.074028}{\emph{Phys. Rev. D} {\bf
  78} (2008) 074028}, [\href{http://arxiv.org/abs/0807.4893}{{\tt 0807.4893}}].

\bibitem{Martinelli:2021frl}
G.~Martinelli, S.~Simula and L.~Vittorio, \emph{{Constraints for the
  semileptonic B\textrightarrow{}D(*) form factors from lattice QCD simulations
  of two-point correlation functions}},
  \href{http://dx.doi.org/10.1103/PhysRevD.104.094512}{\emph{Phys. Rev. D} {\bf
  104} (2021) 094512}, [\href{http://arxiv.org/abs/2105.07851}{{\tt
  2105.07851}}].

\bibitem{Martinelli:2021myh}
G.~Martinelli, S.~Simula and L.~Vittorio, \emph{{Exclusive determinations of
  $\vert V_{cb} \vert $ and $R(D^{*})$ through unitarity}},
  \href{http://dx.doi.org/10.1140/epjc/s10052-022-11050-0}{\emph{Eur. Phys. J.
  C} {\bf 82} (2022) 1083}, [\href{http://arxiv.org/abs/2109.15248}{{\tt
  2109.15248}}].

\bibitem{Martinelli:2021onb}
G.~Martinelli, S.~Simula and L.~Vittorio, \emph{{$\vert V_{cb} \vert$ and
  $R(D)^{(*)}$) using lattice QCD and unitarity}},
  \href{http://dx.doi.org/10.1103/PhysRevD.105.034503}{\emph{Phys. Rev. D} {\bf
  105} (2022) 034503}, [\href{http://arxiv.org/abs/2105.08674}{{\tt
  2105.08674}}].

\bibitem{Martinelli:2022tte}
G.~Martinelli, S.~Simula and L.~Vittorio, \emph{{Exclusive semileptonic B
  \textrightarrow{} \ensuremath{\pi}\ensuremath{\ell}\ensuremath{\nu_\ell} and
  B$_{s}$ \textrightarrow{} K\ensuremath{\ell}\ensuremath{\nu_\ell} decays
  through unitarity and lattice QCD}},
  \href{http://dx.doi.org/10.1007/JHEP08(2022)022}{\emph{JHEP} {\bf 08} (2022)
  022}, [\href{http://arxiv.org/abs/2202.10285}{{\tt 2202.10285}}].

\bibitem{Martinelli:2022xir}
G.~Martinelli, M.~Naviglio, S.~Simula and L.~Vittorio, \emph{{|Vcb|, lepton
  flavor universality and SU(3)F symmetry breaking in
  Bs\textrightarrow{}Ds(*)\ensuremath{\ell}\ensuremath{\nu}\ensuremath{\ell}
  decays through unitarity and lattice QCD}},
  \href{http://dx.doi.org/10.1103/PhysRevD.106.093002}{\emph{Phys. Rev. D} {\bf
  106} (2022) 093002}, [\href{http://arxiv.org/abs/2204.05925}{{\tt
  2204.05925}}].

\bibitem{Boyd:1997kz}
C.~G. Boyd, B.~Grinstein and R.~F. Lebed, \emph{{Precision corrections to
  dispersive bounds on form-factors}},
  \href{http://dx.doi.org/10.1103/PhysRevD.56.6895}{\emph{Phys. Rev. D} {\bf
  56} (1997) 6895--6911}, [\href{http://arxiv.org/abs/hep-ph/9705252}{{\tt
  hep-ph/9705252}}].

\bibitem{Bourrely:2008za}
C.~Bourrely, I.~Caprini and L.~Lellouch, \emph{{Model-independent description
  of B ---\ensuremath{>} pi l nu decays and a determination of |V(ub)|}},
  \href{http://dx.doi.org/10.1103/PhysRevD.82.099902}{\emph{Phys. Rev. D} {\bf
  79} (2009) 013008}, [\href{http://arxiv.org/abs/0807.2722}{{\tt 0807.2722}}].

\bibitem{Ananthanarayan:2012tn}
B.~Ananthanarayan, I.~Caprini and I.~S. Imsong, \emph{{Spacelike pion form
  factor from analytic continuation and the onset of perturbative QCD}},
  \href{http://dx.doi.org/10.1103/PhysRevD.85.096006}{\emph{Phys. Rev. D} {\bf
  85} (2012) 096006}, [\href{http://arxiv.org/abs/1203.5398}{{\tt 1203.5398}}].

\bibitem{Ananthanarayan:2013dpa}
B.~Ananthanarayan, I.~Caprini, D.~Das and I.~Sentitemsu~Imsong,
  \emph{{Parametrisation-free determination of the shape parameters for the
  pion electromagnetic form factor}},
  \href{http://dx.doi.org/10.1140/epjc/s10052-013-2520-9}{\emph{Eur. Phys. J.
  C} {\bf 73} (2013) 2520}, [\href{http://arxiv.org/abs/1302.6373}{{\tt
  1302.6373}}].

\bibitem{Ananthanarayan:2013zua}
B.~Ananthanarayan, I.~Caprini, D.~Das and I.~Sentitemsu~Imsong, \emph{{Two-pion
  low-energy contribution to the muon g\ensuremath{-}2 with improved precision
  from analyticity and unitarity}},
  \href{http://dx.doi.org/10.1103/PhysRevD.89.036007}{\emph{Phys. Rev. D} {\bf
  89} (2014) 036007}, [\href{http://arxiv.org/abs/1312.5849}{{\tt 1312.5849}}].

\bibitem{Ananthanarayan:2018nyx}
B.~Ananthanarayan, I.~Caprini and D.~Das, \emph{{Pion electromagnetic form
  factor at high precision with implications to $a_\mu^{\pi\pi}$ and the onset
  of perturbative QCD}},
  \href{http://dx.doi.org/10.1103/PhysRevD.98.114015}{\emph{Phys. Rev. D} {\bf
  98} (2018) 114015}, [\href{http://arxiv.org/abs/1810.09265}{{\tt
  1810.09265}}].

\bibitem{Lepage:1979zb}
G.~P. Lepage and S.~J. Brodsky, \emph{{Exclusive Processes in Quantum
  Chromodynamics: Evolution Equations for Hadronic Wave Functions and the
  Form-Factors of Mesons}},
  \href{http://dx.doi.org/10.1016/0370-2693(79)90554-9}{\emph{Phys. Lett. B}
  {\bf 87} (1979) 359--365}.

\bibitem{Efremov:1979qk}
A.~V. Efremov and A.~V. Radyushkin, \emph{{Factorization and Asymptotical
  Behavior of Pion Form-Factor in QCD}},
  \href{http://dx.doi.org/10.1016/0370-2693(80)90869-2}{\emph{Phys. Lett. B}
  {\bf 94} (1980) 245--250}.

\bibitem{Chernyak:1977as}
V.~L. Chernyak and A.~R. Zhitnitsky, \emph{{Asymptotic Behavior of Hadron
  Form-Factors in Quark Model. (In Russian)}}, {\emph{JETP Lett.} {\bf 25}
  (1977) 510}.

\bibitem{Farrar:1979aw}
G.~R. Farrar and D.~R. Jackson, \emph{{The Pion Form-Factor}},
  \href{http://dx.doi.org/10.1103/PhysRevLett.43.246}{\emph{Phys. Rev. Lett.}
  {\bf 43} (1979) 246}.

\bibitem{CMD-3:2023alj}
{\scshape CMD-3} collaboration, F.~V. Ignatov et~al., \emph{{Measurement of the
  $e^+e^-\to\pi^+\pi^-$ cross section from threshold to 1.2 GeV with the CMD-3
  detector}},  \href{http://arxiv.org/abs/2302.08834}{{\tt 2302.08834}}.

\bibitem{Lellouch:1995yv}
L.~Lellouch, \emph{{Lattice constrained unitarity bounds for anti-B0
  ---\ensuremath{>} pi+ lepton- anti-lepton-neutrino decays}},
  \href{http://dx.doi.org/10.1016/0550-3213(96)00443-9}{\emph{Nucl. Phys. B}
  {\bf 479} (1996) 353--391}, [\href{http://arxiv.org/abs/hep-ph/9509358}{{\tt
  hep-ph/9509358}}].

\bibitem{Boyd:1994tt}
C.~G. Boyd, B.~Grinstein and R.~F. Lebed, \emph{{Constraints on form-factors
  for exclusive semileptonic heavy to light meson decays}},
  \href{http://dx.doi.org/10.1103/PhysRevLett.74.4603}{\emph{Phys. Rev. Lett.}
  {\bf 74} (1995) 4603--4606}, [\href{http://arxiv.org/abs/hep-ph/9412324}{{\tt
  hep-ph/9412324}}].

\bibitem{Caprini:1997mu}
I.~Caprini, L.~Lellouch and M.~Neubert, \emph{{Dispersive bounds on the shape
  of anti-B ---\ensuremath{>} D(*) lepton anti-neutrino form-factors}},
  \href{http://dx.doi.org/10.1016/S0550-3213(98)00350-2}{\emph{Nucl. Phys. B}
  {\bf 530} (1998) 153--181}, [\href{http://arxiv.org/abs/hep-ph/9712417}{{\tt
  hep-ph/9712417}}].

\bibitem{Bourrely:1980gp}
C.~Bourrely, B.~Machet and E.~de~Rafael, \emph{{Semileptonic Decays of
  Pseudoscalar Particles ($M \to M^\prime \ell \nu_\ell$) and Short Distance
  Behavior of Quantum Chromodynamics}},
  \href{http://dx.doi.org/10.1016/0550-3213(81)90086-9}{\emph{Nucl. Phys. B}
  {\bf 189} (1981) 157--181}.

\bibitem{Simula:2021yvm}
S.~Simula, G.~Martinelli, M.~Naviglio and L.~Vittorio, \emph{{Exclusive B-meson
  semileptonic decays from unitarity and lattice QCD}},
  \href{http://dx.doi.org/10.22323/1.411.0045}{\emph{PoS} {\bf CKM2021} (2021)
  045}, [\href{http://arxiv.org/abs/2203.16213}{{\tt 2203.16213}}].

\bibitem{Bigi:2017njr}
D.~Bigi, P.~Gambino and S.~Schacht, \emph{{A fresh look at the determination of
  $|V_{cb}|$ from $B\to D^{*} \ell \nu$}},
  \href{http://dx.doi.org/10.1016/j.physletb.2017.04.022}{\emph{Phys. Lett. B}
  {\bf 769} (2017) 441--445}, [\href{http://arxiv.org/abs/1703.06124}{{\tt
  1703.06124}}].

\bibitem{Flynn:2023qmi}
J.~M. Flynn, A.~J\"uttner and J.~T. Tsang, \emph{{Bayesian inference for
  form-factor fits regulated by unitarity and analyticity}},
  \href{http://arxiv.org/abs/2303.11285}{{\tt 2303.11285}}.

\bibitem{Buck:1998kp}
W.~W. Buck and R.~F. Lebed, \emph{{New constraints on dispersive form-factor
  parameterizations from the timelike region}},
  \href{http://dx.doi.org/10.1103/PhysRevD.58.056001}{\emph{Phys. Rev. D} {\bf
  58} (1998) 056001}, [\href{http://arxiv.org/abs/hep-ph/9802369}{{\tt
  hep-ph/9802369}}].

\bibitem{Giusti:2017jof}
D.~Giusti, V.~Lubicz, G.~Martinelli, F.~Sanfilippo and S.~Simula,
  \emph{{Strange and charm HVP contributions to the muon ($g - 2)$ including
  QED corrections with twisted-mass fermions}},
  \href{http://dx.doi.org/10.1007/JHEP10(2017)157}{\emph{JHEP} {\bf 10} (2017)
  157}, [\href{http://arxiv.org/abs/1707.03019}{{\tt 1707.03019}}].

\bibitem{Giusti:2018mdh}
D.~Giusti, F.~Sanfilippo and S.~Simula, \emph{{Light-quark contribution to the
  leading hadronic vacuum polarization term of the muon $g-2$ from twisted-mass
  fermions}}, \href{http://dx.doi.org/10.1103/PhysRevD.98.114504}{\emph{Phys.
  Rev. D} {\bf 98} (2018) 114504}, [\href{http://arxiv.org/abs/1808.00887}{{\tt
  1808.00887}}].

\bibitem{Luscher:1985dn}
M.~Luscher, \emph{{Volume Dependence of the Energy Spectrum in Massive Quantum
  Field Theories. 1. Stable Particle States}},
  \href{http://dx.doi.org/10.1007/BF01211589}{\emph{Commun. Math. Phys.} {\bf
  104} (1986) 177}.

\bibitem{Luscher:1986pf}
M.~Luscher, \emph{{Volume Dependence of the Energy Spectrum in Massive Quantum
  Field Theories. 2. Scattering States}},
  \href{http://dx.doi.org/10.1007/BF01211097}{\emph{Commun. Math. Phys.} {\bf
  105} (1986) 153--188}.

\bibitem{Luscher:1990ux}
M.~Luscher, \emph{{Two particle states on a torus and their relation to the
  scattering matrix}},
  \href{http://dx.doi.org/10.1016/0550-3213(91)90366-6}{\emph{Nucl. Phys. B}
  {\bf 354} (1991) 531--578}.

\bibitem{Luscher:1991cf}
M.~Luscher, \emph{{Signatures of unstable particles in finite volume}},
  \href{http://dx.doi.org/10.1016/0550-3213(91)90584-K}{\emph{Nucl. Phys. B}
  {\bf 364} (1991) 237--251}.

\bibitem{Lellouch:2000pv}
L.~Lellouch and M.~Luscher, \emph{{Weak transition matrix elements from finite
  volume correlation functions}},
  \href{http://dx.doi.org/10.1007/s002200100410}{\emph{Commun. Math. Phys.}
  {\bf 219} (2001) 31--44}, [\href{http://arxiv.org/abs/hep-lat/0003023}{{\tt
  hep-lat/0003023}}].

\bibitem{Meyer:2011um}
H.~B. Meyer, \emph{{Lattice QCD and the Timelike Pion Form Factor}},
  \href{http://dx.doi.org/10.1103/PhysRevLett.107.072002}{\emph{Phys. Rev.
  Lett.} {\bf 107} (2011) 072002}, [\href{http://arxiv.org/abs/1105.1892}{{\tt
  1105.1892}}].

\bibitem{Francis:2013fzp}
A.~Francis, B.~Jaeger, H.~B. Meyer and H.~Wittig, \emph{{A new representation
  of the Adler function for lattice QCD}},
  \href{http://dx.doi.org/10.1103/PhysRevD.88.054502}{\emph{Phys. Rev. D} {\bf
  88} (2013) 054502}, [\href{http://arxiv.org/abs/1306.2532}{{\tt 1306.2532}}].

\bibitem{Gounaris:1968mw}
G.~J. Gounaris and J.~J. Sakurai, \emph{{Finite width corrections to the vector
  meson dominance prediction for $\rho \to e^+ e^-$}},
  \href{http://dx.doi.org/10.1103/PhysRevLett.21.244}{\emph{Phys. Rev. Lett.}
  {\bf 21} (1968) 244--247}.

\bibitem{Borsanyi:2016lpl}
S.~Borsanyi, Z.~Fodor, T.~Kawanai, S.~Krieg, L.~Lellouch, R.~Malak et~al.,
  \emph{{Slope and curvature of the hadronic vacuum polarization at vanishing
  virtuality from lattice QCD}},
  \href{http://dx.doi.org/10.1103/PhysRevD.96.074507}{\emph{Phys. Rev. D} {\bf
  96} (2017) 074507}, [\href{http://arxiv.org/abs/1612.02364}{{\tt
  1612.02364}}].

\bibitem{RBC:2018dos}
{\scshape RBC, UKQCD} collaboration, T.~Blum, P.~A. Boyle, V.~G\"ulpers,
  T.~Izubuchi, L.~Jin, C.~Jung et~al., \emph{{Calculation of the hadronic
  vacuum polarization contribution to the muon anomalous magnetic moment}},
  \href{http://dx.doi.org/10.1103/PhysRevLett.121.022003}{\emph{Phys. Rev.
  Lett.} {\bf 121} (2018) 022003}, [\href{http://arxiv.org/abs/1801.07224}{{\tt
  1801.07224}}].

\bibitem{FermilabLattice:2019ugu}
{\scshape Fermilab Lattice, LATTICE-HPQCD, MILC} collaboration, C.~T.~H. Davies
  et~al., \emph{{Hadronic-vacuum-polarization contribution to the
  muon\textquoteright{}s anomalous magnetic moment from four-flavor lattice
  QCD}}, \href{http://dx.doi.org/10.1103/PhysRevD.101.034512}{\emph{Phys. Rev.
  D} {\bf 101} (2020) 034512}, [\href{http://arxiv.org/abs/1902.04223}{{\tt
  1902.04223}}].

\bibitem{Brandt:2013ffb}
B.~B. Brandt, \emph{{The electromagnetic form factor of the pion: Results from
  the lattice}}, \href{http://dx.doi.org/10.1142/S0218301313300300}{\emph{Int.
  J. Mod. Phys. E} {\bf 22} (2013) 1330030},
  [\href{http://arxiv.org/abs/1310.6389}{{\tt 1310.6389}}].

\bibitem{DAgostini:1993arp}
G.~D'Agostini, \emph{{On the use of the covariance matrix to fit correlated
  data}}, \href{http://dx.doi.org/10.1016/0168-9002(94)90719-6}{\emph{Nucl.
  Instrum. Meth. A} {\bf 346} (1994) 306--311}.

\bibitem{mpfun}
D.~H. Bailey, \emph{{MPFUN2020: A thread-safe arbitrary precision package}},
  2020.

\bibitem{Colangelo:2020lcg}
G.~Colangelo, M.~Hoferichter and P.~Stoffer, \emph{{Constraints on the two-pion
  contribution to hadronic vacuum polarization}},
  \href{http://dx.doi.org/10.1016/j.physletb.2021.136073}{\emph{Phys. Lett. B}
  {\bf 814} (2021) 136073}, [\href{http://arxiv.org/abs/2010.07943}{{\tt
  2010.07943}}].

\bibitem{Colangelo:2023rqr}
G.~Colangelo, M.~Hoferichter and P.~Stoffer, \emph{{Puzzles in the hadronic
  contributions to the muon anomalous magnetic moment}},  in \emph{{21st
  Conference on Flavor Physics and CP Violation}}, 8, 2023.
\newblock \href{http://arxiv.org/abs/2308.04217}{{\tt 2308.04217}}.

\bibitem{hyperspherical}
L.~E. Blumenson, \emph{A derivation of n-dimensional spherical coordinates},
  {\emph{The American Mathematical Monthly} {\bf 67} (1960) 63--66}.

\bibitem{Leutwyler:2002hm}
H.~Leutwyler, \emph{{Electromagnetic form-factor of the pion}},  in
  \emph{{Continuous Advances in QCD 2002 / ARKADYFEST (honoring the 60th
  birthday of Prof. Arkady Vainshtein)}}, pp.~23--40, 12, 2002.
\newblock \href{http://arxiv.org/abs/hep-ph/0212324}{{\tt hep-ph/0212324}}.
\newblock \href{http://dx.doi.org/10.1142/9789812776310_0002}{DOI}.

\bibitem{Dudek:2012vr}
J.~Dudek et~al., \emph{{Physics Opportunities with the 12 GeV Upgrade at
  Jefferson Lab}},
  \href{http://dx.doi.org/10.1140/epja/i2012-12187-1}{\emph{Eur. Phys. J. A}
  {\bf 48} (2012) 187}, [\href{http://arxiv.org/abs/1208.1244}{{\tt
  1208.1244}}].

\bibitem{JLAB-E12-06-101}
D.~Gaskell, T.~Horn and G.~Huber, ``{Update on E12-06-101: Measurement of the
  Charged Pion Form Factor to High $Q^2$ and E12-07-105: Scaling Study of the
  L-T Separated Pion Electroproduction Cross Section at 11 GeV}.''
  $\mbox{See~https://www.jlab.org/exp$\_$prog/proposals/19/E12-19-006.pdf}$.

\bibitem{Accardi:2012qut}
A.~Accardi et~al., \emph{{Electron Ion Collider: The Next QCD Frontier}:
  {Understanding the glue that binds us all}},
  \href{http://dx.doi.org/10.1140/epja/i2016-16268-9}{\emph{Eur. Phys. J. A}
  {\bf 52} (2016) 268}, [\href{http://arxiv.org/abs/1212.1701}{{\tt
  1212.1701}}].

\bibitem{Anderle:2021wcy}
D.~P. Anderle et~al., \emph{{Electron-ion collider in China}},
  \href{http://dx.doi.org/10.1007/s11467-021-1062-0}{\emph{Front. Phys.
  (Beijing)} {\bf 16} (2021) 64701},
  [\href{http://arxiv.org/abs/2102.09222}{{\tt 2102.09222}}].

\bibitem{Horn:2016rip}
T.~Horn and C.~D. Roberts, \emph{{The pion: an enigma within the Standard
  Model}}, \href{http://dx.doi.org/10.1088/0954-3899/43/7/073001}{\emph{J.
  Phys. G} {\bf 43} (2016) 073001},
  [\href{http://arxiv.org/abs/1602.04016}{{\tt 1602.04016}}].

\bibitem{FlavourLatticeAveragingGroupFLAG:2021npn}
{\scshape Flavour Lattice Averaging Group (FLAG)} collaboration, Y.~Aoki
  et~al., \emph{{FLAG Review 2021}},
  \href{http://dx.doi.org/10.1140/epjc/s10052-022-10536-1}{\emph{Eur. Phys. J.
  C} {\bf 82} (2022) 869}, [\href{http://arxiv.org/abs/2111.09849}{{\tt
  2111.09849}}].

\bibitem{Chang:2013nia}
L.~Chang, I.~C. Clo\"et, C.~D. Roberts, S.~M. Schmidt and P.~C. Tandy,
  \emph{{Pion electromagnetic form factor at spacelike momenta}},
  \href{http://dx.doi.org/10.1103/PhysRevLett.111.141802}{\emph{Phys. Rev.
  Lett.} {\bf 111} (2013) 141802}, [\href{http://arxiv.org/abs/1307.0026}{{\tt
  1307.0026}}].

\bibitem{Ydrefors:2021dwa}
E.~Ydrefors, W.~de~Paula, J.~H.~A. Nogueira, T.~Frederico and G.~Salm\'e,
  \emph{{Pion electromagnetic form factor with Minkowskian dynamics}},
  \href{http://dx.doi.org/10.1016/j.physletb.2021.136494}{\emph{Phys. Lett. B}
  {\bf 820} (2021) 136494}, [\href{http://arxiv.org/abs/2106.10018}{{\tt
  2106.10018}}].

\bibitem{Gao:2022vyh}
X.~Gao, A.~D. Hanlon, N.~Karthik, S.~Mukherjee, P.~Petreczky, P.~Scior et~al.,
  \emph{{Pion distribution amplitude at the physical point using the
  leading-twist expansion of the quasi-distribution-amplitude matrix element}},
  \href{http://dx.doi.org/10.1103/PhysRevD.106.074505}{\emph{Phys. Rev. D} {\bf
  106} (2022) 074505}, [\href{http://arxiv.org/abs/2206.04084}{{\tt
  2206.04084}}].

\bibitem{LatticeParton:2022zqc}
{\scshape Lattice Parton} collaboration, J.~Hua et~al., \emph{{Pion and Kaon
  Distribution Amplitudes from Lattice QCD}},
  \href{http://dx.doi.org/10.1103/PhysRevLett.129.132001}{\emph{Phys. Rev.
  Lett.} {\bf 129} (2022) 132001}, [\href{http://arxiv.org/abs/2201.09173}{{\tt
  2201.09173}}].

\bibitem{Holligan:2023rex}
J.~Holligan, X.~Ji, H.-W. Lin, Y.~Su and R.~Zhang, \emph{{Precision control in
  lattice calculation of x-dependent pion distribution amplitude}},
  \href{http://dx.doi.org/10.1016/j.nuclphysb.2023.116282}{\emph{Nucl. Phys. B}
  {\bf 993} (2023) 116282}, [\href{http://arxiv.org/abs/2301.10372}{{\tt
  2301.10372}}].

\bibitem{Bharucha:2010im}
A.~Bharucha, T.~Feldmann and M.~Wick, \emph{{Theoretical and Phenomenological
  Constraints on Form Factors for Radiative and Semi-Leptonic B-Meson Decays}},
  \href{http://dx.doi.org/10.1007/JHEP09(2010)090}{\emph{JHEP} {\bf 09} (2010)
  090}, [\href{http://arxiv.org/abs/1004.3249}{{\tt 1004.3249}}].

\end{thebibliography}\endgroup
\bibliographystyle{JHEP}

\end{document}